
%
%

\font\twelverm=cmr10 scaled 1200    \font\twelvei=cmmi10 scaled 1200
\font\twelvesy=cmsy10 scaled 1200   \font\twelveex=cmex10 scaled 1200
\font\twelvebf=cmbx10 scaled 1200   \font\twelvesl=cmsl10 scaled 1200
\font\twelvett=cmtt10 scaled 1200   \font\twelveit=cmti10 scaled 1200

\skewchar\twelvei='177   \skewchar\twelvesy='60


\def\twelvepoint{\normalbaselineskip=12.4pt
  \abovedisplayskip 12.4pt plus 3pt minus 9pt
  \belowdisplayskip 12.4pt plus 3pt minus 9pt
  \abovedisplayshortskip 0pt plus 3pt
  \belowdisplayshortskip 7.2pt plus 3pt minus 4pt
  \smallskipamount=3.6pt plus1.2pt minus1.2pt
  \medskipamount=7.2pt plus2.4pt minus2.4pt
  \bigskipamount=14.4pt plus4.8pt minus4.8pt
  \def\rm{\fam0\twelverm}          \def\it{\fam\itfam\twelveit}%
  \def\sl{\fam\slfam\twelvesl}     \def\bf{\fam\bffam\twelvebf}%
  \def\mit{\fam 1}                 \def\cal{\fam 2}%
  \def\tt{\twelvett}
  \textfont0=\twelverm   \scriptfont0=\tenrm   \scriptscriptfont0=\sevenrm
  \textfont1=\twelvei    \scriptfont1=\teni    \scriptscriptfont1=\seveni
  \textfont2=\twelvesy   \scriptfont2=\tensy   \scriptscriptfont2=\sevensy
  \textfont3=\twelveex \scriptfont3=\twelveex \scriptscriptfont3=\twelveex
  \textfont\itfam=\twelveit
  \textfont\slfam=\twelvesl
  \textfont\bffam=\twelvebf \scriptfont\bffam=\tenbf
  \scriptscriptfont\bffam=\sevenbf
  \normalbaselines\rm}



\def\beginlinemode{\endmode
  \begingroup\parskip=0pt \obeylines\def\\{\par}\def\endmode{\par\endgroup}}
\def\beginparmode{\endmode
  \begingroup \def\endmode{\par\endgroup}}
\let\endmode=\par
{\obeylines\gdef\
{}}
\def\singlespace{\baselineskip=\normalbaselineskip}

\def\oneandahalfspace{\baselineskip=\normalbaselineskip
  \multiply\baselineskip by 3 \divide\baselineskip by 2}
\def\doublespace{\baselineskip=\normalbaselineskip \multiply\baselineskip by 2}

\newcount\firstpageno
\firstpageno=2
\footline={\ifnum\pageno<\firstpageno{\hfil}\else{\hfil\twelverm\folio\hfil}\fi}
\def\toppageno{\global\footline={\hfil}\global\headline
  ={\ifnum\pageno<\firstpageno{\hfil}\else{\hfil\twelverm\folio\hfil}\fi}}
\let\rawfootnote=\footnote              
\def\footnote#1#2{{\rm\singlespace\parindent=0pt\parskip=0pt
  \rawfootnote{#1}{#2\hfill\vrule height 0pt depth 6pt width 0pt}}}
\def\raggedcenter{\leftskip=4em plus 12em \rightskip=\leftskip
  \parindent=0pt \parfillskip=0pt \spaceskip=.3333em \xspaceskip=.5em
  \pretolerance=9999 \tolerance=9999
  \hyphenpenalty=9999 \exhyphenpenalty=9999 }
\def\dateline{\rightline{\ifcase\month\or
  January\or February\or March\or April\or May\or June\or
  July\or August\or September\or October\or November\or December\fi
  \space\number\year}}
\def\received{\vskip 3pt plus 0.2fill
 \centerline{\sl (Received\space\ifcase\month\or
  January\or February\or March\or April\or May\or June\or
  July\or August\or September\or October\or November\or December\fi
  \qquad, \number\year)}}


\hsize=6.5truein
\hoffset=0.05truein    
\vsize=8.9truein
\voffset=0.05truein   
\parskip=\medskipamount
\def\\{\cr}
\twelvepoint            
\doublespace            
\overfullrule=0pt       





\def\title                      
  {\null\vskip 3pt plus 0.1fill
   \beginlinemode \doublespace \raggedcenter \bf}

\def\author                     
  {\vskip 3pt plus 0.25fill \beginlinemode
   \singlespace \raggedcenter}

\def\affil                      
  {\vskip 3pt plus 0.1fill \beginlinemode
   \oneandahalfspace \raggedcenter \sl}

\def\abstract                   
  {\vskip 3pt plus 0.2fill \beginparmode
   \oneandahalfspace ABSTRACT: }

\def\resume                   
  {\vskip 3pt plus 0.2fill \beginparmode
   \oneandahalfspace RESUME: }

\def\endtopmatter               
  {\endpage                     
   \body}

\def\body                       
  {\beginparmode}               

\def\head#1{                    
  \goodbreak\vskip 0.5truein    
  {\immediate\write16{#1}
   \raggedcenter \uppercase{#1}\par}
   \nobreak\vskip 0.25truein\nobreak}

\def\beneathrel#1\under#2{\mathrel{\mathop{#2}\limits_{#1}}}

\def\refto#1{$^{#1}$}    

\def\references         
  {\head{References}      
   \beginparmode
   \frenchspacing \parindent=0pt \leftskip=1truecm
   \parskip=8pt plus 3pt \everypar{\hangindent=\parindent}}

\def\bibliog         
  {      
   \beginparmode
   \frenchspacing \parindent=0pt \leftskip=1truecm
   \parskip=8pt plus 3pt \everypar{\hangindent=\parindent}}

\gdef\refis#1{\item{#1.\ }}                     

\gdef\journal#1, #2, #3, 1#4#5#6{       
    {\sl #1~}{\bf #2}, #3 (1#4#5#6)}        

\def\pr{\journal Phys. Rev., }

\def\prb{\journal Phys. Rev. B, }

\def\prl{\journal Phys. Rev. Lett., }

\def\rmp{\journal Rev. Mod. Phys., }

\def\pl{\journal Phys. Lett., }

\def\jpc{\journal J. Phys. C, }

\def\jetp{\journal Sov. Phys. JETP, }

\def\jetl{\journal Sov. Phys. JETP Letters, }

\def\ssc{\journal Solid State Commun., }

\def\zpb{\journal Zeit. Phys. B., }

\def\jdp{\journal J.  Phys. (Paris), }

\def\jdc{\journal J.  Phys. (Paris) Colloq., }

\def\jpjap{\journal J. Phys. Soc. Japan, }

\def\physica{\journal Physica B, }

\def\phl{\journal Phil. Mag.,}

\def\jmmm{\journal J. Mag. Mat., }

\def\endreferences{\body}

\def\figurecaptions             
  {\endpage
   \beginparmode
   \head{Figure Captions}
}

\def\endfigurecaptions{\body}

\def\tablecaptions             
  {\endpage
   \beginparmode
   \head{Table Captions}
}

\def\endpage                    
  {\vfill\eject}

\def\endpaper                   
  {\endmode\vfill\supereject}

\def\endit
  {\endpaper\end}


\def\heading                            
  {\vskip 0.5truein plus 0.1truein      
   \beginparmode \def\\{\par} \parskip=0pt \singlespace \raggedcenter}

\def\subheading                         
  {\vskip 0.25truein plus 0.1truein     
   \beginlinemode \singlespace \parskip=0pt \def\\{\par}\raggedcenter}

\def\tag#1$${\eqno(#1)$$}

\def\align#1$${\eqalign{#1}$$}

\def\aligntag#1$${\gdef\tag##1\\{&(##1)\cr}\eqalignno{#1\\}$$
  \gdef\tag##1$${\eqno(##1)$$}}

\def\endaligntag{}

\def\overset#1\to#2{{\mathop{#2}^{#1}}}
\def\underset#1\to#2{{\mathop{#2}_{#1}}}


\def\ref#1{Ref.~#1}                     
\def\Ref#1{Ref.~#1}                     
\def\[#1]{[\cite{#1}]}
\def\cite#1{{#1}}
\def\(#1){(\call{#1})}
\def\call#1{{#1}}
\def\taghead#1{}
\def\frac#1#2{{#1 \over #2}}

\def\12{{1\over2}}

\def\sla{\raise.15ex\hbox{$/$}\kern-.57em}
\def\leaderfill{\leaders\hbox to 1em{\hss.\hss}\hfill}
\def\twiddle{\lower.9ex\rlap{$\kern-.1em\scriptstyle\sim$}}
\def\bigtwiddle{\lower1.ex\rlap{$\sim$}}
\def\gtwid{\mathrel{\raise.3ex\hbox{$>$\kern-.75em\lower1ex\hbox{$\sim$}}}}
\def\ltwid{\mathrel{\raise.3ex\hbox{$<$\kern-.75em\lower1ex\hbox{$\sim$}}}}
\def\square{\kern1pt\vbox{
\hrule height 1.2pt\hbox{\vrule width 1.2pt\hskip 3pt
\vbox{\vskip 6pt}\hskip 3pt\vrule width 0.6pt}\hrule height 0.6pt}\kern1pt}
\def\tdot#1{\mathord{\mathop{#1}\limits^{\kern2pt\ldots}}}

\def\pmb#1{\setbox0=\hbox{#1}%
  \kern-.025em\copy0\kern-\wd0
  \kern  .05em\copy0\kern-\wd0
  \kern-.025em\raise.0433em\box0 }

\def \ra{\rangle}
\def\la{\langle}

\def\dg{{^
{\dag}}}

\def\ra{\rangle}
\def\la{\langle}

\def\1{{\bf 1}}
\def\2{{\bf 2}}
\def\l{\lambda}

\def\rarrow{\rightarrow}

\def\ul{\underline}

\def\vk{\vec k}
\def\iom{i \omega _n}
\def\ell{{\it l } {\rm n}}

\def\si{\sigma}

\def\cx2{\sqrt{c^2_x+c^2_y}}

\def\gkk{\gamma _{\vec k}}
\def\gk2{\gkk ^2}
\def\dw{\downarrow}
\def\up{\uparrow}
\def\gtappr{{{\lower4pt\hbox{$>$} } \atop \widetilde{ \ \ \ }}}
\def\ltappr{{{\lower4pt\hbox{$<$} } \atop \widetilde{ \ \ \ }}}

\def\dsp{\displaystyle}
\def\pbar{{\partial\kern-1.2ex\raise0.25ex\hbox{/}}}

\def\up{\uparrow}
\def\dw{\downarrow}

\def\dsp{\displaystyle}

\def\dg{{^{\dag}}}

\def\ra{\rangle}
\def\la{\langle}

\def\1{{\bf 1}}
\def\2{{\bf 2}}
\def\l{\lambda}

\def\rarrow{\rightarrow}

\def\ul{\underline}

\def\vk{\vec k}
\def\vq{\vec q}
\def\iom{i \omega _n}
\def\ell{{\it l } {\rm n}}

\def\si{\sigma}

\def\cx2{\sqrt{c^2_x+c^2_y}}

\def\gkk{\gamma _{\vec k}}
\def\gk2{\gkk ^2}
\def\gtappr{{{\lower4pt\hbox{$>$} } \atop \widetilde{ \ \ \ }}}
\def\ltappr{{{\lower4pt\hbox{$<$} } \atop \widetilde{ \ \ \ }}}

\def\inu{i \nu_n}

\def\refto#1{$^{#1}$}

\def\eps{\epsilon}
\def\3he{{$^3${\rm He}}}


\def\slD{\raise.15ex\hbox{$/$}\kern-.57em\hbox{$D$}}
\def\dsl{\raise.15ex\hbox{$/$}\kern-.57em\hbox{$\Delta$}}
\def\slp{{\raise.15ex\hbox{$/$}\kern-.57em\hbox{$\partial$}}}
\def\nsl{\raise.15ex\hbox{$/$}\kern-.57em\hbox{$\nabla$}}
\def\sla{\raise.15ex\hbox{$/$}\kern-.57em\hbox{$\rightarrow$}}
\def\slla{\raise.15ex\hbox{$/$}\kern-.57em\hbox{$\lambda$}}
\def\gtwid{\raise.3ex\hbox{$>$\kern-.75em\lower1ex\hbox{$\sim$}}}
\def\ltwid{\raise.3ex\hbox{$<$\kern-.75em\lower1ex\hbox{$\sim$}}}

\def\12{{1\over2}}
\def\a{\alpha}
\def\part{\partial}
\def\l{\lambda}

\def\Si{{\Sigma}}

\def\be{{\beta}}

\def\bethlogo{\vbox{\bf \line{\hrulefill}
    \kern-.5\baselineskip
    \line{\hrulefill\phantom{ ELIZABETH A. MASON }\hrulefill}
    \kern-.5\baselineskip
    \line{\hrulefill\hbox{ ELIZABETH A. MASON }\hrulefill}
    \kern-.5\baselineskip
    \line{\hrulefill\phantom{ 1411 Chino Street }\hrulefill}
    \kern-.5\baselineskip
    \line{\hrulefill\hbox{ 1411 Chino Street }\hrulefill}
    \kern-.5\baselineskip
    \line{\hrulefill\phantom{ Santa Barbara, CA 93101 }\hrulefill}
    \kern-.5\baselineskip
    \line{\hrulefill\hbox{ Santa Barbara, CA 93101 }\hrulefill}
    \kern-.5\baselineskip
    \line{\hrulefill\phantom{ (805) 962-2739 }\hrulefill}
    \kern-.5\baselineskip
    \line{\hrulefill\hbox{ (805) 962-2739 }\hrulefill}}}
\def\lisalogo{\vbox{\bf \line{\hrulefill}
    \kern-.5\baselineskip
    \line{\hrulefill\phantom{ LISA R. GOODFRIEND }\hrulefill}
    \kern-.5\baselineskip
    \line{\hrulefill\hbox{ LISA R. GOODFRIEND }\hrulefill}
    \kern-.5\baselineskip
    \line{\hrulefill\phantom{ 6646 Pasado }\hrulefill}
    \kern-.5\baselineskip
    \line{\hrulefill\hbox{ 6646 Pasado }\hrulefill}
    \kern-.5\baselineskip
    \line{\hrulefill\phantom{ Santa Barbara, CA 93108 }\hrulefill}
    \kern-.5\baselineskip
    \line{\hrulefill\hbox{ Santa Barbara, CA 93108 }\hrulefill}
    \kern-.5\baselineskip
    \line{\hrulefill\phantom{ (805) 962-2739 }\hrulefill}
    \kern-.5\baselineskip
    \line{\hrulefill\hbox{ (805) 962-2739 }\hrulefill}}}
\def\ka{{\kappa}}

\def\l{{\lambda}}

\def\low#1{\lower.5ex\hbox{${}_#1$}}
\def\ltwid{\raise.3ex\hbox{$<$\kern-.75em\lower1ex\hbox{$\sim$}}}

\def\om{{\omega}}

\def\psl{\raise.15ex\hbox{$/$}\kern-.57em\hbox{$\partial$}}
\def\partt{\raise.15ex\hbox{$\widetilde$}{\kern-.37em\hbox{$\partial$}}}
\def\parts{\raise.15ex\hbox{$/$}{\kern-.6em\hbox{$\partial$}}}
\def\nablas{\raise.15ex\hbox{$/$}{\kern-.6em\hbox{$\nabla$}}}
\def\partw#1{\raise.15ex\hbox{$/$}{\kern-.6em\hbox{${#1}$}}}

\def\refto#1{$^{#1}$}

\def\si{{\sigma}}

\def\th{{\theta}}
\def\gtappr{{{\lower4pt\hbox{$>$} } \atop \widetilde{ \ \ \ }}}
\def\ltappr{{{\lower4pt\hbox{$<$} } \atop \widetilde{ \ \ \ }}}

\def\topppageno1{\global\footline={\hfil}\global\headline
={\ifnum\pageno<\firstpageno{\hfil}\else{\hss\twelverm --\ \folio
\ --\hss}\fi}}

\def\toppageno2{\global\footline={\hfil}\global\headline
={\ifnum\pageno<\firstpageno{\hfil}\else{\rightline{\hfill\hfill
\twelverm \ \folio
\ \hss}}\fi}}

\catcode`@=11
\newcount\tagnumber\tagnumber=0

\immediate\newwrite\eqnfile
\newif\if@qnfile\@qnfilefalse
\def\write@qn#1{}
\def\writenew@qn#1{}
\def\w@rnwrite#1{\write@qn{#1}\message{#1}}
\def\@rrwrite#1{\write@qn{#1}\errmessage{#1}}

\def\taghead#1{\gdef\t@ghead{#1}\global\tagnumber=0}
\def\t@ghead{}

\expandafter\def\csname @qnnum-3\endcsname
  {{\t@ghead\advance\tagnumber by -3\relax\number\tagnumber}}
\expandafter\def\csname @qnnum-2\endcsname
  {{\t@ghead\advance\tagnumber by -2\relax\number\tagnumber}}
\expandafter\def\csname @qnnum-1\endcsname
  {{\t@ghead\advance\tagnumber by -1\relax\number\tagnumber}}
\expandafter\def\csname @qnnum0\endcsname
  {\t@ghead\number\tagnumber}
\expandafter\def\csname @qnnum+1\endcsname
  {{\t@ghead\advance\tagnumber by 1\relax\number\tagnumber}}
\expandafter\def\csname @qnnum+2\endcsname
  {{\t@ghead\advance\tagnumber by 2\relax\number\tagnumber}}
\expandafter\def\csname @qnnum+3\endcsname
  {{\t@ghead\advance\tagnumber by 3\relax\number\tagnumber}}

\def\equationfile{%
  \@qnfiletrue\immediate\openout\eqnfile=\jobname.eqn%
  \def\write@qn##1{\if@qnfile\immediate\write\eqnfile{##1}\fi}
  \def\writenew@qn##1{\if@qnfile\immediate\write\eqnfile
    {\noexpand\tag{##1} = (\t@ghead\number\tagnumber)}\fi}
}

\def\callall#1{\xdef#1##1{#1{\noexpand\call{##1}}}}
\def\call#1{\each@rg\callr@nge{#1}}

\def\each@rg#1#2{{\let\thecsname=#1\expandafter\first@rg#2,\end,}}
\def\first@rg#1,{\thecsname{#1}\apply@rg}
\def\apply@rg#1,{\ifx\end#1\let\next=\relax%
\else,\thecsname{#1}\let\next=\apply@rg\fi\next}

\def\callr@nge#1{\calldor@nge#1-\end-}
\def\callr@ngeat#1\end-{#1}
\def\calldor@nge#1-#2-{\ifx\end#2\@qneatspace#1 %
  \else\calll@@p{#1}{#2}\callr@ngeat\fi}
\def\calll@@p#1#2{\ifnum#1>#2{\@rrwrite{Equation range #1-#2\space is bad.}
\errhelp{If you call a series of equations by the notation M-N, then M and
N must be integers, and N must be greater than or equal to M.}}\else%
 {\count0=#1\count1=#2\advance\count1
by1\relax\expandafter\@qncall\the\count0,%
  \loop\advance\count0 by1\relax%
    \ifnum\count0<\count1,\expandafter\@qncall\the\count0,%
  \repeat}\fi}

\def\@qneatspace#1#2 {\@qncall#1#2,}
\def\@qncall#1,{\ifunc@lled{#1}{\def\next{#1}\ifx\next\empty\else
  \w@rnwrite{Equation number \noexpand\(>>#1<<) has not been defined yet.}
  >>#1<<\fi}\else\csname @qnnum#1\endcsname\fi}

\let\eqnono=\eqno
\def\eqno(#1){\tag#1}
\def\tag#1$${\eqnono(\displayt@g#1 )$$}

\def\aligntag#1\endaligntag
  $${\gdef\tag##1\\{&(##1 )\cr}\eqalignno{#1\\}$$
  \gdef\tag##1$${\eqnono(\displayt@g##1 )$$}}

\def\eqalignno#1{\displ@y \tabskip\centering
  \halign to\displaywidth{\hfil$\displaystyle{##}$\tabskip\z@skip
    &$\displaystyle{{}##}$\hfil\tabskip\centering
    &\llap{$\displayt@gpar##$}\tabskip\z@skip\crcr
    #1\crcr}}

\def\displayt@gpar(#1){(\displayt@g#1 )}

\def\displayt@g#1 {\rm\ifunc@lled{#1}\global\advance\tagnumber by1
        {\def\next{#1}\ifx\next\empty\else\expandafter
        \xdef\csname @qnnum#1\endcsname{\t@ghead\number\tagnumber}\fi}%
  \writenew@qn{#1}\t@ghead\number\tagnumber\else
        {\edef\next{\t@ghead\number\tagnumber}%
        \expandafter\ifx\csname @qnnum#1\endcsname\next\else
        \w@rnwrite{Equation \noexpand\tag{#1} is a duplicate number.}\fi}%
  \csname @qnnum#1\endcsname\fi}

\def\ifunc@lled#1{\expandafter\ifx\csname @qnnum#1\endcsname\relax}

\let\@qnend=\end\gdef\end{\if@qnfile
\immediate\write16{Equation numbers written on []\jobname.EQN.}\fi\@qnend}

\catcode`@=12

\catcode`@=11
\newcount\r@fcount \r@fcount=0
\newcount\r@fcurr
\immediate\newwrite\reffile
\newif\ifr@ffile\r@ffilefalse
\def\w@rnwrite#1{\ifr@ffile\immediate\write\reffile{#1}\fi\message{#1}}

\def\writer@f#1>>{}
\def\referencefile{
  \r@ffiletrue\immediate\openout\reffile=\jobname.ref%
  \def\writer@f##1>>{\ifr@ffile\immediate\write\reffile%
    {\noexpand\refis{##1} = \csname r@fnum##1\endcsname = %
     \expandafter\expandafter\expandafter\strip@t\expandafter%
     \meaning\csname r@ftext\csname r@fnum##1\endcsname\endcsname}\fi}%
  \def\strip@t##1>>{}}

\def\citeall#1{\xdef#1##1{#1{\noexpand\cite{##1}}}}
\def\cite#1{\each@rg\citer@nge{#1}}	

\def\each@rg#1#2{{\let\thecsname=#1\expandafter\first@rg#2,\end,}}
\def\first@rg#1,{\thecsname{#1}\apply@rg}	
\def\apply@rg#1,{\ifx\end#1\let\next=\relax
\else,\thecsname{#1}\let\next=\apply@rg\fi\next}

\def\citer@nge#1{\citedor@nge#1-\end-}	
\def\citer@ngeat#1\end-{#1}
\def\citedor@nge#1-#2-{\ifx\end#2\r@featspace#1 
  \else\citel@@p{#1}{#2}\citer@ngeat\fi}	
\def\citel@@p#1#2{\ifnum#1>#2{\errmessage{Reference range #1-#2\space is bad.}%
    \errhelp{If you cite a series of references by the notation M-N, then M and
    N must be integers, and N must be greater than or equal to M.}}\else%
 {\count0=#1\count1=#2\advance\count1
by1\relax\expandafter\r@fcite\the\count0,%
  \loop\advance\count0 by1\relax
    \ifnum\count0<\count1,\expandafter\r@fcite\the\count0,%
  \repeat}\fi}

\def\r@featspace#1#2 {\r@fcite#1#2,}	
\def\r@fcite#1,{\ifuncit@d{#1}
    \newr@f{#1}%
    \expandafter\gdef\csname r@ftext\number\r@fcount\endcsname%
                     {\message{Reference #1 to be supplied.}%
                      \writer@f#1>>#1 to be supplied.\par}%
 \fi%
 \csname r@fnum#1\endcsname}
\def\ifuncit@d#1{\expandafter\ifx\csname r@fnum#1\endcsname\relax}%
\def\newr@f#1{\global\advance\r@fcount by1%
    \expandafter\xdef\csname r@fnum#1\endcsname{\number\r@fcount}}

\let\r@fis=\refis			
\def\refis#1#2#3\par{\ifuncit@d{#1}
   \newr@f{#1}%
   \w@rnwrite{Reference #1=\number\r@fcount\space is not cited up to now.}\fi%
  \expandafter\gdef\csname r@ftext\csname r@fnum#1\endcsname\endcsname%
  {\writer@f#1>>#2#3\par}}

\def\ignoreuncited{
   \def\refis##1##2##3\par{\ifuncit@d{##1}%
     \else\expandafter\gdef\csname r@ftext\csname
r@fnum##1\endcsname\endcsname%
     {\writer@f##1>>##2##3\par}\fi}}

\def\r@ferr{\endreferences\errmessage{I was expecting to see
\noexpand\endreferences before now;  I have inserted it here.}}
\let\r@ferences=\references
\def\references{\r@ferences\def\endmode{\r@ferr\par\endgroup}}

\let\endr@ferences=\endreferences
\def\endreferences{\r@fcurr=0
  {\loop\ifnum\r@fcurr<\r@fcount
    \advance\r@fcurr by1\relax\expandafter\r@fis\expandafter{\number\r@fcurr}%
    \csname r@ftext\number\r@fcurr\endcsname%
  \repeat}\gdef\r@ferr{}\endr@ferences}


\let\r@fend=\endpaper\gdef\endpaper{\ifr@ffile
\immediate\write16{Cross References written on []\jobname.REF.}\fi\r@fend}

\catcode`@=12

\def\reftorange#1#2#3{$^{\cite{#1}-\setbox0=\hbox{\cite{#2}}\cite{#3}}$}

\citeall\refto		
\citeall\ref		%
\citeall\Ref		%
\ignoreuncited
\tolerance=5000

\def\vk{{\rm  \vec k}}
\def\vq{{\rm \vec q}}
\vskip 0.2truein
\title Odd Frequency Pairing in the Kondo Lattice
\author P. Coleman$^1$, E. Miranda $^1$ and A. Tsvelik $^{2}$

\affil $^1$ Serin Physics Laboratory
Rutgers University
PO Box 849
Piscataway NJ 08855

\affil $^2$ Dept. of Physics
Oxford University
1 Keble Road
Oxford OX1 3NP
UK

\abstract{ We discuss the possibility that heavy fermion superconductors
involve odd-frequency triplet pairing.  A key technical innovation in
this discussion is a Majorana representation for the local moments
which enables the $n_f=1$ constraint in the Kondo lattice to be
handled without a Gutzwiller approximation.  Our mean field theory for
odd frequency pairing involves a condensation of local moments and
conduction electrons, and is characterized by a spinor order
parameter.  This is a stable realization of odd frequency triplet
pairing. It predicts a Fermi surface of gapless quasiparticles, whose
spin and charge coherence factors vanish {\sl linearly} in energy.
The unusual energy dependence of coherence factors leads to a $T^3$
NMR relaxation rate that coexists with a linear specific heat. A
prediction of the theory is that a Korringa relaxation will fail to
develop in heavy fermion superconductors, even in the limit of strong
pair-breaking and severe gaplessness.  }

\vskip 0.2 truein PACS Nos. 75.20.Hr, 75.30.Mb, 75.40.Gb

\endtopmatter
\body
\noindent{\bf 1. Introduction}

\taghead{1.}
Heavy fermion superconductivity has attracted
great interest in recent years as a candidate
for electronically mediated pairing.\refto{jon,aeppli3}
Six heavy fermion metals are  superconducting at
room pressure:
$CeCu_2Si_2$\refto{steglich}, $UBe_{13}$\refto{ube13}, $UPt_3$\refto{upt3},
$URu_2Si_2$\refto{uru2si2}, $UNi_2Al_3$ and $UPd_2Al_3$.\refto{steglich2}
These metals
contain a dense array of magnetic rare earth or
actinide ions that
collectively participate in the formation of the superconducting state.
Many properties set these systems apart from traditional
superconductors.
In particular, power laws
in the specific heat, thermal conductivity, NMR relaxation rate and acoustic
attenuation
all point to the existence of gap nodes,  and have been interpreted
in terms of gap zeroes along lines of the Fermi surface. Furthermore,
each of these superconductors appears to  coexist with some measure
of antiferromagnetic order.

Existing phenomenological models of
heavy fermion superconductivity treat it as a
pairing process involving the pre-formed heavy f-quasiparticles.
\reftorange{miyake}{varma, volovic,norman}{Upt3theory}
The strong repulsive interactions between these f-quasiparticles
favor
the development of
nodes in the pair wavefunction,
as suggested by the preponderance of power laws. Theoretical work on heavy
fermion superconductivity
has focussed largely on the
possibility of momentum anisotropy in the gap function
$\Delta_{\vk}$ as the origin of this node formation.
The
simplest candidates for gap functions with nodes
are odd-parity triplet pairing, $\Delta_{\vk} = - \Delta_{-\vk}$, or even
parity d-wave pairing.
Two points appear to favor the latter possibility:

\item{ $\diamond$ }d-wave pairing is favored by the
antiferromagnetic interactions that are characteristic
to heavy fermion compounds.\refto{miyake,bealmonod}

\item{$\diamond$ }lines of gap zeroes inferred from many power-law
properties of the condensed state, e.g $T^3$ dependence
of the $NMR$ relaxation rate, $T^2$ dependence
of specific heat and thermal conductivity, effectively rule out
triplet odd-parity pairing.
Simple symmetry arguments show that
odd-parity triplet would give rise to a gap vanishing
at points, rather than lines on the Fermi surface,
in the presence of strong spin-orbit scattering.\refto{blount,volovic}

Unfortunately, there are several observations
that do not fit naturally into the d-wave scenario.
One  puzzling observation is the persistence of the
$T^3$ NMR relaxation rate in heavy fermion superconductors
with very large
densities of gapless excitations and
correspondingly  ``heavy'' linear components to their specific
heat.\refto{upt3cv,uthbe13cv}
This gaplessness has been
attributed to
pair breaking by  resonantly scattering off
non-magnetic defects.\refto{pethick}
Remarkably, the constant
density of quasiparticle states \underbar{never} appears as an
observable Korringa  NMR  relaxation.  In $UPt_3$ and $U_{1-x}Th_xBe_{13}$ for
example,\refto{upt3cv,uthbe13cv} the linear specific heat
is of the same order as the normal phase value,
yet there are more than two
decades\refto{upt3nmr,uthbe13nmr} of $T^3$ spin relaxation.
$$
\eqalign{
&\ \left. {C_v\over T} =   \gamma  + B T
\ \ \qquad\qquad\right\}{\rm linear\  term\  from\  pair\  breaking.}\cr
        &\ \ \  \qquad\ \vdots\cr
&\left.  { 1 \over T_1 T}  =   {?}
+ D T^2\ \qquad\qquad\right\}
{\rm no \ linear\  term\  from
\  pair \ breaking?}\cr
\ &\cr}
\eqno(help)
$$
In conventional gapless superconductivity,
\reftorange{schloussky}{maki2}{hirschfeld}
spin coherence factors are unity
at the Fermi energy. The robustness
of the $T^3$ NMR signal suggests vanishing  spin coherence
factors: a feature not easily accommodated by a conventional pairing
hypothesis.

The d-wave scenario
is also unable
to explain the isotropy of the $H-T$ phase diagram of
$UPt_3$.\refto{Upt3theory,machida}  $UPt_3$ has \underbar{three}
separate low temperature flux phases that have been interpreted in terms of
anisotropic pairing.
There is a two-stage phase transition
at zero field associated with the symmetry breaking effects of the weak heavy
fermion
antiferromagnetism, and perhaps also, a recently discovered incommensurate
charge
density wave.\refto{hayden2} The d-wave scenario supposes a gap function that
transforms under
a two-dimensional representation of the point-group:
though this picture can  account for the two stage transition,
it  predicts a two-phase flux lattice
for  all orientations of the applied field.

Finally of course,
the d-wave pairing picture of heavy fermion superconductivity
makes no reference to the close link between heavy fermion superconductivity
and
magnetism. Typically, the entropy associated with the superconducting phase
$$
S_{sc} = \left.\int_0^{T_c} dT { C_V (T) \over T}  =
C_V [T_c]\right|_n
$$
is a significant proportion of the ${ \rm R  ln}2$ entropy associated with the
quenching of
the low lying doublets: in this sense heavy fermion superconductivity is a {\sl
spin ordering}
process, involving the magnetic, rather than the charge degrees of freedom of
the f-electrons.
Experimentally,  $\ul{\sl each}$ heavy fermion superconductor
appears to have a coexistent antiferromagnetic order. In the recently
discovered $1-2-3$
compound $UPd_2Al_3$, an ordered moment of $0.8\mu_B$
coexists\refto{upd3al2mom} with the
superconductivity.
In $URu_2Si_2$, there is also evidence for a large moment-free
order parameter
that breaks time reversal and translation symmetries.\refto{us, buyers}
Unlike the well-known Chevrel phases,\refto{chevrel}
this moment shares the same magnetic degrees of freedom
that are involved in pairing.  It is rather difficult
to account for coexistent magnetism and superconductivity
in terms of two weakly coupled order parameters.

These difficulties motivate us
to reconsider the way in which heavy fermion superconductors
develop nodes in the pair wavefunction.
Past analyses of heavy fermion superconductivity have focussed
on the spatial anisotropy. In this paper we explore a new avenue,
examining the possibility of pair condensation into a state
where the pair wavefunction has {\sl odd} temporal parity.
\reftorange{berezinskii}{abrahams,emery}{kirkpatrick}
In this hypothetical state, pairing is retarded and
the pair wavefunction contains a {\sl node in time}.\refto{previous}

Berezinskii\refto{berezinskii} first pointed out that a
general pairing hypothesis must consider the symmetry of the
pair wavefunction under frequency inversion.
Let us denote the pair wavefunction
$$
\bigl[\ul{F}(\ka)\bigr]_{\alpha \beta} =  \la \psi_{\a}(\ka )
\psi_{\be}(-\ka)\rangle
\eqno(pair)
$$
Here $\la\dots\ra$ denotes the time-ordered expectation value, and
we use a four-vector notation $\ka \equiv ( \vk, \om)$. Since the Fermi
operators
anticommute, the pair wavefunction satisfies
$$
\ul{F}(\ka) = - \ul{F}^{\rm T}(-\ka)
\eqno(pair2)
$$
where $[\ul{F}^{\rm T}]_{\alpha \beta} =[\ul{F}]_{ \beta \a}$ denotes the
wavefunction with spin indices transposed.
Now if we assume that the state breaks neither time reversal
symmetry, nor spatial parity, then the pair wavefunction must
have distinct spatial, temporal and spin parity. Let
$({\rm S,\ P,\ T\ } = \pm 1)$ be the parities of the pair wavefunction
under the interchange of spin, space or time coordinates respectively, i.e.
$$
\ul{F}(\vk, \om)
=\left\{\eqalign{&{\rm S}\ul{F}^{\rm T}(\vk, \om)\cr
&{\rm P}\ul{F}(-\vk, \om)\cr
&{\rm T}\ul{F}(\vk, -\om)\cr
}\right.\qquad\qquad ({\rm S,\ P,\ T\ } = \pm 1)
\eqno(symmetries)
$$
then the total antisymmetry of the pair wavefunction implies that
the combined product of all three parities
must equal $-1$:
$$
{\rm SPT} = -1
\eqno(thefacts)
$$
Superconductors with an antisymmetric spin wavefunction, $S=-1$ are ``singlet''
superconductors,
$$
\ul{F}(\ka) = i\ul{\si}_2 F_s(\ka)\quad\qquad (\ul{F}=-\ul{F}^{\rm T})
\eqno(singlet)
$$
whereas superconductors with a symmetric spin wavefunction, $S=+1$ are
``triplet''
$$
\ul{F}(\ka) = i\ul{\si_2 \vec \si} \cdot \vec F_t( \ka )
\quad\qquad (\ul{F}=+\ul{F}^{\rm T})\eqno(triplet)
$$
In conventional superconductivity, $T=+1$, so that the spatial
parity of singlet and triplet states is even and odd respectively.
Berezinskii has argued that
symmetry also permits the
possibility of odd frequency
pairing where
$$
{\rm T}= -1 \qquad {\rm P} =  \left\{
\eqalign{
&+1\qquad\qquad(  {\rm triplet}\ \  \{S,P,T\} =\{+,+,-\} )
\cr
&-1\qquad\qquad(  {\rm singlet}\ \  \{S,P,T\} =\{-,-,-\} )
\cr}\right.\eqno(odd)
$$
Odd frequency, {\sl even parity} triplet pairing was first considered  by
Berezinskii
in the context of $He-3$.
A renaissance of interest in these types of states
has been prompted by the work of Balatsky and Abrahams, who
are the first to discuss the possibility of odd
frequency, {\sl odd parity} singlet pairing.\refto{abrahams}

Historically, odd frequency pairing has not enjoyed a great
deal of attention. One reason for this lack of attention
is that the simplest
odd frequency paired state is \underbar{unstable},
with a negative Meissner phase stiffness.
For example, in the s-wave triplet
state, (S, P$=+1$) the momentum dependence of the gap function
vanishes, and the  London Kernel is
formally identical to s-wave singlet pairing
$$
\Lambda
= {\pi Ne^2 T\over  m}  \sum_{\om_n ^2 + \Delta_n^2 >\ 0}
{ \Delta_n^2 \over ( \Delta_n^2 + \om_n^2 )^{3 \over 2}}\qquad
\qquad\biggl(\Delta_n = \Delta(\iom)\biggr)
\eqno(stiff5a)
$$
Since an odd gap function where $\Delta_n= -\Delta_{-n}$ must also
satisfy the analyticity requirement
$\Delta_n= \Delta^*_{-n}$, this implies $\Delta_n$ is
purely imaginary. Thus
$\Delta_n^2 < 0$
and the stiffness is negative.
Loosely interpreted, this negative stiffness suggests
the microscopic phase of the order parameter likes to
``coil up'', breaking translation symmetry and assuming
a staggered configuration.

We  shall argue that
the Kondo effect between a conduction
sea and
local moments in heavy fermion metals provides an ideal source
of retarded scattering for odd frequency pairing.
In the normal state,
this
retardation generates
resonant  bound states between the conduction electrons
and local moments, quenching the moments
and forming the heavy quasiparticles.
In the superconducting state,  the  resonant Kondo scattering
acquires a pairing
component that results in even parity, odd frequency triplet pairing
of the conduction electrons.
This state develops a
phase stiffness by the simultaneous condensation of
the local
moments and the conduction electron pair
degrees of freedom.
The equal time order parameter is
a matrix
correlating these two degrees of freedom
$$
\la \tau_{\a}(x)S^{\be}(x)
\ra
= g{\cal M}_{\alpha}^{\ \beta}(x)
\qquad\qquad(\alpha, \beta = 1,2,3)\eqno(op)
$$
Here
$\vec S[x_j]$ denotes the local moment spin at site j;
$\vec \tau[x]$ is the
conduction electron ``isospin'', whose $z$ component describes
the number density, and  transverse components describe the pairing
$$
\eqalign{
\tau_3=& {1 \over 2}(\rho(x)-1)\cr
\tau_+(x) = & \psi_{\up}^{\dg}(x)\psi_{\dw}^{\dg}(x)\cr}
\eqno(isospin)
$$
The quantity $g$ defines the magnitude of the order parameter.
$\ul{{\cal M}}$ is an orthogonal matrix whose rows
define  an orthogonal triad of
unit vectors $\hat d_{\l}$,
$$
\ul{\cal M}(x)
= \pmatrix{
\hat d_1(x) &\cr
\ \hat d_2(x)&\cr
\ \hat d_3(x)& \cr}.
\eqno(staggering)
$$
Stabilization of the odd-frequency paired state is achieved by
{\sl staggering} this order parameter: in a simple model,
 $\hat d_1$ and $\hat d_2$ are staggered commensurately with the lattice.
A ``composite'' order parameter of this form has been recently
suggested, in the context of the two-channel Kondo model, as an order
parameter for odd frequency pairing, by Emery and
Kivelson.\refto{emery}

Within our theory, the microscopic manifestation
of this type of pairing is
an  anomalous self-energy in the triplet channel
with a pole at zero frequency
$$
\ul{\Delta}(\kappa)
= i \si_2
\ul{ d}_c \left({V^2 \over 2 \omega}\right)\qquad\qquad(\ul{d}_c= [\hat d_1
+ i \hat d_2]\cdot\vec \si)
\eqno(crazy)
$$
A spinless component of the conduction electron band
decouples from this singular pairing field, giving rise
to surfaces of gapless excitations. Spin and charge
coherence factors of these  quasiparticles
vanish linearly with the energy on the Fermi surface,
$$
\left.\eqalign{
&\la \vk' \vert \rho_{\vk - \vk '} \vert \vk \ra \cr
&\la \vk' \vert S^z_{\vk - \vk '} \vert \vk \ra \cr}\right\}
\propto \om \qquad\qquad ( \omega = E_{\vk'} + E_{\vk})
\eqno(coherence)
$$
creating the unusual circumstance where a flow of quasiparticles
transmits heat without passage of charge or spin.
These unusual
coherence factors lead to power laws
in the nuclear magnetic relaxation
$$
{1 \over T_1 } \propto T^3 \eqno(power)
$$
that coexist with a  linear
specific heat capacity.


An essential part
of our analysis is a  second quantized
description of the local moments that avoids
constraints.
Existing  treatments factorize the spin variable in
terms of spin $1/2$ fermions.
$$
\vec S =  f\dg_{\a} \bigl[{ \vec \si \over 2} \bigr]_{\alpha \beta }
f_{\be}
\eqno(abrik)
$$
This approach
requires a constraint
$n_f=1$ to
impose the condition $S={1\over 2}$, which
is the origin of additional complications.
In many
practical applications, the constraint is weakened,
imposing it at the mean field or the
Gaussian level of approximation.
Here, we  employ
real, or ``Majorana'' fermions to represent spins.
Perhaps the most famous example of a Majorana fermion is a single
Pauli spin operator. Recall that
$$
\{\si_a \ , \si_b \} = 2 \delta_{ab}
$$
so the fermions
$$
\vec \eta = { 1 \over \sqrt{ 2} }  \vec \si
\eqno(major1)
$$
also satisfy a canonical anticommutation algebra
$\{ \eta_a , \eta_b\} = \delta _{ a b }$.
Indeed, for  {\it any} such
triplet of Majorana fermions, it follows that
the ``Spin operators''
$$
\vec S = -{ i \over 2 } \vec \eta \times  \vec \eta
\eqno(major2)
$$
simultaneously satisfy {\it both} the spin algebra,
{\it and} the constraint $S^2=3/4$.\refto{history}
The generalization of this result to a lattice of spins,
employs a vector of Majorana fermions
$\eta_j^a = ( \eta^a_j) \dg$, ($a= 1,\ 2,\ 3$)
for each
site j where,
$$
\{\eta^a_i, \eta^b_j\} =
\delta_{ij}\delta^{ab}
\eqno(comm)
$$
from which the spin operator at each site is constructed
$$
\vec S_j = -{i \over 2} \vec \eta_j \times \vec \eta_j\eqno(spin2)
$$
These spin operators behave as independent
spin $1/2$ operators.

Our Majorana representation of spins provides a natural
lattice generalization of
{\sl anticommuting} Pauli operators.
On a lattice, we may represent Majorana fermions in terms of a set of
$N/2$ independent complex fermions that span half the Brillouin zone:
$$
\vec \eta_j = {1 \over \sqrt{ N}}
 \sum_{\vk \in {\rm {1/2} \
 B. Z.}}
\left\{\vec \eta_{\vk} e^{i \vec k \cdot \vec R_j}+
\vec \eta\dg_{\vk} e^{-i \vec k \cdot \vec R_j}\right\}
\eqno(realtocomplex)
$$
These complex fermions obey
canonical commutation relations
$\{\eta^{a}_{\vk}, \eta^{b}_{\vk'}\dg\} = \delta^{ab} \delta_{\vk \vk'}$;
opposite halves of the Brillouin zone are related as
complex conjugates:
$
\bigl[\eta^a_{\vk}\bigr]\dg =  \eta^a_{-\vk}
$.
Since the Fock space
is spanned by $3N/2$ complex  Fermi operators,
it is
$2^{N/2}$ times larger than a Hilbert space of $N$ commuting
spin $1/2$ operators.
The spin algebra {\sl and} the condition
$S=1/2$
are satisfied between all states of the Fock space,
thus the anticommuting representation
{\it replicates} the spin Hilbert space  $2^{N/2}$ times.\refto{weyl}
We may then represent the
partition function of an electronic system containing
$N$ spins as an {\it unconstrained} trace over the independent
Fermi fields
$$
Z= { 1 \over 2^{N/2}}
{ \rm Tr}\left\{ e^{-\beta H[\vec S_j]
}\right\} \qquad\qquad (\vec S_j \rarrow
-{i \over 2} \vec \eta_j \times\vec \eta_j)
\eqno(partn)
$$
where the formal normalization factor associated with
the replication of states has been added.

Our basic model for a heavy fermion system is an  $S=1/2$ Kondo lattice
with a single band interacting with one local f-moment
$\vec S_j$ in each unit cell. In a real Kondo lattice, the local moments
are strongly spin-orbit coupled into a state of definite $J$.
We shall assume that the low lying spin excitations are described by
a Kramers doublet, where a low energy $S=1/2$ Kondo model
becomes more appropriate.
For simplicity, we shall ignore
the anisotropies that are necessarily present in a real heavy fermion
system.
In our model, the latent
superconducting pairing is driven by the on site Kondo interactions,
and the state that forms exhibits a coexistence of magnetism and
superconductivity.

The simplified  isotropic  Kondo lattice model that we shall
use  is then written
$$
H= H_{c} + \sum_j H_{int}[j]\eqno(strip)
$$
where
$$H_c = \sum_{\vk} \eps_{\vk} \psi\dg_{\vk}\psi_{\vk}\eqno(band)
$$
describes the conduction
band, and $\psi\dg_{\vk}= ( \psi\dg_{\vk \up} ,\psi\dg_{\vk \dw} ) $  is
a conduction electron spinor. The exchange interaction at each site $j$
is written in a tight binding
representation as
$$
H_{int}[j]=
J(\psi\dg_{j\alpha} \vec \si_{\alpha\beta} \psi_{j \beta}) \cdot \vec S_j
\eqno(kondo)
$$
When written in terms of the Majorana fermions, this term becomes
$$
H_{int}[j]=  - {J \over 2}
\psi\dg_j[\vec \si \cdot\vec \eta_j]^2 \psi_j,\qquad\qquad\psi_j\equiv\left(
\matrix{\psi_{j \up}\cr
\psi_{j \dw}\cr}
\right)
\eqno(exch2)
$$
where we have used the result
$i\vec \si.(\vec \eta \times \vec \eta) = [\vec \eta \cdot
\vec \si]^2 - {3\over 2}$ to simplify the interaction, absorbing the
bilinear term as  a redefinition
of the chemical potential.
This simple form of the interaction can be rewritten in a suggestive
form, by defining the composite spinor operator
$$
\hat V_j=\left(\matrix{\hat V_{j\up} \cr \hat V_{j\dw}}\right)=-{J \over 2} [
\vec
 \si \cdot \vec \eta_j]\psi_j
\eqno(spinor2a)
$$
The Kondo interaction is then the ``square'' of this operator:
$$
H_{int}[j] = -{2 \over J} \hat V\dg_j \hat V_j\eqno(suggestive)
$$
suggesting that in the lattice, we should consider the possibility of
states where the local moment and electron spins condense together
to develop a vacuum expectation value of this spinor quantity
$$
\left(\matrix{V_{j\up} \cr  V_{j\dw}}\right)=
\left(\matrix{\la \phi\vert \hat V_{j\up}\vert \phi\ra \cr
\la\phi\vert \hat V_{j\dw}\vert \phi\ra}\right)
\eqno(basicfact2)
$$
This order parameter
transforms as a spin $1/2$ object, so changes in its sign
correspond to physical rotations of the condensate
by $\2 \pi$.
Defects in the spinor field are then disclination lines or ``$Z_2$ vortices'',
around which the phase of the
spinor order parameter changes by $\pi$ (Fig. 1).
The  gauge equivalent
integral of the vector
potential around a $Z_2$ defect is
$$
{e \over \hbar} \int \vec A \cdot dx = \pi \eqno(z2defect)
$$
so the flux quantum of a ``charge e'' spinor is the same as a
charge ``2e'' scalar:
$$
\Phi_o = \int \vec A \cdot dx = {\hbar \over e}\pi = { h \over 2 e}
\eqno(z2defectb)
$$
In our model there is a microscopic
``$Z_2$'' gauge symmetry
$$\vec \eta_j \rarrow \pm \vec \eta_j\eqno(z2)
$$
$V_j$ transforms in the same way as the Majorana fermions
and thus its phase is defined to within $\pm \pi$.
Physical quantities
involve the combinations of the square of $V_j$
at each site and are $Z_2$ invariants.
The three $Z_2$ invariant  quantities that this
defines are just the components of the matrix ${\cal M}(x)$
$$
\eqalign{
\la \rho(x) \vec S(x)\ra=&g V(x)\dg\vec \si V(x)\cr
\la \psi_{\up}^{\dg}(x)\psi_{\dw}^{\dg}\vec S(x)\ra =&g V^{\rm T}(x) i \si_2
\vec \si V(x) \cr
g\equiv&{ J^2\over 2}\cr}
\eqno(axes2)
$$
Under a phase change of $\pi$ in the spinor field, the axes of
the composite order parameter rotate through $2 \pi$.


The outline of this paper is as follows:

\item{\bf 2.}Development of a path integral
formulation of the Kondo lattice,
demonstrating how the simplest decoupling procedure leads
to an odd frequency paired state.

\item{\bf 3.}Discussion of the quasiparticle
excitations and coherence factors.

\item{\bf 4.}Calculation of
the mean field thermodynamics in this paired state.

\item{\bf 5.}Computation of the Meissner stiffness of this phase, and form
of the Landau-Ginzburg theory.

\item{\bf 6.}Effect of vanishing coherence factors on
local magnetic and charge response.

\item{\bf 7.}Interplay with magnetism.

\item{\bf 8.}Critique and discussion: possible application to the theory of
heavy fermion superconductivity.

\noindent Certain formal arguments, not
pertinent to the main flow of ideas have been reproduced in the
the appendices. In appendix A, we
show how the Majorana representation is related to the Abrikosov
pseudofermion representation. In appendix B we
give some  examples of the application of the Majorana
representation to simple spin models, showing the relation to the Jordan
Wigner transformation in the one dimensional Heisenberg model.

\noindent{\bf 2. Path Integral Representation of the Kondo Lattice Model}

\taghead{2.}

To develop a ``toy model'' for the odd-paired state, we
focus our attention on a stripped-down Kondo lattice model, with the
Hamiltonian described in \(strip) to \(exch2)
$$
H= \sum_{\vk} \eps_{\vk} \psi\dg_{\vk}\psi_{\vk} - {J \over 2} \sum_j
\psi\dg_j[\vec \si \cdot\vec \eta_j]^2 \psi_j\eqno(kl1)
$$
We have suppressed both the momentum
dependence and anisotropy of the coupling.  In a real heavy fermion system,
we envisage that the spin
indices would refer to the conserved pseudospin indices of the
low lying Kramers doublets.

To illustrate the calculations in this section, we shall use Feynman
diagrams, as shown in Fig. 2. The bare propagator for the conduction
electrons is represented by a solid arrow, the bare propagator for
the Majorana fermions by a dashed line, without an arrow.
$$
\eqalign{
\relbar\joinrel\relbar\quad
\relbar\joinrel\relbar\quad
\relbar\joinrel\relbar\quad
\relbar\joinrel\relbar\quad
=&
\la \eta^{a}(\ka) \eta^{b '}(-\ka) \ra_o = \delta^{a b}
\left( { 1 \over \iom}\right)\cr
\relbar\joinrel
\relbar\joinrel
\relbar\joinrel
\relbar\joinrel
\relbar\joinrel\triangleright\joinrel
\relbar\joinrel
\relbar\joinrel
\relbar\joinrel
\relbar\joinrel
\relbar\joinrel\quad
=&{\delta_{\si \si'}\over \iom - \eps_{\vk}}
\cr}
\eqno(props)
$$
The product form of the exchange interaction \(exch2)
clearly suggests a decoupling in terms of the spinor variable
$$
V_j=\left(\matrix{V_{j\up} \cr V_{j\dw}}\right)=-{J \over 2} \la [ \vec
 \si \cdot \vec \eta_j]\psi_j\ra
\eqno(spinor2)
$$
corresponding to the bound state of an electron and a local moment.

With this point in mind, we
now write the partition function as a path integral,
$
Z= \int_{\rm P}
e^{- \int_0^\beta {\cal L(\tau)}d\tau }
$
where
$$
{\cal L}(\tau) = \sum_{\vk }
\psi_{\vk} \dg\partial_{\tau} \psi_{\vk}+
\sum_{\vk \in {1 \over 2} BZ}\vec \eta_{\vk} \dg\partial_{\tau}\vec
\eta_{\vk}
+H_c + \sum_j H_{int}[j]
\eqno(lag2)
$$
and we have
factorized  the interaction in terms of a fluctuating
two-component spinor $V\dg_j=(V^*_{\up}, V^*_{\dw})$
$$
H_{int}[j]=
\psi\dg_j ( \vec \si \cdot \vec \eta_j)V_j
+V\dg_j
( \vec \si \cdot \vec \eta_j)\psi_j + 2|V_j|^2/J
\eqno(int2bc)
$$
For later purposes, it is particularly useful for us to introduce
a Balian Werthammer four-spinor notation, defining
$$
\eqalign{
\Psi_j =&
\left(\matrix{\psi_j \cr  -i \si_2\psi^*_j\cr
}
\right)
=
\left(\matrix{
&\psi _{j\up} \cr
&\psi _{j\dw}\cr
-&\psi\dg_{j \dw}
\cr
&\psi
\dg_{j \up}
\cr}
\right)\cr
{\cal V}_j =&
\left(\matrix{V_j \cr  -i \si_2V^*_j\cr
}
\right)=
\left(\matrix{
&V _{j\up} \cr &V_{j\dw}\cr
-&V^{*}_{j \dw}\cr
&V^*_{j\up}\cr}
\right)}\eqno(bignam)
$$
The lower two entries of each spinor are the time reversed pairs
of the upper two entries.
In terms of  these spinors, the conduction electron Hamiltonian
is
$$
H_c= \sum_{\vk \in {1 \over 2} {\rm B.Z.}}
\Psi\dg_{\vk}( \epsilon_{\vk} -\mu)\ul{\tau}_3\Psi_{\vk}
$$
where
\def\ldash{\joinrel\relbar\joinrel\relbar\joinrel\relbar\joinrel\relbar}
\def\2mat#1#2#3#4{\left[\matrix{
\matrix{#1 \cr{\ldash}
\cr #3}\left| \matrix{
 #2 \cr {\ldash}\cr #4}\right.} \right]}
\def\ldas{\joinrel\relbar\joinrel\relbar\joinrel\relbar}
\def\9mat#1#2#3#4{\left[\matrix{
\matrix{#1 \cr{\ldas}
\cr #3}\left| \matrix{
 #2 \cr {\ldas}\cr #4}\right.} \right]}
$$
(\ul{\tau}_1,\ \ul{\tau}_2, \ \ul{\tau}_3)=\left(
\9mat{\quad}{\ul{
1}} {\ul{1}} {\quad} ,
\9mat{\quad}{\ul{i}} {- \ul{i} } {\quad}
,\9mat{\ul{1}}{\quad} {\quad} {-\ul{1}} \right)
\eqno(s1)
$$
denotes the triplet of isospin operators. Note the definition of $\ul{\tau}_2$.
The
factorized interaction can also be written
\def\cV{{\cal V}}
$$
H_{int}[j]=
{1 \over 2}\left[\Psi\dg_j ( \vec {\ul{\si}} \cdot \vec \eta_j){\cal V}_j
+\cV\dg_j
( \vec {\ul{\si}} \cdot \vec \eta_j)\Psi_j\right] + \cV_j\dg\cV_j/J
\eqno(int3)
$$
where,
$$
\ul{\vec\si} \equiv \vec \si\otimes 1 =
\2mat {\vec \si}{\qquad}{\qquad}{\vec \si}
\eqno(s11)
$$
denotes the spin operator in the Balian Werthammer notation.

We are particularly interested in expanding around
static mean field configurations where the amplitude of
$V_j$ is constant,
$$
{\cal V}_j = {V\over \sqrt {2}}{\cal Z}_j,
\qquad\qquad
V_j = {V\over \sqrt{2}} \pmatrix{ z_{j \up} \cr
z_{j \dw}}  \qquad\qquad(z_j\dg z_j=1)\eqno(spinortry)
$$
This choice of mean field theory is equivalent to a resummation of
the interaction lines in the pairing channel between the conduction
and Majorana fermions, leading to a
saddle point condition
for the anomalous
average of Majorana and conduction electrons, as illustrated diagrammatically
in  Fig. 3(i). The development of this anomalous average leads to self-energy
insertions in the conduction electron lines (Fig. 3(ii)).
 From this diagram, it is evident that the conduction electron self
energies are bilinear forms in the spinor $V_j$, and are hence
{\sl invariant} under the  $Z_2$ gauge symmetry.

We can actually find a class of degenerate mean field solutions
by arbitrarily reversing the sign of $V_j\rarrow m_j V_j$
($m_j = \pm$)
at any site.
For each choice of sign, there are $2^{N/2}$ equivalent
ways of
choosing the independent Majorana creation operators in momentum space,
($\eta\dg_{-\vk} = \eta_{\vk}$),
there are thus $2^{N-N/2}= 2^{N/2}$ independent
degenerate saddle point solutions for each static solution $\{{\cal V}_j\}$.
Each saddle point is physically identical, so
we may absorb the $2^{N/2}$ normalization in the partition
function by restricting our attention to one representative saddle
point
$$
Z = { 1 \over 2^{N/2}}\sum_{\{m_j\}}Z[\{m\}] = Z[\{m\}]\vert_{m_j=1}
\eqno(gauge)
$$
In this way, we fix the gauge for the local $Z_2$ invariance.

On a bipartite lattice, it is convenient to carry out a gauge
transformation that moves the origin of momentum space
to the Brillouin zone center, by defining
$$
\eqalign{
z_j = &
e^{i \left({\theta_j\over 2}\right)}\tilde z_j
\qquad\qquad ({\cal Z}_j = e^{i ({\theta_j\over 2})\ul{\tau}_3}
\tilde {\cal Z}_j)
\cr
\theta_j = &  \vec Q \cdot \vec R_j \cr
\Psi_j=&  e^{i \left({\theta_j\over 2}\right)\ul{\tau}_3}\tilde\Psi_j}
\eqno(shifto)
$$
The conduction electron Hamiltonian can then be written
$$
H_c=
\sum_{\vk \in{1 \over 2} {\rm B. Z.}}
\Psi \dg _ {\vk} \bigl[
\tilde \eps_{\vk}  - \mu_{\vk} \ul{\tau}_3 \bigr]\Psi_{\vk}
\eqno(kinen)
$$
where
$$
\eqalign{
\tilde \eps_{\vk}=& {1 \over 2}\bigl[\eps_{\vk - \vec Q /2 }-\eps_{\vk+
\vec Q/2}\bigr]\cr
\mu_{\vk}=& \mu-{1 \over 2}\bigl[\eps_{\vk - \vec Q /2 }+\eps_{\vk+ \vec
Q/2}
\bigr]\cr}\eqno(def6)
$$
and we have suppressed the tildes on the electron operators.
For a simple bipartite
tight-binding lattice, taking
$\vec Q=( \pi, \pi, \pi)$, then $\mu_{\vk} = \mu$ and $\tilde {\eps}_{\vk}
= \eps_{\vk - \vec Q /2 }$. With this choice of gauge, the conduction
electron kinetic energy is manifestly particle-hole symmetric.

The lowest energy
mean field solution is obtained for a spatially uniform $\tilde{\cal V}_j
= {\cal V}_o$.
With this choice, the admixture between conduction electrons and
local moments is described by the mean field Hamiltonian
$$
H_{mix}= \sum_{\vk \in {1 \over 2} { \rm BZ}}
\bigl[\Psi\dg_{\vk} ( \ul{\vec\si} \cdot \vec \eta_{\vk})
{\cal V}_o + {\cal V}\dg_o(\vec \eta\dg_{\vk} \cdot \ul{\vec\si}
)
\Psi_{\vk}\bigr]
\eqno(mfham)
$$
To gain further insight into the nature of this state, let us
``integrate out'' the Majorana fermion degrees of freedom.
Within a diagrammatic approach, this corresponds to introducing
a self-energy
into the conduction electron propagators, as shown in
Fig. 3(ii). The
effective action for the conduction electrons contains this self
energy:
$$
S_{\rm eff} = \sum_{{\vk \in {1 \over 2} {\rm BZ}, \iom}} \Psi\dg(\ka)
\bigl[
\om - \tilde \eps_{\vk} - \mu_{\vk}\ul{\tau}_3 - \ul{\Sigma}(\kappa)
\bigr]
\Psi(\ka)
\eqno(effact)
$$
$\ul{\Sigma}(\kappa)$ describes the resonant scattering
through zero energy spin states, and is given by
$$
\ul{\Sigma}(\ka)
=
\ul{\si}^{\l}
\ul {\cal M}
\ul{\si}^{\l '}
\la \eta^{\l}(\ka) \eta^{\l '}(-\ka) \ra
\eqno(sen)
$$
where the matrix $\ul{\cal M}$ is formed from the square of the spinor
order parameter

\def\clz#1{{\cal Z}_{#1}}
\def\clzd#1{{\cal Z}\dg_{#1}}
$$
{\cal M}_{\alpha \be}[j] ={\cal V}_{\a}[j]
{\cal V}\dg_{\be}[j]= {V^2 \over 2} \clz \a [j] \clzd \be [j]
$$
The general expansion
of ${\cal M}$ is
$$
\ul{\cal M}  =  {1 \over 4} V^2 \bigl[ \ul{1} + d_{ab}\si^a
\otimes
\tau^b
\bigr]
$$
where $d_{ab}[j]={1 \over 2}{\cal Z}\dg[j]\si^a\otimes \tau^b {\cal Z}[j]$.
The columns of the matrix $\ul{d}$ define a triad of orthogonal
unit vectors $[\hat d^{\lambda}]_a= d_{a \lambda}$, ($\l=1,2,3$)
$$
\eqalign{
 \hat d^1_j + i \hat d_j^2=&
z^T_j i \si_2 \vec \si z_j
\cr
\hat d^3[j] =& z\dg_j\vec \si z_j\cr}
\eqno(triad)
$$
that set the orientation of the order parameter
in spin space.
The resulting conduction electron self-energy is proportional to $1/\omega$
$$
\ul{\Sigma}(\ka)=  {V^2 \over \om} \ul{\cal P}
\eqno(self)
$$
where the projection operator
$$
 \ul{\cal P} = V^{-2}\si^{\l}{\cal M}\si^{\l}=  {1 \over 4}\left[
3(\ul{1}) -  d_{ab} \si^a \otimes \tau^{b}\right]
\qquad\qquad
(\ul{\cal P}^2 = \ul{\cal P} )
\eqno(sen3b)
$$
The anisotropic component of the self-energy
$\ul{\Sigma}^{an}(\kappa)={V^2 \over 4 \om}[{\cal P} - {3 \over 4}]$
contains ``anomalous'' components, and may be written
$$
\ul{\Sigma}^{an}(\om)= - {1 \over 2}\biggl[
\bigl(\vec B_j(\om) \cdot\vec\si \bigr)\tau_3 + \bigl(
\vec \Delta\dg(\om)\cdot
\vec\si\bigr) {\tau_{+} \over 2} +
\bigl( \vec \Delta(\om)\cdot
\vec\si\bigr){\tau_{-} \over 2}\biggr]
\eqno(pairing)
$$
We interpret the quantities
$$
\eqalign{
\vec B(\om) = &{\Delta({\om})}
 \hat d^3\cr
\vec \Delta(\om) =&{\Delta({\om})}( \hat d^1+ i \hat d^2)\cr
}\qquad\qquad\biggl(\Delta(\om) ={V^2 \over 2 \om}\biggr)\quad
\eqno(fields)
$$
as resonant exchange and triplet pairing fields, respectively.
This is a realization of odd frequency pairing in the triplet
channel.
Unlike earlier realizations of odd frequency triplet pairing,
\refto{berezinskii,abrahams}
the gap function diverges at zero frequency.
Such resonant contributions
to the self-energy are well known within mean field treatments of the
Kondo lattice, but here the resonant scattering acquires additional,
anisotropic pairing terms associated with the pair condensate.

The projective form of the  pairing self-energy means that
not all components of the conduction fluid experience
the resonant scattering.  This is more easily seen by decomposing the
conduction electron operators into four Majorana components
\def\psk{\psi_{\vk}}
\def\pskd{\psi\dg_{\vk}}
\def\psd #1{\psi^{#1\dagger}_{\vk}}
\def\psu #1{\psi^{#1}_{\vk}}
$$
\psk = { 1 \over \sqrt{2}}\bigl[
\psu 0 + i \vec \psk \cdot \vec \si
\bigr]z_o
\eqno(decomp1)
$$
The ``vector'' components of the conduction electron are projected out
by the operator ${\cal P}$.   By
substituting  into the mean field Hamiltonian, we see
that only these  components of the conduction electrons couple to the
resonant scattering potential
$$
H_{mix} =
\sum_{\vk \in { 1 \over 2 } B.Z.}
- i V \bigl[
\vec \pskd \cdot\vec  \eta_{\vk} - { \rm  H . C. } \bigr]
\eqno(hammy1)
$$
The zeroth component $\Psi_o=(\ul{1} - \ul{\cal P})\Psi$
does not couple directly
to the resonant scattering potential, leading to a gapless
quasiparticle mode.

We may gain insight into the meaning of the order parameter matrix
$\ul{\cal M}$ by making the
identification
$$
V_j \equiv -{J \over 2}\vec\si \cdot\vec \eta_j \psi_j
\eqno(repl)
$$
or
$$
{\cal V}_j \equiv -{J \over 2}\ul{\vec\si} \cdot\vec \eta_j \Psi_j
\eqno(repl2)
$$
inside the path integral.
The expectation value of the matrix $\cal{M}$ is then the
irreducible part of the corresponding product of operators.
In particular,
$$
\la {\cal V}_j\dg \tau^a \si^b {\cal V}_j\ra =
{J^2 \over 4 } \la \Psi_j\dg \tau^a ( \vec \si \cdot \eta_j)
\si^b
( \vec \si \cdot \eta_j)\Psi_j\ra_{\rm I}
\eqno(equivo)
$$
where ``I'' denotes the irreducible part.
Using the identity $( \vec \si \cdot \eta_j)\vec \si
( \vec \si \cdot \eta_j)= [-i\vec \eta_j \times \vec \eta_j- \vec \si/2]$
it follows that
$$
\la {\cal V}_j\dg \tau^a \si^b {\cal V}_j\ra =
{J^2  }\la
\tau^{a}[\vec x_j]S^b[\vec x_j]
\ra
\eqno(equiv1)
$$
where $\vec S(x_j)\equiv\vec S_j$ is the local moment at site j
and
$$
\tau^a[\vec x] =  {1 \over 2}\Psi\dg(x) \ul{\vec\tau} \Psi(x)
\eqno(joint)
$$
is the conduction electron ``isospin''.
This order parameter represents a bound-state between the local
moments  and the conduction electron charge and pair degrees of
freedom. The composite order parameter
$$\vec d_c(x)=\hat d^2(x) - i \hat d^2(x)= {2J^2 \over V^2} \la \psi_{\dw}(x)
\psi_{\up}(x)\vec S(x)\ra\eqno(pairo)
$$
represents the development of a {\it joint} correlation between the conduction
electron singlet pair density and the local moment spin density.\refto{emery}
Clearly, this state breaks (i) electron gauge symmetry,
(ii) spin rotation symmetry and (iii) time reversal symmetry.
Despite these features it does not necessarily follow that the
state formed has either an ordered moment, or an equal time pairing
field.

\noindent{\bf 3. Excitation spectrum and quasiparticles}
\taghead{3.}

Let us now examine the nature of the excitation spectrum in this
odd paired state.  Let us begin by rewriting the
mean field
Hamiltonian in terms of the Majorana components \(decomp1), then

$$
H =
\sum_{\vk \in { 1 \over 2 } B.Z.}
\eps_{\vk} \bigl[\psd 0 \psu 0 + \vec \pskd \cdot \vec \psk \bigr]
- i V \bigl[
\vec \pskd \cdot\vec  \eta_{\vk} - { \rm  H . C. } \bigr]   - \mu n_{\vk}
\eqno(hammy)
$$
where
$$
n_{\vk}
= i \bigl[
\psd 0 \psu 3 + \psd 1 \psu 2 -( {\rm H. C. })
\bigr]
\eqno(charge)
$$
is the total charge operator.
Let us consider the special case of $\mu=0$, when the spectrum of
the zeroth component remains unrenormalized.
In this special case, the Hamiltonian can be written in terms
of quasiparticle operators as follows
$$
\eqalign{
H =& H_o + H_g\cr
H_o=& \sum_{\vk\in { 1 \over 2} { \rm B.Z.}}\eps_{\vk} a\dg_{\vk \ 0}
a_{\vk \ 0}
\cr
H_g=&
\sum_{\vk,\ a=(1,2,3)}E_{\vk}a\dg_{\vk a}a_{\vk a} \cr}
\eqno(qps)
$$
The first term describes a gapless ``Majorana'' conduction band
that spans the half  Brillouin zone where $\eps_{\vk} > 0$. The second
term describes a gapped band with excitation energies
$$
E_{\vk}=
{ \eps_{\vk} \over 2} \pm \sqrt{
\left({\eps_{\vk} \over 2}\right)^2
 + V^2}
\eqno(gapped2)
$$
This band spans the entire Brillouin zone, since it incorporates
three Majorana conduction and three Majorana spin fermions.
The basic character of the quasiparticle spectrum is unchanged when
we consider finite deviations from particle-hole symmetry $\mu_{\vk}\neq 0$.
There are two important features (Fig. 4):

\item{\bf 1.} A three-fold degenerate gapful excitation centered around
$\vec k= - \vec Q/2$.
In the vicinity
of the gap
$$
H_g\sim \sum_{\vk \sim -\vec Q/2, \ a=(1,2,3)}
\left[
\Delta_g + { (\vec k+ \vec Q/2)^2 \over m^*} \right]a\dg_{\vk a} a_{\vk
a}\eqno(lowl)
$$
where $\Delta_g \sim V^2/D$ and
$m^* = {D\over \Delta_g}m$ and $m$ is the conduction band mass
at the band edge. Just above the gap, the quasiparticles have almost
no conduction character: correlation functions of the local moments
are then determined through the relation
$$
\eta^b_j = {1 \over \sqrt{ N}}
 \sum_{\vk \in {\rm {1/2} \
 B. Z.}}
\left\{ a_{\vk b} e^{i \vec k \cdot \vec R_j}+
a^{\dagger}_{\vk b} e^{-i \vec k \cdot \vec R_j}\right\}
\eqno(lowl2)
$$

\item{\bf 2.} A neutral Majorana band, located around
$\vec k =0$.

To gain more insight into these excitations, let us consider
the conduction electron propagator
$$
{\cal G}^{-1}(\om) =
\bigl[
\om - \tilde \eps_{\vk} - \mu_{\vk}\ul{\tau}_3 - \ul{ \Sigma(\kappa)}
\bigr]
\eqno(prop2)
$$
Choosing $z_o= \pmatrix{ 1 \cr 0 }$, then $
 (\hat d^1 ,\hat d^2 ,\hat d^3)=
(\hat x, \hat y , \hat z) $, so that $d_{ab} = \delta_{ab}$ and
$$
\ul{\Sigma}(\ka)=  {V^2 \over \om} \ul{\cal P}; \qquad
 \ul{\cal P} =  {1 \over 4}\left[
3(\ul{1}) -   \si^a\otimes \tau^{a}\right]
\eqno(sen3bb)
$$
It is useful to define ``up'' and ``down'' spin projection operators
$$
\eqalign{
P_{\up} = & { 1 \over 2 }  \bigl[  1 +  \si_3 \otimes \tau_3 \bigr]\cr
P_{\dw} = & 1 - P_{\up} = { 1 \over 2 }  \bigl[  1 -  \si_3
\otimes \tau_3 \bigr]\cr
}
\eqno(project)
$$
We can use these operators to project out the
``up'' and ``down'' electron propagators. ($G_{\si}P_{\si} = {\cal G }
P_{\si})$)
$$
\eqalign{
G_{\up} =& [ ( \om - \tilde \eps_{\vk} - \Delta_{\om} ) + \mu \tau_3 +
\Delta_{\om}\tau_1
] ^{-1}
\cr
G_{\dw}=& [ ( \om - \tilde \eps_{\vk} - 2\Delta_{\om} ) + \mu \tau_3 ] ^{-1}
\cr
}\qquad\qquad\biggl[\Delta_{\om} = \Delta(\om) \biggr]
\eqno(ppgtor)
$$
It is also useful to evaluate the determinants
$$
\eqalign{
{\rm det}[G^{-1}_{\up}(\vk ,\om)] =&[( \om - \tilde \eps_{\vk} - \Delta_{\om}
)^2 - \mu^2 - \Delta_{\om}^2]\cr
{\rm det}[G^{-1}_{\dw}(\vk,\om)] =&[( \om - \tilde \eps_{\vk} - 2\Delta_{\om}
)^2 - \mu^2 ]\cr}
\eqno(poles2)
$$
Zeroes of these functions determine the quasiparticle excitation energies
$\om_{\vk\si}$: ${\rm det}[G^{-1}_{\si}(\vk, \om_{\vk \si})]=0$ (Fig. 5).

The ``down''  propagator contains
no pairing terms, and describes
a gapful band of quasiparticles with excitation energies
\def\epm{ \tilde \eps_{\vk} - \mu}
$$
\om_{\vk} =
{ \epm \over 2} \pm \sqrt{
\left({\epm \over 2}\right)^2
 + V^2}
\eqno(gapped)
$$
This spectrum closely resembles the large $N$ solution
to the particle-hole symmetric Kondo model, with a hybridization gap
$2 \Delta_g$.\reftorange{lacroixins}{millisins}{doniachins}

The ``up'' electron propagator describes
a band of odd-frequency paired electrons. The poles of this propagator
$$
G_{\up}(\om)= {  ( \om - \tilde \eps_{\vk} - \Delta_{\om} ) - \mu \tau_3-
\Delta_{\om} \tau_1  \over  [(
\om - \tilde \eps_{\vk} - \Delta_{\om} )^2 - \mu^2 - \Delta_{\om}^2
]}\eqno(uppr)
$$
at $\om=\om_{\vk}$ are determined by the cubic equation
${\rm det}[G_{\up}^{-1}(\vk, \om_{\vk})]=0$ (see \(poles2)).
Solving for the conduction electron energy as a function of $\om_{\vk}$,
$\eps_{\vk}=\eps_{\pm}(\om_k)$, we
find

$$
\eps_{\pm}(\om_{\vk})=
[ \om_{\vk} - \Delta(\om_{\vk})] \pm
{\rm sgn(\om_{\vk})}\sqrt{ [\Delta^2(\om_{\vk}) + \mu^2 ]}
\eqno(det)
$$
which defines  two branches of the ``up'' quasiparticle excitation spectrum.
The (-) branch is gapful
with a gap $2 \Delta_g \sim {2V^2\over D}
\left(1- {\mu^2\over D^2}\right)^{-{1}}$,
where $D$ is the conduction electron
half-bandwidth. The (+) branch is gapless, corresponding to the Majorana
component of the conduction sea that decouples from the resonant
scattering center.
The quasiparticle density of states corresponding to these two
branches is
$$
\eqalign{
N_{\pm}^*(\om) = &{\rho \over 2}Z_{\pm}^{-1}(\om)\cr
Z_{\pm}^{-1}(\om)=& {{\rm d} \eps_{\pm}(\om) \over { \rm d }\om} =
\left\{
1 + { \Delta_{\om} \over \om} \biggl[
1 \mp { 1 \over \sqrt{
1 + \bigl( {\mu \over \Delta_{\om} } \bigr)^2}}
\biggr]
\right\}\cr}
\eqno(dens)
$$
At low energies
$$
\om_{\vk (+)} =
\eps_{\vk}\bigl[
1 + { \mu ^2 \over V^2 }\bigr]^{-1}
\eqno(gapl)
$$
giving an enhanced density of states $N_+^*(0) = {\rho \over 2{\rm Z}_O}$
where ${\rm Z}_O= \biggl[
1 + {\mu^2 \over V^2 }\biggr]$,  at the Fermi surface.

Let us explicitly construct these gapless quasiparticles.
It is convenient to
split the Hamiltonian into ``up'' and ``down'' spin parts
$H= H_{\up } + H_{\dw}$, where
$$
\eqalign{
H_{\dw}=& \sum_{\vk} \left\{
(\eps_{\vk} - \mu )
 \psi\dg_{\vk \dw}
 \psi_{\vk \dw} +  V  \bigl[
 \psi\dg_{\vk \dw}\eta_{\vk \dw} +
({\rm H. C. })\bigr] \right\}\cr
\eta_{\vk \dw} = &{ 1 \over \sqrt{2} }
\bigl[
\eta_{\vk}^ {\rm 1} + i \eta_{\vk }^{{\rm 2}}  \bigr]\cr }
\eqno(downers)
$$
describes the  hybridized band of unpaired ``down'' electrons,  and
\def\l2dash{\joinrel\relbar\joinrel\relbar\joinrel\relbar\joinrel\relbar
\joinrel\relbar\joinrel\relbar\joinrel\relbar\joinrel\relbar
\joinrel\relbar\joinrel\relbar
\joinrel\relbar\joinrel\relbar}
\def\3mat#1#2#3#4{\left[\matrix{
\matrix{\cr#1 &
\cr{\ }\cr
{\l2dash}
\cr #3}\left|\matrix{
#2
\cr {\ldash}\cr #4}\right.} \right]}
$$
\eqalign{
H_{\up} =&\sum_{\vk \in { 1 \over 2 } B.Z.}
A\dg_{\vk}\ul{h}_{\vk} A_{\vk}\cr
\ul{h}_{\vk}
=&
\3mat {{\eps_{\vk}-\mu \ul{\tau_3}
}\ \ }{{\matrix{ {{V \over \sqrt2}}&\cr
{{-{V \over \sqrt2}}}&\cr{\qquad}\cr}}}{{V \over \sqrt2}\quad -{V \over
\sqrt2}}{0}
\cr
A\dg_{\vk} =& ( \psi\dg_{\vk \up}, \psi_{- \vk \up} ,
\eta^{3\dagger}_{\vk})\cr}
\eqno(hammy2)
$$
describes the paired ``up'' electrons.
In terms of the quasiparticle operators
$$
H_{\up} = \sum_{\vk} \om_{\vk \alpha} a\dg_{\vk \alpha} a_{\vk \alpha}
\eqno(qpdec)
$$
where only positive energies enter into the Hamiltonian. Here
$\a = 0,3$ denotes the gapless and gapful excitation branch respectively.
The quasiparticle operators for the gapless ``up'' electrons can be constructed
from a generalized Bogoliubov transformation
$$
a\dg_{\vk 0} = \sqrt{Z_{\vk }}
\biggl(
u_{\vk}\psi\dg_{\vk \up} + v_{\vk} \psi_{-\vk \up}
\biggr)+ \sqrt{(1 - Z_{\vk})}\eta^{3 \dagger}_{\vk}
\eqno(qp2)
$$
The eigenvector $\phi_{\vk }$  containing the Bogoliubov
coefficients
satisfies
$$
\ul{h}_{\vk} \phi_{\vk }= \om_{\vk 0} \phi_{\vk }
\eqno(expand)
$$
Eliminating $Z_{\vk }$ and
substituting back into \(expand) then gives
$$
\ul{ G}^{-1}_{\up}(\vk, \om_{\vk}) \pmatrix{
u_{\vk} \cr
v_{\vk } \cr}
=0
\eqno(exp2)
$$
where $\ul{ G}^{-1}_{\up}(\ka)$ is taken from \(ppgtor).
Diagonalizing this eigenvalue equation gives
$u_{\vk }= u^2(\om_{\vk 0 })$,
$v_{\vk }= v^2(\om_{\vk 0})$, where
$$
\eqalign{
u^2(\om)
 =&{ 1 \over 2}
\left[
1 +{}{  \mu \over \sqrt{
\Delta_{\om}^2 + \mu^2 }}
\right]\cr
v^2(\om)=&{ 1 \over 2}
\left[
1 -{}{  \mu \over \sqrt{
\Delta_{\om}^2 + \mu^2 }}
\right]\cr}
\eqno(bog)
$$
where the energies are given by  \(det). $Z_{\vk }= Z_{+}(\om_{\vk+})$
takes the form
given in \(dens). Energy, rather than momentum-dependent Bogoliubov
coefficients are a characteristic of odd-frequency pairing.

Let us now consider the charge and spin coherence factors of these gapless
excitations. Within the gapless ``up'' band
only the conduction charge and spin
operators contain diagonal matrix elements. Suppose we attempt to excite
quasiparticles out of the ground-state, by coupling to a charge, or
spin excitation; the relevant diagonal
matrix elements are
$$
\la\vk^-
\vert \left\{ \matrix{ \rho_{\vq} \cr \si^z_{\vq}  \cr}\right\}
\vert  \vk^+ \ra =
\la \vk^-
\vert
\bigl\{
\psi\dg_{\vk^- \up} \psi_{\vk^+ \up} -
\psi_{-\vk^- \up} \psi\dg_{-\vk^+ \up} \bigr\}\vert   \vk^+ \ra
\eqno(coherence1)
$$
where $\vk^{\pm} = \vk \pm \vec q /2$
and $\vert \vk^{\pm}\ra = a\dg_{\vk^{\pm}}
\vert 0 \ra$ denotes the state
with one quasiparticle added to the mean field ground-state.
In terms of the Bogoliubov coefficients this is
$$
\la\vk^-,\vert \left\{ \matrix{ \rho_{\vq} \cr \si^z_{\vq}  \cr}\right\}
\vert\vk^+\ra =
 \sqrt{Z_{\vk^+} Z_{\vk^-}}
\biggl[
u_+u_- - v_+ v_-
\biggr]
\eqno(coherence2)
$$
where $u_{\pm} = u_+(\om_{\vk^{\pm}})$ and $v_{\pm} = v_+({\om_{\vk^{\pm}} })$
are the Bogoliubov factors in the gapless band ($+$).
On the Fermi surface, since $u=v={1 \over \sqrt{2}}$, this coherence factor
\underbar{\sl vanishes}.  Away from the Fermi  surface, the spin/charge
coherence factor grows linearly with energy
$$
\la\vk^-
\vert \left\{ \matrix{ \rho_{\vq} \cr \si^z_{\vq}  \cr}\right\}
\vert \vk^+
\ra =
(\om_{\vk^+}+\om_{\vk^-})
 \biggl( {\mu \over V^2+ \mu^2}\biggr)
\quad \left(\om_+,\om_- << \Delta_g \right)
\eqno(coherence2.5)
$$
(In the special case of particle-hole symmetry, these
coherence factors vanish \underbar{throughout} the gap.)
In a similar fashion, we may examine quasiparticle  components of the
charge and spin, given by
$$
\eqalign{
\left\{\matrix {Q \cr
\si^z}
\right\}_{\vk}=&\lim_{\vq \rarrow 0}
\la\vk^- \vert \left\{ \matrix{ \rho_{\vq} \cr \si^z_{\vq}  \cr}\right\}
\vert\vk^+\ra =
Z_{\vk}
\biggl[
u_{\vk}^2 - v_{\vk}^2
\biggr]\cr
=&
 {\rm Z}(\om_{\vk}) { \mu \over \sqrt{ \Delta^2(\om_{\vk })
+ \mu^2 }}\cr
\sim &{2 \mu \om_{\vk} \over{ \mu^2 + V^2 }} \qquad(\om_{\vk}<< \Delta_g)}
\eqno(coherence3b)
$$
 From these results, we conclude that there is no way to couple
via charge or spin probes to the quasiparticle excitations
at the Fermi surface. These gapless excitations
are devoid
of charge, or spin quantum numbers on the Fermi surface.
This dramatic effect is a direct consequence of the resonant pairing
and the pole in the gap function. So long as this pole is maintained,
the coherence factors will identically vanish on the Fermi surface.
Note however, that these quasiparticles  can still carry
entropy, and in this sense can be regarded as thermal quasiparticles.

It is particularly instructive to examine the {\sl local} conduction
electron propagator and the pair wavefunction in this simple
mean field theory. The local propagator for the paired ``up'' electrons
is
$$
\ul{G}_{\up}(\om)
= \sum_{\vk} \ul{ G}_{\up}(\vk,\om )  = \rho \int_{-D}^{D} {d \eps }
{ G}_{\up}(\eps,\om )
\eqno(local)
$$
Carrying out the integral over the conduction electron energies
we find
$$
{1 \over  \pi \rho}{\rm Im} \biggl[\ul{ G}_{\up}(\om-i \delta)\biggr] =
\left\{
\eqalign{   &\ul{1} \qquad\ \ \qquad\qquad\qquad\qquad\qquad( \vert \om\vert >
\Delta_g)\cr
{ 1 \over 2}
&\left[
1 + {\rm sgn(\om)}{ \Delta ( \om) \tau_1 + \mu \tau_3 \over \sqrt{
\Delta(\om)^2 + \mu^2 }}
\right]
\ \ ( \vert \om\vert < \Delta_g)\cr} \right.
\eqno(localprop)
$$
Loosely speaking, the  electrons are normal outside the gap region
and become paired at energies less than the gap $\Delta_g$.
The spectral function can be rewritten in terms of
the energy dependent Bogoliubov coefficients derived in \(bog)
$$
{1 \over \pi \rho}{\rm Im}\biggl[\ul{{ G}}_{\up}(\om-i \delta)
\biggr] = \left[\matrix{
u_{\om}^2 & u_{\om}v_{\om}\cr
u_{\om}v_{\om}& v^{2}_{\om}\cr
}
\right]\eqno(bogprop)
$$
where the coefficients $u_{\om}$ and $v_{\om}$ are evaluated in
the gapless band.

Finally, we may construct the
pair wavefunction from
the off-diagonal components of this spectral function
$$
\eqalign{
\la \psi_{\up}(x ,i \om_n) \psi_{\up}(x , -i\om_n) \ra
=& {\rho\over 2}\int_{-D}^D {d \om \over \pi}{ 1 \over i \om_n - \om}
{ \Delta ( \om){\rm sgn}(\om)\over
\sqrt{\Delta(\om)^2 + \mu^2 }}\cr
=&-i \om_n \rho\int_{0}^D {d \om \over \pi}{ 1 \over [ \om_n^2 +\om^2]}
{ \Delta ( \om)\over
\sqrt{\Delta(\om)^2 + \mu^2 }}\cr}
\eqno(pairing2)
$$
thereby
explicitly displaying the odd-frequency character of the pairing.

\noindent{\bf 4. Mean Field Thermodynamics}

\taghead{4.}

Next, we discuss the mean field thermodynamics.
The mean field free energy per site is written in terms
of the conduction electron propagators as
$$\eqalign{
F_{MF} = &
{ V^2 \over J} -{T\over 2} \sum_{\ka}
{\rm Tr}
l { \rm n}\biggl[
{ \cal G}^{-1} ( \ka)\biggr]\cr
= &
{ V^2 \over J} -{T\over 2} \sum_{\ka}
l { \rm n}\left\{
{\rm det} \biggl[
{ \cal G}^{-1} ( \ka)\biggr]
\right\}\cr}
\eqno(mf1)
$$
where the determinant can be expanded in terms of the ``up'' and ``down''
components of
the propagator \(ppgtor)
$$
{\rm det} \biggl[
{ \cal G}^{-1}(\ka) \biggr]
={\rm det} \bigl[
{  G}_{\up}^{-1}(\ka) \bigr]
{\rm det} \bigl[
{ G}_{\dw}^{-1}(\ka) \bigr]
\eqno(mf2)
$$
Differentiating with respect to the order parameter, yields
the mean field equation
$$
{ 1 \over J}+
{T\over 2} \sum_{\ka}
{ 1 \over \iom}\biggl\{ {\dsp
{
\iom - \tilde \eps_{\vk} \over
{\rm det} \bigl[
 {  G}_{\up}^{-1}(\ka) \bigr]
}
+
{2(\iom - \tilde \eps_{\vk}-2\Delta_n) \over
{\rm det} \bigl[
 { G}_{\dw}^{-1}(\ka) \bigr]
}
}\biggr\}=0
\eqno(mf3)
$$
where the denominators in these equations are given in \(poles2).
For a constant conduction electron density of states $\rho$, we
may replace
$$
\sum_{\vk}\{ \dots \} \longrightarrow \rho \int_{-D}^{D} d\eps \{ \dots \}
= \rho \int {d z \over 2 \pi i}
\Theta\bigl[ z \bigr]\{ \dots \}
\eqno(replace)
$$
where the contour integral proceeds clockwise around the branch cut in
the function
$$
\Theta
\bigl[ z \bigr] = l {\rm n}
\left[
{z - D \over -D -z }
\right]
\eqno(cutoff)
$$
The energy integrals can then be performed by closing the contour
around the poles in the Green functions, which are located at
$$
\eqalign{\eps_{\up \alpha}(\omega) =&
\om - \Delta_{\om} + \alpha \ {\rm sgn}(\om)\sqrt{\Delta_{\om}^2 + \mu^2}
\cr
\eps_{\dw \alpha}(\omega) =&
\om - 2\Delta_{\om} + \alpha  \mu\cr}
\qquad\qquad(\alpha= \pm)
\eqno(pole2)
$$
for the up and down electrons  respectively.
Carrying out the complex integral then gives
$$
{1\over J} ={\pi T\over 2} \sum_{{\iom \atop\a = \pm}}
{1
\over \iom}\left\{
F_{\up\a}(\iom)
\biggl(
{ 1 \over 2} -
{\a\Delta_n \over 2 \sqrt{ (\Delta_n^2 + \mu^2)}}
\biggr)
+
 F_{\dw \a}(\iom)\right\}
\eqno(mf5)$$
where $F_{\si\a}(z) = \rho \Theta\bigl[
\eps_{\si \alpha}(z)\bigr]/\pi$ and
we have used the notation $\Delta_n\equiv \Delta(\iom)$.
Carrying out the Matsubara sums then yields
$$
{ 1 \over J} = \sum_{\a = \pm}\int {d \om \over 4\om}
{ {\rm th}({\beta\om\over 2})
}
\left\{
{F^{''}_{\up\a}(\om^+)\over 2}
 \biggl(
1-{\a\vert \Delta_{\om} \vert \over
 \sqrt{ (\Delta_{\om}^2 + \mu^2)
}}
\biggr)
+
F^{''}_{\dw\a}
(\om^+)
 \right\}
\eqno(mf6)
$$
where ${\om^+=\om+i\delta}$.
The functions $F^{''}_{\si \a}(\om^+)
\equiv {\rm Im}F_{\si\a}(\om^+)$ count the number of
up and down excitation branches at frequency $\om$.
Ignoring the small differences between the up and down
spin excitation gaps,
$$
F^{''}_{\si \a}(\om^+)=\rho \left\{
\eqalign{\theta(D- \vert \om\vert)\qquad\qquad ( \si = \up \a = +)\cr
\theta(D- \vert \om\vert) -\theta(
\Delta_g - \vert \om \vert )\qquad\qquad ({\rm otherwise})\cr}\right.
\eqno(states)
$$
for the gapless and gapful branches, respectively.
This simple mean field theory then gives  rise to a phase transition at a
temperature
$$T_c\sim D \exp \biggl[-{ 1 \over (3/2J \rho)}\biggr]= T_K
\exp \biggl[-{ 1 \over (6J \rho)}\biggr]
$$
without the formation of an intermediate heavy fermion phase,
where $T_K =D \exp \bigl[-{ 1 \over (2J \rho)}\bigr]$ is the single
ion Kondo temperature.
At $\mu=0$, the gapless excitation branch
of the spectrum does not contribute to
mean field equation.  At finite $\mu$, the gapless branch develops
a small linear coherence factor, and we see that this has the effect
of suppressing the transition temperature. For all values of
$\mu$ however, the form of the mean field $\Delta_g(T)$ quite closely
resembles that of a singlet BCS superconductor.

The precise relation
between the single
ion Kondo temperature and $T_c$ is not reliably predicted by the
mean field theory.
Our path integral approach amounts to a ``Hartree'' decoupling
of the interaction. Had we chosen a more conventional
diagrammatic approach, carrying out
a ``Hartree-Fock'' decoupling
of the original Hamiltonian, writing
$$
\la \vec \eta \psi_j \ra = -{1 \over 2J}\vec \si \pmatrix{ V_{j\up} \cr
V_{j\dw}\cr}\eqno(junk2)
$$
so that
$$
-i{J\over 2} {\psi\dg_j \vec \eta }
\times
{\vec \eta \cdot \vec \si \psi_j}
-i{J\over 2} \psi\dg_j \vec \eta\times\vec \eta \cdot \vec \si
\psi_j\longrightarrow -{3 \over 2J} V_j\dg V_j
\eqno(hfcd)
$$
then the mean field Hamiltonian would have become
$$
H_{int}[j]=\longrightarrow
\psi\dg_j ( \vec \si \cdot \vec \eta_j)V_j
+V\dg_j
( \vec \si \cdot \vec \eta_j)\psi_j + 3|V_j|^2/2J
\eqno(int2)
$$
so that for this scheme, $T_c^{MF} = T_K$, to logarithmic accuracy.
The path integral approach recovers  the ``Fock'' contributions
to the pairing as a leading order component
of the RPA fluctuation corrections to the
mean field transition temperature (Fig. 6).
Since fluctuation effects will suppress $T_c$ in either scheme,
the particular choice of mean field theory is somewhat arbitrary, and
will not matter at the next level of approximation.

If we take our mean field theory literally,
we see that from the point of view of the original conduction band, the
transition into the odd frequency state can occur for arbitrarily
weak coupling constant, taking advantage of the Kondo effect to
produce a logarithmic divergence in the pairing channel.
Significant pair breaking effects will of course come from the fluctuations.
In the one impurity problem, there are infrared divergences in the Gaussian
fluctuations that suppress the mean field transition temperature to
zero.
However, (unlike the large $N$ approach)
in the lattice model
there is $\ul{\rm no}$ continuous gauge symmetry
so the development of a gap in the spectrum
will cut off the one-impurity infrared divergences, preserving a finite
temperature transition.

Finally, we note that expanding the Free energy at low temperatures yields a
linear
specific heat proportional to the density of states in the gapless
band
$$
\eqalign{
\gamma=& { \pi^2 k_B^2 \over 3}{\rho \over 2} \left[
1+ \biggl( {\mu \over V} \biggr)^2\right]\cr
=& { \pi^2 k_B^2 \over 3} {1 \over 2} \left[
\rho + { 1 \over 2 \Delta_g}\biggl( {\mu \over D} \biggr)^2\right]\cr}
\eqno(sheat)
$$
Depending on the ratio $(\mu/D)$, this linear specific heat can range from
a value characteristic
of the free conduction band,
$(\mu/D) \sim 0$
to a value more characteristic of a heavy
fermion metal $\sim {1 \over \Delta_g}$ for $(\mu/D) \sim 1$ (Fig. 7).

\noindent{\bf 5. Rigidity of the odd frequency paired state}

\taghead{5.}

One of the key issues associated with odd frequency pairing, is whether
it leads to a real superconducting Meissner effect.
Past attempts to
construct an odd frequency paired state within an Eliashberg formalism,
have experienced difficulty in producing a state with a positive
phase stiffness\refto{private} and a finite London penetration depth.  To begin
our discussion, we
first discuss the form of the long-wavelength effective action.

\noindent{\bf 5. (a) Long-wavelength action}

In general, the mean field free energy will depend on gradients of the
order parameter field, and the form of the applied vector potential.
Let us now consider slow deformations of the order parameter
$$
z(x) = g(x) \pmatrix{1 \cr 0 }\eqno(deform2)
$$
where
$$
g(x) = e^{\bigl[i {\vec \theta(x) \cdot \vec s
}\bigr]}
=
\pmatrix{
z_{\up} & -z^*_{\dw} \cr
z_{\dw} &  z^*_{\up} \cr
}\qquad\qquad \vec s = \biggl({ \si \over 2}\biggr)
\eqno(deform)
$$
is an $SU(2)$ rotation matrix.
The rate of  rotation is given by
$\vec \nabla g = g \vec \om$, where
$$
\vec \om=
 g^{-1} \vec \nabla g ={i}\vec \om_{\lambda} s^{\lambda}
\eqno(rotation)
$$
is decomposed in terms of its components $\vec \om_{\lambda}(x)$ along the
principle axes $\hat d_{\lambda}$ of the order parameter.
The leading quadratic terms in the
gradient expansion of the Free energy about the uniform
mean field theory are then
$$
F  =
\sum_{\lambda = 1,3}{\rho_{\lambda}
\over 2} \int {
\vec \omega_{\lambda}^2 }
d^3  x
\eqno(lwact)
$$
Here the stiffnesses $\rho_{\lambda}$
for slow twists about each principle
axis are analogous to the moments of inertia of a top.

To include the effects of an external magnetic field, we introduce
a finite vector potential by an appeal
to gauge invariance. Our original model is  gauge invariant under the
transformation
\def\lrow{\longrightarrow }
$$
\eqalign{
\psi(x) \lrow
& e^{i \phi(x) } \psi(x)\cr
z(x)\lrow & e^{i \phi(x)}z(x)\cr
\vec A(x)
\lrow & \vec A(x) + \biggl({\hbar\over e}\biggr)
\vec \nabla \phi(x)
\cr}
\eqno(ginv2)
$$
so that
$$
\eqalign{
g(x)\lrow & g(x)e^{i \phi(x)\si_3}\cr
}
\eqno(ginv3)
$$
This means that  the long wavelength action must be invariant under the
transformation
$$
\eqalign{
\vec \omega_3 \lrow & \vec \omega_3 + 2\vec \nabla  \phi(x)\cr
{e \over \hbar} \vec A(x) \lrow &{e \over \hbar} \vec A(x)
+ \vec \nabla \phi(x)\cr}
\eqno(ginv4)
$$
In other words, a uniform vector potential $\vec A$ is  equivalent to a
uniform rotation rate $-{2e\over \hbar} \vec A$
about the $\hat d_3$ axis.
The gauge invariant form of the Free energy is then
$$
F  =
{1
\over 2} \int d^3  x
\biggl[
\rho_{m}\vec \omega_{\perp}^2 +
\rho_s\bigl(\vec \omega_{3}- {2e \over \hbar} \vec A\bigr)^2
+ {B^2 \over  \mu_o}\biggr]
\eqno(lwact2)
$$
In terms of the vector $\hat n \equiv \hat d_3 = z\dg \vec \si z$, this action
can also
be written
$$
F  =
{1
\over 2} \int d^3  x
\biggl[
\rho_{m}\bigl(\nabla \hat n\bigr)^2  +
\rho_{s}
\bigl(\vec \omega_{3}- {2e \over \hbar} \vec A\bigr)^2
+
{B^2 \over  \mu_o}\biggr]
\eqno(lwact3)
$$
The term $\rho_s$ is the Meissner stiffness of the superconductor,
whereas the term $\rho_{m}$ can be regarded as a ``spin stiffness''
of the triplet paired state.
It is at first surprising that a charge $1e$ spinor order parameter
can give rise to a  charge $2e$ coupling between gradients of the phase
and the vector potential. We can resolve this apparent paradox by noting that
that  $z(x)$ is also a spin $1/2$ object, thus a phase change
$$
\pmatrix{1 \cr 0} \lrow e^{i \phi}\pmatrix{1\cr 0}
$$
corresponds to a rotation through $\theta= 2 \phi$.
A rotation through $2 \pi$ leads to a sign change in $z(x)$,
and in physical configurations, $z(x)$ must be continuous up to
a sign $\pm 1$.  The gauge invariant
coupling between $\phi$  and $\vec A$ is then
$$
(\nabla \phi- {e \over \hbar} \vec A)^2 \lrow
{1 \over 4}(\nabla \theta- {2e \over \hbar} \vec A)^2
$$
and hence the coupling between physical rotations and the vector
potential is a charge $2e$ coupling, as in conventional superconductivity.
Note finally, that
if we include spin anisotropy into the original Hamiltonian,
then this will tend to align the order parameter, for example, through the
inclusion of a term of the form
$$
F_a = -{ \rho_{m} \over l_o^2 }
\int d^3  x
(\hat d_3\cdot \hat z)^2
\eqno(aniso)
$$
On length scales $l > l_o$, the system behaves as a
conventional Landau-Ginzburg theory.

\noindent{\bf 5. (b) Computation of the Meissner Stiffness}

To compute the Meissner and spin stiffnesses
in this gradient expansion,
we consider a configuration with a uniform rotation about
the principle axes $\hat d_a$
$$
z_j= e^{[i \vec \omega\cdot \vec R_j {\rm S}^a] }z_j^0\qquad\qquad\biggl(
S^a \equiv { \si^a \over 2}\biggr)
\eqno(slowrot)
$$
We may absorb this uniform rotation into a gauge transformation of the
the conduction electrons through the replacement of
the conduction electron kinetic energy by
$$
\eqalign{
\tilde \eps_{\vk}
\lrow
\tilde \eps_{\vk - \vec \omega {\rm S}^a} =& \tilde \eps_{\vk} + h({\vk})
\cr
\ul{h}({\vk})  = - \om_{\mu}\nabla_{\mu} \tilde \eps_{\vk}\ul{\rm S}^a
+ { 1 \over 8}\om_{\mu}\om_{\nu}\nabla^2_{\mu\nu } \tilde \eps_{\vk}\cr}
\eqno(gtransf2)
$$
Here we have expanded the kinetic energy to quadratic order in the
twist.  The effect of the twist in the phase can then be included into
the electronic Green function by
$$
-{\cal G}^{-1}(\ka ) \lrow
-{\cal G}^{-1}(\ka ) + h({\ka})
\eqno(gtransform)
$$
The Free energy of the system in the presence of a uniform twist
can be calculated from the trace of the conduction electron
propagator, as follows
$$
\eqalign{
F[\vec \om_a]-F[0] =&
-{ T \over 2} \sum_{\ka}
{\rm Tr}\biggl\{l{\rm n}\bigl[-{\cal G}^{-1}(\ka ) + h({\ka})\bigr] -
l{\rm n}\bigl[-{\cal G}^{-1}(\ka ) \bigr]\biggr\}\cr
=&
-{ T \over 2} \sum_{\ka}
{\rm Tr}\biggl\{l{\rm n}\bigl[1 - {\cal G}(\ka) h({\ka})\bigr] \biggr\}\cr
}\eqno(freeshift)
$$
Expanding the logarithm to quadratic order gives
$$\eqalign{
F[\vec \om]-F[0] =& {\rho_a \vec \omega^2 \over 2}\cr
\rho^a=&
{T \over 8} \sum_{\ka}
\bigl({\nabla^2 \tilde \epsilon_{\vk} \over 3}\bigl)
{\rm Tr}\biggl[{\cal G}(\ka)\biggr]
+ \nabla {\tilde \eps_{\vk}}^2
{\rm Tr}\biggl[{\cal G}(\ka)\si^a
{\cal G}(\ka)\si^a
\biggr]
}\eqno(stiff3)
$$
The first term in $\rho^a$ can be integrated by parts
to yield
$$
\rho^a =
{T \over 8} \sum_{\ka}
(\vec \nabla \tilde \eps_{\vk})^2
{\rm Tr}\biggl[
{\cal G}(\ka)\si^a
{\cal G}(\ka)\si^a
-{\cal G}(\ka)^2
\biggr]
\eqno(stiff4)
$$
for the stiffness about the $\vec d_a$ axis.
For our simple model, $\rho^1= \rho^2 = \rho_{m}$, and in the special
case where $\mu = 0$ all three stiffnesses are equal.
We shall explicitly focus on the stiffness about the $\hat n \equiv \hat
d_3 $ axis, which is associated with the London Kernel
$$
\eqalign{
{\partial^2 F \over \partial A_{\nu}\partial A_{\mu}} =& Q_{\mu \nu}\cr
Q_{\mu \nu} =& 4e^2 \rho_3\delta_{\mu \nu}\cr}\eqno(kernel)
$$
We can separate the trace into ``up'' and `` down'' components. The
``down'' component is unpaired and explicitly vanishes. The ``up'' component
gives
$$
\rho_3 =
-{T \over 4} \sum_{\ka}
\vec v_{\vk} ^2
{
2 \Delta^2 _n \over
\bigl[
( \iom - \Delta_n - \tilde\eps_{\vk} ) ^2 - \Delta_n^2 - \mu^2
\bigr]^2}\ \ \qquad(\vec v_{\vk} = \vec \nabla \tilde \eps_{\vk})
\eqno(theworks)
$$

In  a conventional
BCS theory, it is sufficient to impose a low energy cutoff
on the frequency sum,
$$
\vert \om_n \vert \le \Lambda
$$
after which, the conduction electron band-width can be taken to
infinity. We are unable to take this continuum limit, for we
must maintain the value of the excitation gap of the
unpaired down electrons  $\Delta_g \sim  V^2 /D$: this
means that we must maintain a finite band electron cutoff.
(In other words, the
number of electrons per local moment  $\sim N(0)D$ must remain finite.)
Since our mean field theory will not be accurate at frequency scales
that are large compared with the Kondo temperature, we shall
choose a frequency cutoff that is intermediate between the conduction
electron bandwidth and the Kondo temperature
$$
T_K <<   \Lambda << D
$$
With this choice, we are able to replace the
density of states by its value at the Fermi surface:
the energy integral is then carried out in the same fashion as section (4.),
replacing
$$
\sum_{\vk}\{ \dots \} \longrightarrow \rho \int_{-D}^{D} d\eps \{ \dots \}
= \rho \int {d z \over 2 \pi i}
\Theta\bigl[ z \bigr]\{ \dots \}
\eqno(replace2)
$$
where the contour integral proceeds clockwise around the branch cut in
the function
$
\Theta
\bigl[ z \bigr] = l {\rm n}
\left[
{z - D \over -D -z }
\right]
$.
The energy integrals can then be performed by closing the contour
around the poles in the Green functions, which are located at
$$
\eps_{\up \alpha}(\omega) =
\om - \Delta_{\om} + \alpha \ {\rm sgn}(\om)\sqrt{\Delta_{\om}^2 + \mu^2}
\qquad\qquad(\alpha= \pm)
\eqno(pole23)
$$
Carrying out the contour integral in $z= \epsilon$ then gives
$$
\eqalign{
\rho_3 =& {\rho v_F^2 T\over 24}  \sum_{ \iom , \ \alpha}
{\alpha \Delta_n^2 \over ( \Delta_n^2 + \mu^2 )^{3 \over 2}}
\Theta[\epsilon_{\alpha}(\iom)]\cr
=& {\rho v_F^2 \over 24}  \sum_{ \alpha} \int {d \omega \over 2 \pi}
{\rm Im} \{\Theta[\epsilon_{a}(\om+i \delta)]\}{\rm th}[\beta \om /2]
{ \a \Delta_{\om}^2 \over ( \Delta_{\om}^2 + \mu^2 )^{3 \over 2}}\cr}
\eqno(stiff5)
$$
These two poles cancel one another's contributions, except in the
gap region $\vert \om \vert < \Delta_g$, where the gapless
excitation branch contributes a finite amount to the stiffness.
Our final result for the London
stiffness  is then
\def\th{{\rm th}\biggl[{\beta \om\over 2}\biggr]}
$$
Q = 4 e^2 \rho_3 = {Ne^2 \over 4 m}  \int_0^{\Delta_g} {d \omega}\th
{ \Delta_{\om}^2 \over ( \Delta_{\om}^2 + \mu^2 )^{3 \over 2}}
\ \ \qquad(\Delta_{\om} = {V^2 \over  2 \om})
\eqno(stiff6)
$$
where we have set ${N \over m} \equiv
{2 \rho v_F^2 \over 3}$. By making the low temperature expansion
$$
\th = {\rm sgn}(\om) + { \pi^2 {\rm T}^2 \over 3} \delta'(\om)+ O(T^4)
\eqno(expando)
$$
the temperature dependence of the Meissner stiffness and penetration
depth become
$$
{1  \over \lambda^2_L(T)}=
{1 \over (\lambda_L^o)^2}    \left(
1 - F({\mu  \over D}){ \pi^2 {\rm T}^2 \over 3\Delta_g^2}
\right)
$$
where ${4 \pi} Q(T) =  [\lambda_L(T)]^{-2}$ defines the
London penetration depth and
$$
F(x) = 2x^2 \left[1 - { 1 \over \sqrt{1 + 4x^2}}\right]^{-1}
\eqno(funny)
$$
This $T^2$ variation of the penetration depth is similar to that
expected for point-nodes in a conventional pairing scenario.
In the special case of $\mu=0$, the Meissner stiffness is simply
$$
Q =  {Ne^2 \over  m} \left( {
\Delta_g^2 \over 4 V^2 }\right) \approx
{Ne^2 \over  m} \left( {
\Delta_g  \over 4 D }\right) \eqno(hard)
$$
where we have set $\Delta_g = {V^2 /  D  }$.
The stiffness of the order parameter is thus finite, but
suppressed by a factor
of $Z=\Delta_g/D$ compared with a conventional metal.
Loosely speaking,
we may consider this to be an effect of the condensation of heavy fermions,
whose effective mass is enhanced by a factor $m^*/m\sim 1/Z$,
and whose rigidity
is then depressed by the factor $m/m^*\sim Z$.

The staggered
phase of the order parameter plays a critical role in developing
this finite stiffness. This point is illustrated in Fig. 8.
The ``uniform'' odd-frequency
triplet state ($\vec Q=0$) is unstable and its energy may
be monotonically reduced by twisting the order parameter until
the stable minimum at $\vec Q = (\pi,\pi,\pi) $ is reached.

The ``spin stiffness'' $\rho_{m}=\rho_{1,2}$ for twisting the order
parameter about the $\hat d_{\perp}$ axes can be calculated in a similar
fashion.
When $\mu=0$, the system is particle-hole symmetric, and
$\rho_{m} = \rho_3$ as given above.
Like the superfluid stiffness $\rho_s$, contributions to the stiffness come
predominantly from the neutral excitation band inside the gap, though
the formal expression for $\mu \ne 0$
is more complicated, and shall not be given here.

\noindent{\bf 5. (c) Collective Modes}

%
%
%
To end our discussion on the long-wavelength properties, we
should like to briefly mention the collective modes of the condensate.
Let us generalize the effective action to incorporate the leading
order time dependence of the pairing field
$$
\eqalign{S  = &
{1
\over 2} \int dt d^3  x
\left[\biggl\{
\chi_m\bigl(\partial_t\hat n\bigr)^2
-\rho_{m}\bigl(\nabla \hat n\bigr)^2 \biggr\}\right.\cr +&
\left.\biggl\{\chi \bigl(\omega_{3}^0-{2e \over \hbar}
V\bigr)^2  -
\rho_{s}
\bigl(\vec \omega_{3}- {2e \over \hbar}
 \vec A\bigr)^2 \biggr\}
+
\biggl({(E/c)^2 -B^2
\over  \mu_o}\biggr)\right]}
\eqno(lwact33)
$$
where $\chi_m$ denotes the magnetic, or spin susceptibility, and
$\chi$ denotes the charge susceptibility.
An applied vector potential is gauge equivalent to a rotation about
the $\hat d^3$ axis, and must be included with the kinetic terms to
maintain the gauge invariance, as explained above.
The spin and charge susceptibilities are given
by $\chi=\chi_m=\rho$. In the absence of a coupling to the electromagnetic
field, this action would give rise to a collective phase mode, and
a spin-wave mode, with velocities
$$
v_{spin}^2\sim v_{phase}^2 \sim v_F^2 \left( { T_K \over D}\right)
\eqno(collective)
$$
Of course, phase modes are gauged away as fluctuations of the electromagnetic
field, converting the
phase mode into a longitudinal plasmon mode as part of the Meissner
effect, following the well-known
Anderson-Higgs mechanism.\refto{anderson} The spin-wave mode cannot
be gauged away in this fashion and is {\sl unscreened}, leading
to gapless collective spin modes in the superconducting state that
coexist with the superconductivity.
 From the velocity of the spin-wave excitations, we can see that these
modes cross into the bottom of the quasiparticle continuum at a wave vector
$$
q_o\sim \sqrt{{\Delta_g\over D}} a^{-1}\eqno(qzero)
$$
where $a$ is the lattice parameter. This
is much smaller than the size of the Brillouin zone so
long-wavelength
fluctuations of the order parameter will not lead to a dramatic
reduction in its magnitude. On length scales shorter than
$$
\xi\sim a \sqrt{{D \over \Delta_g}}\eqno(phasecoher)
$$
and at frequencies greater than
$$
\om\sim \Delta_g
$$
this system will behave much in the way of a single Kondo impurity.
The development of coherence on longer length-scales provides
a vital cut-off to the infrared fluctuations that destroy the
condensate in a single impurity model.\refto{nickcat, colemanlong}

\noindent{\bf 6. Effect of coherence factors on local Response Functions}
\taghead{6.}

The unusual nature of the coherence factors in the quasiparticle
excitations have interesting consequences for the low frequency
response of this system.  Of particular interest here, are the local
dynamical spin susceptibility  and charge susceptibilities
$$\eqalign{
\chi^c(\om) =&
 -i\int_0^{\infty} dt \la [ \rho_c(t) , \rho_c(0)] \ra e^{i \om t}\cr
\chi^s_{ab}(\om)
=& -i\int_0^{\infty} dt \la [ S_a(t) , S_b(0)] \ra e^{i \om t}\cr}
\eqno(dyns)
$$
These functions are directly related to the ultrasonic attenuation
and the NMR relaxation rate, $1/T_1$
$$
\eqalign{
\a_s(T) =& \lambda_1 \lim_{\om\rarrow 0} \left[{
{\chi^{c''}}(\om) \over \om} \right]\cr
{1 \over T_1}({\hat b}) =& \lambda_2 \lim_{\om\rarrow 0} \left[{
{\chi_{+-}^{''}(\om) \over \om}} \right]\qquad\qquad
\chi_{+ -}(\om) = {\rm Tr} \bigl[(\ul{1} - \hat b \hat b)\ul{\chi}^s(\om)]
\cr}\eqno(real)
$$
associated with the conduction electrons. Let us focus on contributions
to these response functions derived from
the gapless excitations in the ``up'' spin band of
our toy model.

The imaginary part of the local spin or charge
response function of these excitations is given by
$$
\eqalign{{\chi^c(\om)\over \om}=&
{4\chi^s_{zz}(\om)\over \om} \cr
=&\pi
\sum_{\vk_1,  \ \vk_2}
\vert\la \vk_1 \vert
\rho_{\vk_2 - \vk_1}
\vert\vk_2 \ra
\vert^2
{
\bigl[
f(E_{\vk_1}) - f(E_{\vk_2})
\bigr]\over \om}
\delta[(E_{\vk_2}-E_{\vk_1})-\om]\cr}
\eqno(impart)
$$
(The only component of the susceptibility
matrix which couples to the low energy quasiparticles is
$\chi^s_{zz}$.)
Since the coherence factors grow linearly
in the energy
$$
\la\vk^-
\vert \left\{ \matrix{ \rho_{\vq} \cr \si^z_{\vq}  \cr}\right\}
\vert \vk^+
\ra \sim \sqrt{Z_1Z_2}
(\om_{\vk^+}+\om_{\vk^-})
 \biggl( {\mu \over V^2}\biggr)
\eqno(coherenc2)
$$
The low energy form of this response function is given by
$$
\eqalign{
{\chi^c(\om)\over \om}=&
{4\chi^s_{zz}(\om)\over \om}=\pi
\int dE N^*(E_+)Z(E_+)
N^*(E_-)Z(E_-)\left\{ \dots\right\}\cr
\left\{ \dots\right\}= &
\biggl[
{f(E_{-}) - f(E_{+})
\over \om}\biggr]
\left[{(E_++E_-)\mu \over V^2}\right]^2
\cr}
\eqno(coherenc3)
$$
where $E_{\pm} = E \pm \om/2$.
The density of states of the quasiparticles is $N^*(E) = { ({\rho/
2}) Z^{-1}(E)}$, thus at low temperatures and frequencies
$$
\chi^c(\om)/\om=
4\chi^s_{zz}(\om)/\om =
{ \pi \rho^2 \mu^2 \over 4} \left(
{ \om^2 + (2 \pi T)^2 \over V^4}
\right)\eqno(cohernc4)
$$
This quadratic temperature and frequency dependence
of the local charge and spin responses results from the unique
energy dependence of the spin and charge matrix elements,
and the neutrality and spinless character of excitations at
the Fermi surface.
In a more conventional superconductor, where gapless excitations
carry charge and spin, these
matrix elements would be unity at the Fermi energy, and this kind
of quadratic behavior can only be produced by a {\sl linear}
density of states.  A quadratic growth of the
dynamic spin and charge susceptibility results in characteristic
$T^3$ response of the NMR relaxation rate (Fig. 9).
$$
{ 1 \over T_1T} \propto
\left({\mu \over D}\right)^2
\left({T\over T_K}\right)^2
\eqno(nmrfact)
$$
and a $T^2$ response of the ultrasonic attenuation rate.
$$
\alpha_s(T) \propto  \biggl({\mu \over
D}\biggr)^2 \biggl({T \over T_K}\biggr)^2.\eqno(usoundfact)
$$
This property of the odd-frequency paired state is of particular
interest, because power-law behavior of the above variety
is observed in heavy fermion systems. Conventionally,
it is ascribed to d-wave pairing and gaps vanishing
along lines of the Fermi surface. Our odd-paired state offers
the interesting alternative description of this as a matrix
element effect.

As a more specific illustration of this effect, let us examine
the detailed temperature dependent NMR relaxation effect in our
toy model.
The complete local dynamical spin susceptibility has contributions from both
the
conduction electrons and the local spins
$$\eqalign{
\chi_{ii}(\om) = &-i \int {dt} \exp(-i\om t) \theta (t) \la [ S_i(t),
S_i(0) ] \ra; \cr
\vec S = &\sum_i \left[ \vec S_c + \vec S_f \right].
}
\eqno(Dynspin)
$$
In general, the local moment spin operator mixes
the gapless and gapful excitations and only the conduction electron
spin operator in the $\hat z$ direction
is able to create gapless excitations. Let us examine the
corresponding conduction electron spin response in Matsubara
frequency,
given simply by
$$
\chi^s_{zz}(\iom) = \la S^c_{z}(\iom) S^c_{z}(-\iom) \ra = {1 \over 4}
\sum_{\si = \pm}\la \rho_{c\si}(\iom) \rho_{c\si}(-\iom) \ra
\eqno(condresp)
$$
We may now write
$$
\chi^s_{zz}(\iom) = -{T\rho^2 \over 16} \sum_{\inu} {\rm Tr}
\left[ {\cal G}_{\up}(\inu + \iom) \tau_3 {\cal G}_{\up}(\inu)\tau_3 \right].
$$
where we have used  the summations over momentum to replace the ``up''
conduction
propagators by the corresponding local propagator \(localprop)
$$
{1 \over \rho}\sum_{\vk} {\cal G}_{\vk\up}(\om + i\delta) = {1\over 2}
\sum_{\alpha} \rho_{\up \alpha}(\om) \biggl\{ -1 +
{\alpha ( - \Delta_{\om} \tau_1 - \mu \tau_3 ) \over E_{\om}}
\biggr\}.
$$
where
$$\eqalign{
E_{\om} = {\rm sgn}(\om)
\sqrt{ \Delta_{\om}^2 + \mu^2};\cr
\rho_{\up \alpha}(\om) = (1/\rho) {\rm Im}[F_{\up\alpha}(\om + i\delta)].
}
$$

This spectral density can be used to carry out the Matsubara
frequency convolution. We can consider the $\alpha = +$ term only,
since it corresponds to the gapless excitations. The final expression
for the imaginary part of the low-energy spin response is
$$
\eqalign{{\chi^{s''}_{zz}(\om) \over \om} =& - {\pi \rho^2 \over 32}
\int dx \rho_{\up +}(x) \rho_{\up +}(x-\om)
{f(x) - f(x-\om) \over \om}\biggl\{\dots\biggr\}\cr
\biggl\{\dots\biggr\}=&
\biggl\{ 1 - {\Delta_x \Delta_{x-\om} - \mu^2 \over E_{x} E_{x-\om}
}\biggr\} \cr}\eqno(finally)
$$

At low energies, the energy
dependence of the magnetic relaxation is governed by the linear energy
dependence of the spin coherence factors.  The low energy imaginary
part of the susceptibility is hence quadratic in the external
frequency,
$$
{\chi^{s''}_{zz}(\om) \over \om} = {\pi \rho^2 \over 48} \biggl({\mu \over
D}\biggr)^2 \biggl({\om \over T_K}\biggr)^2.
$$
This in turn leads to an NMR relaxation rate that is  proportional to
$T^3$
$$
{1 \over T_1 T} = \lambda_1 {\pi \rho^2 \over 48} \biggl({\mu \over
D}\biggr)^2 \biggl({2 \pi T \over T_K}\biggr)^2.
$$

All other parts of the spin response function lead to a gapped
contribution to the magnetic relaxation rate.
Furthermore, in the undoped case,
the mid-gap coherence factors are identically zero, and the
imaginary part of the spin and charge response functions are zero
throughout the gap (Fig. 10).

The key difference between the d-wave and odd frequency description,
is that the former relies on a node in momentum space, whereas the
latter relies on a node in frequency space. For this reason,
we expect that odd-frequency pairing is rather insensitive
to elastic scattering. This point can be illustrated in the following
general way. Let us consider a general odd-frequency paired state,
where the pairing self-energy takes the form
$$
\eqalign{
\ul{\Si}(\om;x,x')=&
\zeta(x){V_o^2 \over \om}
{\cal P}(x)\delta_{x x'}
\cr
 \ul{\cal P} =& {1 \over 4}\left[
3(\ul{1}) -  d_{ab} (x)\si^a\otimes \tau^{b}\right];
\cr}
\eqno(fluctuation)
$$
here the amplitude $V_o^2\zeta(x)$ of the resonant scattering
and the orientation $d_{ab} ={ 1\over 2}{\cal Z}\dg(x)\si_a \tau_b
{\cal Z}(x)$
of the triplet order parameter may be site dependent.
The component of the conduction electron states which
does not directly couple to the resonant pairing, $\Psi_o(x)$
may be projected out of the conduction electron spinor as follows
$$
\eqalign{
\Psi_o(x)= &\ul{p}(x)\Psi(x)\cr
\ul{p}(x)= &\ul{1} -\ul{\cal P}(x)={1 \over 4}\left[
(\ul{1}) + d_{ab} (x){\si}^a\otimes {\tau}^{b}\right]\cr}
\eqno(zeroproj)
$$
This component experiences an indirect effect of the resonant
pairing
through mixing with the directly scattered components.

Consider a general conduction electron band with disorder,
described by the Hamiltonian
$$
H_c={ 1\over 2}\sum\Psi\dg(x){\cal H}(x,x')\Psi(x')
\eqno(disordercond)
$$
Let us project out  the parts of the
conduction electron Hamiltonian that couple directly, or
indirectly to the resonant scattering, writing
$$
\2mat {h_{x x'}}  {\alpha_{x,x'}}{\alpha\dg_{x,x'}} {H_{x x'}}=
\2mat {p(x)} {\  }{\  } {{\cal P}(x)}
{\cal H}(x,x')
 \2mat {p(x)} {\  }{\  } {{\cal P}(x)}
\eqno(splitter)
$$
The conduction electron Green function can then be written
$$
\ul{\cal G}(\om)=
\left[
\om - \ul{{\cal H}}-
 {V_o^2 \over \om}\ul{\zeta}\ul{{\cal P}}
\right]^{-1}\qquad\qquad(\ul{\zeta}_{x,x'}=\zeta(x)\delta_{x,x'})
\eqno(greenery)
$$
where all subscripts have been omitted.
The projected component of the conduction electron propagator
for those states that do not directly couple to
the resonant scattering is then given by
$$
\ul{G}^{-1}_o(\om) =
\om - \ul{h}- \ul{\a} \biggl(
\om - H - {V_o^2 \over \om}\ul{\zeta}
\biggr) ^{-1}\ul{\a}\dg
\eqno(projgreen)
$$
Though these states do not directly couple to the
resonant scattering, they couple indirectly because of the
off-diagonal terms in the Hamiltonian that mix them with the
resonantly scattered states.  At low energies, the
resonant scattering dominates the ``self-energy'' correction
in the propagator, which then becomes
$$
\eqalign{
\ul{G}^{-1}_o(\om) =&
\ul{Z}^{-1}\om- \ul{h}
\cr
\ul{Z}^{-1}=&\biggl[1+ {\ul{\a}\ul{\zeta}^{-1}
\ul{\a}\dg\over V_o^2}\biggr]}
\eqno(projgreen4)
$$
Notice that the effect of the resonant scattering is to introduce
a wavefunction renormalization into the propagator. Low
energy eigenstates are set by the determinantal equation
$$
Det[\ul{G}_o^{-1}]=0
\eqno(determ)
$$
Clearly then, if there are zero energy eigenstates of the projected Hamiltonian
$$
\ul h(x,x') \xi_{\lambda}(x') =0\eqno(zeromodes)
$$
then these will give rise to zeroes in this determinant.
In other words: the projective character of the resonant
scattering means that the indirectly coupled
zero energy states form zero energy excitations of the \underbar{complete}
Hamiltonian.
Suppose we define the Majorana conduction electron states
$$
a_{\lambda}\dg= \sum_x \psi_o(x)\xi_{\lambda}(x)
\eqno(zero2)
$$
Then their propagator will be given by
$$
\la a(\om) a\dg(\om)\ra = \xi\dg_{\lambda}\ul{G}_o(\om) \xi_{\lambda}
={ Z_{\lambda}\over \om}
\eqno(zeroprop)
$$
where the pole strength $Z_{\lambda}$ is
$$
Z_{\lambda}=
\biggl[1+ \xi\dg_{\lambda}{\ul{\a}\ul{\zeta}^{-1}\ul{\a}\dg\over V_o^2}
\xi_{\lambda}\biggr]^{-1}
\eqno(projgreen5)
$$
Thus the zero energy eigenstates of the complete Hamiltonian
will have the form
$$
\tilde a_{\lambda}\dg= \sqrt{Z_{\lambda}}
\sum_x \psi_o(x)\xi_{\lambda}(x)+ \dots
\eqno(zero6)
$$
where the residual part of strength $\sqrt{(1-Z_{\lambda})}$ is  carried
entirely by the Majorana spin fermions. In general then, the off-diagonal
coupling between the indirectly, and directly scattered states leads
to  a reduction in the conduction character of the Majorana zero modes,
and a corresponding enhancement of the density of gapless excitations.
$$
N^*(0)=\left({\rho \over 2}\right)
\la Z^{-1}_{\lambda}\ra
_{\lambda}
\eqno(densren)
$$
The important point however, is that despite these effects,
the gapless excitations remain Majorana fermions: the spin and charge
operators are completely off-diagonal at the Fermi surface,
and coherence factors must consistently vanish in this region.

To provide a specific example, consider the generalization
of our toy model with a random
chemical potential
$$
H_c=
{1 \over 2}
\sum_{\vk}
\Psi\dg_{\vk} \tilde \eps_{\vk} \psi_{\vk} -{1 \over 2}
\sum_x\Psi\dg(x)\mu(x)\ul{\tau}_3 \Psi(x)
\eqno(randompot)
$$
The chemical potential term can be identified as the off-diagonal
coupling, whereas the kinetic energy term commutes with the projection.
The $Z$-factor for a gapless plane wave
$\xi_{\vk}(x)={ 1 \over \sqrt{N}}e^{i \vk \cdot\vec x}
{\cal Z}_{\vk}$
is then
$$
Z_{\vk}^{-1}= \left(
1 + \la \mu^2(x)\ra /V_o^2 \right)\eqno(randommu)
$$
which gives rise to an enhancement of the gapless density of states
given by
$$
N^*(0) = {\rho \over 2}\left(
1 + \la \mu^2(x)\ra /V_o^2 \right)\eqno(enhance3)
$$
where for weak scattering, we have ignored the fluctuations in
the strength of the resonant scattering potential that will be
induced by the disorder. Thus we see that disorder {\sl enhances}
the density of states but sustains the
electric and magnetic neutrality of the Fermi surface.

\noindent{\bf 7. Interplay with magnetism}

In this section we discuss some of the magnetic aspects of the
odd-frequency state.
Even in our toy model, where we have not included any detailed
effects of band-structure or anisotropy, there are a variety of
locally stable phases  where the order parameter is commensurately
staggered. Quite generally, as we now show, the odd-frequency
state will develop an ordered magnetic moment, aligned parallel
to the $\hat d^3$ vector, giving rise to coexistence of antiferromagnetism
and superconductivity.

\noindent{\bf 7(a) Pair-Spin Correlations}

We begin our discussion by returning to the simplest
example, where the $\hat d^3$ vector is uniformly oriented,
giving rise to a state with ferromagnetic correlations.
We will generalize our
discussions to a more realistic  antiferromagnetically ordered case
at the end of the section.
When the conduction band is half filled, and hence completely
particle-hole symmetric, the
odd-frequency  paired state is  magnetically isotropic
and both static  magnetic order and static pair correlations
are absent $\la \vec S \ra =0$, $\la \vec \tau_j\ra = 0$
at  half-filling. In this state, there is long range order with
an  order parameter
$$
\la S^a(x) \tau^b(x) \ra = 2 {V^2 \over J^2} \left( \hat d_b (x)\right)^a.
$$
where $\vec S(x)=\vec S_c(x)+\vec S_f(x)$ is the total moment at site $x$,
and $\vec \tau(x)$ is the conduction electron isospin at site $x$.
Odd frequency triplet pairing is thus seen to strongly
couple spin and  pair correlations. Clearly however,
these cross-correlations induce  anomalous
response functions, coupling the development of charge correlations
to the application of a magnetic field or the development
of magnetic correlations to the application of a chemical potential
or pairing field. To study this effect, we introduce the
spin-charge susceptibility
$$
\chi^{ab}(\ka) = \la S^a(\ka) \tau^b( - \ka)\ra
$$
where the static susceptibility $\ul{ \chi}(0)$ is of particular
interest.
In the vicinity of half filling, the state is completely isotropic
in spin and isospin space, and we then expect
$$
\chi^{ab} = \chi_o d^{ab}
\eqno(phsym)
$$
Thus, once
the system is doped ($\mu \ne 0$), the presence of a cross-correlation
between the charge and spin degrees of freedom leads to the development
of a magnetic moment and strong anisotropy
$$
\vec M \sim \chi_o \hat d^3\mu
$$
We may calculate $\chi_o$ very simply as follows. Let us choose
$\hat d^3 = \hat z$. The coupling of the magnetic field
is given by
$$\eqalign{
H_{B} = - \sum_j\vec B\cdot  \biggl\{
\psi\dg(j)\bigl[{ \vec \si\over 2}\bigr] \psi(j) -  {i\over 2} \vec
\eta_j\times \vec \eta_j\biggr\}
}
\eqno(Bfield)
$$
Thus the total magnetic
moment in the $\hat z$ direction per site
is
$$
\eqalign{
M= &
\la {1 \over 2}(n_{c \up} - n_{c \dw}) - i \eta_1(j) \eta_2(j) \ra \cr
= &\la {1 \over 2} (n_{c \up} - n_{c \dw}) + (1/2 - n_{\eta \dw})\ra\cr}
\eqno(moment2)
$$
here $n_{c\si}$,  ( $\si = \up , \dw $)
is the number of ``up'' or ``down'' conduction electrons per site and following
the notation of \(downers),
$n_{\eta\dw} = {1 \over N_{sites}}\sum_{\vk}\eta_{\vk \dw}\dg \eta_{\vk \dw}$
is
the number  of ``down'' Majorana fermions per site.  Now note that the
down Majorana electrons are hybridized  with the down conduction electrons
to produce a completely filled hybridized band, with one electron
per unit cell. Thus in the ground-state, $\la n_{\eta} + n_{c \dw}\ra =1$
per site, so
$$
M_z = {1 \over 2}\la n_{c \up} +n_{c \dw} \ra
- {1 \over 2}
\eqno(moment3)
$$
per site. Since the local moments leave the
conduction electron density essentially unchanged,
$\la n_{c \up} +n_{c \dw} -1 \ra  = 2 \rho \mu$, giving
$$
M_z =  \rho \mu
\eqno(mom4)
$$
and
$$
\chi_o =   \rho + O(T_K/D)
\eqno(thesusc)
$$
A departure from particle hole symmetry $\delta \mu$ thus generates
an ordered (ferromagnetic) moment of strength $$
\vec M(x)=
\delta \mu \rho
\hat d^3(x)\eqno(flucmag)$$

In a similar fashion,
an application of a magnetic field
will influence the charge and pair correlations:
a field along the $\hat d_3$ direction
will develop a charge density,
$$
\rho(x) = \rho  B_z(x)
$$
More remarkably, a {\sl transverse}
magnetic field $\vec B = B_{\perp} \hat d^{l}$, ($l=1,2$)
will induce  a conventional pairing field of magnitude
$$
\la \psi_\up\dg \psi\dw\dg \ra  \sim  \rho_o  B_{\perp}.
$$
This  leads to
a possible field dependent Josephson coupling  with conventional
superconductors.

Even in this ferromagnetically ordered state, there are strong
antiferromagnetic correlations. To illustrate this point, consider
the case where $\mu=0$. Here, the main contribution to the magnetic
susceptibility is provided by the local moments, and may be calculated
from the spatial correlations of the Majorana fields.  Low frequency
properties of the local moments are determined by excitations across
the indirect gap at wave vectors $\vec q = \pm \vec Q/2$. In the vicinity
of the gap, the Majorana fermions can be expanded as shown in \(lowl) and
\(lowl2). In momentum space, the Majorana propagators have the form
$$
\eqalign{
\la\eta^a(\ka)\eta^b(-\ka)\ra =&\delta_{ab}D(\ka)\cr
D(\ka)= &{ ( \iom - \eps_{\vk}) \over
[\iom  ( \iom - \eps_{\vk}) - V^2 ]}\cr}
\eqno(prop)
$$
so that $$
{\rm Im} D(\vk,\om-i\delta) = {V^2 \over V^2 + \om^2} \pi \delta(\om -
\om_{\vq})
\eqno(specfn)
$$
The spin correlations are determined from the product of two Majorana
propagators:
$$
\la S^a(q)S^b(-q)\ra =\delta^{ab}\chi(q)=
\delta^{ab} {{\rm T} \over 2}
\sum_{\ka} D(q/2+ \ka) D(\ka -q/2)\eqno(chigive)
$$
At low frequencies, this is dominated by
governed by excitations in
the vicinity of the indirect gap,
the quasiparticle spectrum is parabolic $\om_q =\pm{\Delta_g + {{(q\pm\vec
Q/2)^2}\over 2m^*}}$, where $m^*={D\over T_K}m$, and $m$ is the electron
mass at the band-edge.  Using this parabolic approximation to carry out the
momentum space integrals gives
$$
\eqalign{{\rm Im}\chi(\vec q ,\om+i \delta)
\approx &\delta^{ab} { (m^*)^{3/2}\over (2 \pi)^2 }
\left[
{V^2 \over V^2 + (\om/2)^2}\right] \sqrt{\om -  \Delta_{\vq}}
\Theta(\om- \Delta_{\vq})\cr
(\Delta_{\vq} = &2 \Delta_g + {(\vec q-\vec Q)^2\over 4m^*})\cr
}
\eqno(prop22)
$$
In other words, independently of the ordered magnetic moments,
there is a large amount of spin-fluctuation spectral
weight above the superconducting gap at the antiferromagnetic
zone vector. This general feature survives when
we come to consider more general, staggered configurations of $\hat d^3(x)$.

Let us now consider the possibility of more general, staggered configurations
of $\hat d^3$.  Take the more general ansatz for the mean field order parameter
$$
\eqalign{
z(\vec x)= &
e^{-i
{1 \over 2}\vec x \cdot({\vec Q}+ \vec P  \ul{\si_1})}
z_o\qquad\qquad z_o \equiv
\left(\matrix{1 \cr 0\cr}\right)
\cr
\ul{\cal M}(x)
= &\pmatrix{
\hat d_1(x) &\cr
\ \hat d_2(x)&\cr
\ \hat d_3(x)& \cr}.
= \pmatrix{
\hat d_1\cos [\vec Q\cdot \vec x]
 &\cr
\ \hat d_2\cos [(\vec Q+ \vec P)\cdot \vec x]&\cr
\ \hat d_3\cos [(\vec P\cdot \vec x)]& \cr}.}
\eqno(staggering2)
$$
where $\vec P$ and $\vec Q$  are commensurate vectors.

As before, we can redefine the conduction electron states to
take account of the staggered order,
$$
\eqalign{
z_j = &
e^{{i\over 2} \vec x_j\cdot\left(\vec Q  + \vec P \ul{\si}_1\right)}
\tilde z_j
\qquad\qquad ({\cal Z}_j =
e^{{i\over 2} \vec x_j\cdot\left(\vec Q\ul{\tau}_3  + \vec P \ul{\si}_1\right)}
\tilde {\cal Z}_j)
\cr
\Psi_j=&  e^{{i\over 2} \vec x_j\cdot\left(\vec Q\ul{\tau}_3  + \vec P
\ul{\si}_1\right)}
\tilde\Psi_j}
\eqno(shifto2)
$$
The conduction electron Hamiltonian can then be written
$$
H_c=
\sum_{\vk \in{1 \over 2} {\rm B. Z.}}
\Psi \dg _ {\vk} \bigl[
(\eps[\vk-{\vec Q/2}\ul{\tau}_3-{\vec P/2}\ul{\si}_1]  - \mu)\ul{\tau}_3
\bigr]\Psi_{\vk}
\eqno(kinen2)
$$
where the kinetic energy term can be expanded as
$$
\eqalign{
[\eps(\vk &-{\vec Q/2}\ul{\tau}_3-{\vec P/2}\ul{\si}_1)- \mu]\ul{\tau}_3\cr=&
{1 \over 4}\sum_{\alpha, \be = \pm 1}
\{(\ul{\tau}_3 + \a)(1 + \beta \ul{\si}_1)
\eps[\vk - \a { \vec Q /2} - \be {\vec Q/2}]\}- \mu\ul{\tau}_3
\cr=&
[\tilde \eps^o_{\vk}+
\tilde\eps^1_{\vk}  \ul{\si}_1] - [\mu_{\vk}^0 +
\mu_{\vk}^1\ul{\si}_1]\ul{\tau}_3\cr}
\eqno(exps)
$$
 From the discussion of section 3, we know that gapless
excitations will develop on the ``Fermi surfaces'' described
by
$$
\tilde\eps^0_{\vk} = {1 \over 4}
\sum_{\alpha, \be = \pm 1}
\a
\eps[\vk - \a { \vec Q /2} - \be {\vec P/2}] =0
\eqno(zeros3)
$$
To make our example more specific, consider the case where
$$
\eqalign{
\eps_{\vk} =& -2t[c_x+c_y+c_z] \qquad\qquad(c_l = \cos[k_l], l=1,2,3)\cr
\vec Q=&(\pi,\pi,\pi), \qquad\vec P=(\pi,0,0)\cr
}\eqno(ex3)
$$
corresponding to a staggered $\hat d^3$ vector in the
x direction.
In this case,
$$
\eqalign{
\tilde\eps^o_{\vk} =&-2t[s_y+s_z]\qquad\qquad(s_l = \sin[k_l], l=1,2,3)\cr
\tilde\eps^1_{\vk}=&2tc_x\cr
\mu^o_{\vk} =& \mu,\qquad\qquad \mu^1_{\vk} = 0\cr}\eqno(ex4)
$$
The gapless modes lie on a tube with a square cross-section $[s_y+s_z]=0$
and the spectrum is given by
$$
\eqalign{
{\rm Det}&\biggl[
\ul{G}^{-1}_{\up}(\om)\ul{G}^{-1}_{\dw}(\om) -[\eps^1_{\vk}]^2\ul{1}\biggr]
=0\cr
\ul{G}^{-1}_{\up}(\om)=&\bigl[
\om - \eps^o_{\vk}-\Delta_{\om}(1 - \tau_1) + \mu\tau_3\bigr]\cr
\ul{G}^{-1}_{\dw}=&\bigl[
\om - \eps^o_{\vk}-2\Delta_{\om} + \mu\tau_3\bigr]\cr}
\eqno(spec4)
$$
After a short calculation the corresponding mass renormalization
of the gapless quasiparticles
is found to be
$$
{m\over m^*} = \left[ 1 + \biggl({\mu^2 + (2tc_x)^2 \over V^2}\biggr)\right]
\eqno(enhance)
$$
In Fig. 11, we show the mean-field ground-state energy as a function
of $\vec P = (P_x,P_y,P_z)$, clearly showing the development of local
minima at the commensurate points in the Brillouin zone.
At each of these points the odd-frequency state will develop a staggered
magnetization with wave vector $\vec P$ and approximate magnitude
$M\sim \rho\mu$.
In practice, we would argue that the small differences in
energy between these different commensurate states will depend
on several factors that are not included in the toy model.
It is interesting to note that in two dimensions, the toy model
predicts that a $\vec P=(\pi,0) $ state is more stable than the
$\vec P = (\pi, \pi)$ state. The important point however, is
that the mean field energy of these magnetic phases is
close in energy  to the uniform state and furthermore, is locally
stable. We may conclude that this type of odd frequency pairing
can homogeneously coexist with antiferromagnetism.  The local
moments participate in both the  spin and the pair condensate.

\break
\noindent{\bf 8. Critique and Discussion}

Our paper has presented a ``toy'' realization
of odd frequency pairing, with the aim of elucidating its
key properties.
In this section,
we  discuss the odd frequency state in a more
general setting and examine the possibility that this
kind of paired state might be applicable to heavy fermions.

One of the most dramatic features of the theory
is the projective character of the  resonant pairing self-energy:
$$
\eqalign{
\ul{\Sigma}(\om) =& {V^2 \over \om}\ul{\cal P}\cr
( \ul{\cal P} =& {1 \over 4}\left[
3(\ul{1}) -  d_{ab} \si^a \otimes \tau^{b}\right])\cr}
\eqno(prject)
$$
Can we understand this feature in a more general context,
outside the restrictive realm of the Kondo model, and our
Majorana treatment?
Let us consider the possible extension
of our odd paired state within an Anderson model for
heavy fermions, with an on-site repulsion term
$$
H_I= {1 \over 2}U(n-1)^2
$$
at each magnetic site.
Highly correlated states minimize this on-site interaction energy,
tending to produce local moment states where $n=1$.
It is quite useful to examine
this constraint in terms of the correlations between the charge
(isospin) and spin of the localized states. If we
expand the localized f-electron in terms of its four real
components

\def\zspinor{\pmatrix{z_{\up} \cr z_{\dw}} }

$$
\pmatrix{f_{\up} \cr f_{\dw}} =
 {1 \over \sqrt 2}
\bigl[
f^0 + i \vec f\cdot\vec \si \bigr] z
\eqno(deco)
$$
where, $z=\zspinor $ is a unit spinor,
then  the interaction may be written as
a symmetric  product of all four fields.\refto{affleck1}
$$
\eqalign{
H_I = & U[(\tau_3)^2-(S_z)^2]+ {U \over 4}\cr
=& {U}[ (i f^0 f^1)(i f^2 f^3) + { 1 \over 4}]\cr }\eqno(facts3)
$$
On a lattice, the interaction between the electrons can then be written
$$
S_I = -U\sum_{\kappa}\left\{  f^{0}_{\ka_o} f^1_{\ka_1} f^2_{\ka_2}
f^3_{\ka_3}\right\}\delta_{\ka_o + \ka_1+ \ka_2 + \ka _3}\eqno(facts3.2)
$$
where we use the Fourier transformed operators
$$
f^a(\ka)=
{1 \over \sqrt{\beta N}}\sum_{j}\int d\tau e^{-i\ka .X}f(X),
\qquad \ka \cdot X\equiv \vk\cdot \vec R_j-\omega_n\tau.
\eqno(fourierspace)
$$
and it is understood that all time coordinates are ultimately to
be time-ordered.

A novel way to reduce the on-site correlation energy is to develop a correlated
state where certain Majorana  components of the f-state are
{\sl absent}.
Remarkably, the operator that projects out the zeroth
component at wave vector $\a$
$$
f^0(\ka) = {1 \over \sqrt{2}}\biggl[ z\dg f(\ka) + f\dg(-\ka) z \biggr]
\eqno(pro)
$$
is a {\sl one particle} operator
$$
{\rm p}_{\ka} = f_o[\ka]f_o[-\ka]=
{ 1 \over 4} {\cal F}\dg_{\ka} \biggl[\ul{1} + d_{ab}\si_a \otimes
\tau_b \biggr]{\cal F}_{\ka}
$$
where $d_{ab}= {1 \over 2} { \cal Z} \dg \si_a \otimes \tau _b { \cal Z}$,
$${\cal Z }= \pmatrix{ z \cr - i \si_2 z^* \cr},
$$
and $${\cal F}_{\ka}=\pmatrix{ f_{\ka}\cr - i \si_2 f^*_{-\ka} \cr},\qquad
f_{\ka}=
\pmatrix{ f_{\ka\up}\cr  f_{\ka\dw} \cr} $$
are the Balian Werthammer four spinors for the f-state and z spinor.
The residual ``vector'' ($1,2,3$)  components of the f-states are
projected out by  the one particle operator
$$
\hat {\cal P}_{\ka} = \sum_{j=1,3}f_j[-\ka]f_j[\ka]=
1 - {\rm p}_{\ka} =  {\cal F}_{\ka}\dg \ul{\cal P}{\cal F}_{\ka}
$$
The Majorana character of this operator implies  that it is
asymmetric
$$
\hat {\cal P}_{\ka} = - {\cal P}_{-\ka} \eqno(oddish)
$$

The
self-energy term  that will
selectively decouple Majorana components of the f-electrons will have the
general form
$$
\eqalign{
\ul{\Si}(\ka)
= &\Delta(\ka) \ul{\cal P}\cr
S_I=& \sum_{\ka} {\cal F}\dg(\ka)\ul{\Si}(\ka)
{\cal F}(\ka)=
 \sum_{\ka}
\Delta(\ka)\hat{\ul{\cal P}}_{\ka}\cr}
\eqno(projj)
$$
Since $\hat {\cal P}_{\ka}$ is an odd function of $\kappa$, it follows that
$$\Delta(\ka) = - \Delta(-\ka). \eqno(finalfacts)
$$
If the physics is local in time,
then the frequency dependence of $\Delta(\ka)$ can be dropped,
leading to p-wave triplet pairing.
However,  if the  momentum
dependence of $\Delta(\ka)$ is even, the frequency dependence
is automatically {\sl odd}, leading to
odd frequency pairing.
In the simplest s-wave version of this
pairing, the physics is spatially local, so that $\Delta(\ka) =\Delta(\om)$.
This establishes  an intimate connection between the
projection of Majorana degrees of freedom from the ground-state
and the development of nodes in the wavefunction: when the physics is
local, this projection results in {\sl node in time}, and the development
of odd-frequency pairing.

A general spectral decomposition of $\Delta(\om)$ will always contain a zero
frequency pole
$$
\eqalign{
\Delta(\om) = &{Z \over \omega} + \omega \int { d \nu \over \pi}
{ A(\nu) \over \om^2 - \nu^2 }\cr
{1 \over \pi}Im\bigl[\Delta(\om - i \delta) \bigr]=&
Z\delta(\om) +  A(\om)\cr}
\eqno(specdecomp)
$$
This pole is a  unique feature of odd-frequency pairing: it
suppresses  the ``vector'' components
of the f-electron from the low energy excitations,
gaining correlation energy and decoupling a band of
gapless singlet excitations.
Our simple
mean field theory can be viewed as a dominant pole approximation
to the  pairing field.
A pressing need for the near future is to show that
such general constructions can lead to stable Eliashberg-type
treatments of more general models, such as the finite $U$ Anderson
lattice.

In its current form, our prototype for odd frequency pairing is too
simplistic to account for details of heavy fermion behavior.
We should like to list some important issues which need to be
addressed in future developments:

\item{$\bullet$}{\bf Magnetism}. The toy model has shown
that odd-frequency pairing has a propensity to coexist with
magnetism. A more realistic model will need to take explicit account
of the RKKY interactions and their role in establishing
the detailed superconducting order.

\item{$\bullet$}{\bf Normal Phase}. The normal phase of
heavy fermion superconductors, with its profusion of Fermi liquid
features, does not appear in the toy model. The Majorana formalism
may be a poor starting point to recover the normal phase properties,
and this suggests that we should seek a way to obtain the odd-frequency
paired state within perturbation theory for the finite $U$ Anderson model,
 or perhaps the large
$N$ approach to the heavy fermion problem.

\item{$\bullet$}{\bf Anisotropy}. Measurements of the gap and
ultrasound absorption\refto{shivaram,louis} in $UPt_3$
show the presence of anisotropy in the order parameter and have been
traditionally interpreted within a d-wave pairing scenario.
These
results {\sl do not} reveal the temporal parity
of the paired state, but tend to reinforce the
conclusion that momentum anisotropy can not be ignored in a more
advanced version of the model. Indeed, there is no reason not
to contemplate the possibility of odd frequency d-wave pairing.

\item{$\bullet$} {\bf Power Laws}. Power laws in the temperature
dependence of the specific heat and NMR relaxation rate
of heavy Fermion compounds
develop much closer to $T_c$ than any simple
mean field theory can account for. One possibility is
that dynamic
pair-breaking effects have suppressed
$T_c$ significantly below the gap. Odd frequency pairing
accounts for the finite linear specific heats in heavy fermion
superconductors in terms of a band of excitations with vanishing
coherence factors. At present, the toy model is {\sl unable} to account
for the $T^2$ term in $C_V$ that is also seen. Fluctuation
effects need to be examined carefully.

We should like to spend a moment discussing
the long-wavelength properties of odd-frequency triplet paired
states. The intimate relation between spin and pair degrees of freedom
in this kind of state leads to rather interesting consequences
in the Landau-Ginzburg theory.
Suppose one considers the simple long-wavelength action discussed in
section (5)
$$
F  =
{1
\over 2} \int d^3  x
\biggl[
\rho_{\perp}\bigl(\nabla \hat n\bigr)^2  +
\rho_{s}
\bigl(\vec \omega_{3}- {2e \over \hbar} \vec A\bigr)^2
+
{B^2 \over  \mu_o}\biggr] -{ \rho_{s} \over l_o^2 }
\int d^3  x
(\hat d_3\cdot \hat z)^2
\eqno(lwact3b)
$$
then the anisotropy plays a vital role in establishing the
topological stability of persistent currents.
Unlike a conventional superconductor, the supercurrent
is linked to the spin order and involves
all three Euler angles of the order parameter.
To see this, it is
instructive to consider a loop of superconductor of length $L$, threaded by
a solenoid.  The supercurrent around the loop is given by
$$
\eqalign{
\vec j_s =& {2e\over \hbar}
\rho_s (\vec \om_3 - {2 e\over \hbar} \vec A )\cr
\vec \om_3 = &\nabla \phi + \cos \theta \vec \nabla \psi \cr
}
\eqno(Scurrent)
$$
where $(\phi,\theta,\phi)$ are the Euler angles defining the
orientation of the triad $d_{ab}$.
Notice that unlike a conventional superconductor, the supercurrent
involves \underbar{both} the U(1) phase $\phi$ and the
orientation of the magnetic vector $\hat d$, defined by
$(\theta,\psi)$.
For a conventional superconductor
the total phase change around the loop is a topological invariant
$$
\Delta \phi =  \int \vec dr\cdot \vec \nabla \phi=2 \pi n
\eqno(looper)
$$
that  is unchanged
upon application
of a flux through the solenoid, leading to a linear
relation $j_s = -{4\pi e\over \hbar L} \rho_s  {\Phi \over \Phi_o}$
between the enclosed flux $\Phi$ and the supercurrent density
$j_s$.  In this superconductor, the
analogous integral around the current loop is
$$
\Delta \phi =
\int \vec dr \cdot\left[\nabla \phi + \cos \theta \vec \nabla \psi
\right]
= 2 \pi n + \Omega
\eqno(looper2)
$$
The second term
$$
\Omega= \int \vec dr \cdot\left[ \cos \theta \vec \nabla \psi \right]
= \int_S d\vec S \cdot (\epsilon_{ab}\nabla_a\hat n
\times \nabla_b\hat n )\eqno(solid)
$$
is the solid angle subtended by the $\hat n$ vector
around the loop: this is \underbar{not} an invariant, and can change by
multiples of $4\pi$ to relax the current. Unlike the $U(1)$ superconductor,
the only stable vortex configuration involves a net phase change
of $2 \pi$ around the loop: this is the so called ``$Z_2$'' vortex
of an SO(3) order parameter, and it has the property that two such
vortices can be adiabatically deformed back to the vacuum (Fig. 12).
Each $Z_2$ vortex pair reduces the effective flux through the solenoid
by $2$ flux quanta, thus
the non-linear current will be given by
$$
j=
{2 e\over \hbar L} \rho_s \biggl[2 \pi {\Phi \over \Phi_o} -4 \pi n_Z\biggr]
\eqno(nlin)
$$
where the number of $Z_2$ vortex pairs is
$$
n_Z = {\rm Int}\biggl[{\Phi \over 2\Phi_o} +{1\over 2}
\biggr]\eqno(nflux)
$$
where ${\rm Int}(x)$ denotes the largest integer smaller than x.
In this way, the  current density around the loop can never exceed
$$
j_{o} = {(2 \pi)^2 \over L\Phi_o} \rho_s
$$
and  will never become macroscopic.
Thus, without anisotropy, the critical current for the odd-frequency
state is {\sl zero}.
Furthermore, if the flux through the solenoid is
$\Phi= \int V(t) dt$, then  the response to an oscillatory
electric field will not occur at the driving frequency. This will
eliminate the low frequency linear Meissner response to macroscopic fields,
removing the pole in the optical conductivity and producing
an {\sl apparent violation} of the linear response optical sum rule.
This type of behavior is most likely to occur
in the vicinity of particle-hole symmetry, and suggests that
this half filled state will more closely resemble an insulator,
rather than a superconductor. This may  be
an interesting way of thinking about Kondo insulators\refto{aeppli3},
where an anomalous reduction in the low frequency oscillator strength of the
optical conductivity has recently been reported.\refto{schlessinger}
As anisotropy is increased,
a macroscopic Free energy barrier will have to be crossed in order to
add pairs of $Z_2$ vortices to the superconducting state, restoring
the linear Meissner response. This
will lead to a non-trivial dependence of the critical current
on anisotropy.

Pending further theoretical work, it appears that there may be some
useful experiments on heavy fermion superconductors that could
help to compare the d-wave and odd-frequency scenarios. A key
issue is to verify the relation between the gapless excitations
and the
NMR relaxation rate, most notably to confirm the presence or absence
of a Korringa term in the relaxation rate in severely gapless
heavy fermion superconductors.
Another area of fruitful investigation
concerns the field dependence of the proximity effect. Negative
proximity effects have been observed between $UBe_{13}$ and $Ta$
superconductors.\refto{proximity}
If the symmetry of heavy fermion superconductors has a different
temporal parity, then we expect the application of a magnetic
field to enhance the coupling between the two order parameters,
leading to a strong reduction of the negative proximity effect in
a field, and a strong field dependence of the Josephson current.

In conclusion, we have presented a stable realization of
odd frequency triplet pairing in a Kondo lattice model for heavy fermions.
Under rather general conditions, the odd frequency
state that forms has a gapless singlet mode of quasiparticles.
Spin and charge coherence factors for these quasiparticles
grow linearly in their energy. Our pairing hypothesis
provides an alternative  explanation of various power laws in
heavy fermions in terms of a vanishing of coherence factors
at the Fermi energy, rather than a vanishing of the density of states.
We have  conjectured that this may explain the absence of
a Korringa law in the NMR, even when the superconductor is
highly gapless.
Odd-frequency triplet superconductivity appears to be able to
coexist with magnetism, and in our simple toy model, ferromagnetic
order coexists with the pairing.  We think our results are
encouraging enough to prompt efforts to develop a description of
odd-frequency pairing within more general models, and to consider
seriously the possibility that this provides a viable alternative
to the d-wave pairing hypothesis in heavy fermion superconductors.

We would particularly like to thank E. Abrahams
and
P. W. Anderson for discussions related to this work.
Discussions with N. Andrei,
A. V. Balatsky, D. Khmelnitskii, G. Kotliar, L. Ioffe,
G. Lonzarich, D. Maclaughlin  and A. Ramirez are also gratefully acknowledged.
Part of the work was supported by NSF grants DMR-89-13692 and  NSF 2456276.
E. M. was supported by a grant from CNPq, Brazil.

\taghead{A.}

\noindent{\bf APPENDIX A.  Majorana Representation of Spins}

In this section,  we present
a derivation of the Majorana representation
that provides a link with the Abrikosov
fermion representation and illustrates how the constraint is
avoided by the uniform replication of the spin Hilbert space.
We begin by noting that for any two component electron spinor there
are two operators of interest: the ``spin''
$$
\vec s = f\dg_{\a} \biggl[{\vec \sigma\over 2}\biggr]_{\alpha \be} f_{\be}
\eqno(appm)
$$
and the ``isospin''
$$
\vec \tau = \tilde f\dg_{\alpha} \biggl[{\vec \sigma\over 2}\biggr]_{\alpha
\be}
\tilde f_{\be},\eqno(appn)
$$
where we have introduced the Nambu spinor
$$
\tilde f =
\left(\matrix{f_{\up} \cr f\dg_{\dw} \cr
}
\right).\eqno(appo)
$$
These operators are independent $[s_a,\tau_b]=0$ and
each satisfy an $SU(2)$ algebra
$$
[s_i,s_j] = i \epsilon^{ijk} s_k, \qquad \qquad
[\tau_i,\tau_j] = i \epsilon^{ijk} \tau_k,\eqno(appp)
$$
In the subspace where the spin is finite, the isospin is zero, and vice
versa. The
{\it sum} of both operators
$$
\vec S = \vec s + \vec \tau,\eqno(appq)
$$
satisfies an $SU(2)$ algebra, and is either equal to the
``spin'' or ``isospin'', depending on
which component of the Fock space is projected.
For any interacting system of
electrons containing $N$ local moments, we may write the partition
function as a constrained trace
$$
Z= { \rm Tr}\biggl[\prod_{j} P_{j}^{q_{j}} \exp (-\beta H[\vec S_j]
)\biggr], \qquad\qquad(q_j={s, \tau})\eqno(appr)
$$
with
$$
\vec S_j  =
f\dg_{j \alpha} \biggl[
{\vec \sigma
\over 2}
\biggr]_{\alpha \beta} f_{j \beta}
+ \tilde f\dg_{j \alpha}
\biggl[{\vec \sigma
\over 2}\biggr]_{\alpha \beta}
\tilde f_{j \beta},\eqno(apps)
$$
where the projection operator $P_{j}^{q_j}$
projects the
``spin'' or ``isospin'' component
of the Fock space at site j
$$
\left. \eqalign{P_j^{\tau}=&(n_j-1)^2 \cr
 P_j^{s}=&(n_{j\up}-n_{j\dw})^2\cr}\right\}
\qquad P_j^{s}+P_j^{\tau}=1 \eqno(appt)
$$
There are then $2^N$ ways of choosing the projection
operators: each
choice projects a replica of the spin Hilbert space
with precisely the same partition function.  Summing over
all replicas we can write
$$
Z= {1 \over 2^N} \sum_{q_{j} = s, \tau} { \rm Tr}\biggl[\prod_{j} P_{j}^{q_{j}}
\exp (-\beta H[\vec S_j]
)\biggr],\eqno(appu)
$$
The sum over all $2^N$ projectors is the identity operator
$$
\sum_{q_{j} = s, \tau} \prod_{j} P_{j}^{q_{j}}
= \prod_{j}\biggl[ P_{j}^{s} + P_{j}^{\tau}\biggr]=1\eqno(appv)
$$
and hence the replicated partition function can be written as an {\it
unconstrained}
trace, with each local moment represented as a sum of the
Pauli spin and isospin.
$$
Z= {1 \over 2^N} { \rm Tr}\biggl[\exp (-\beta H[\vec S_j]
)\biggr],\eqno(appw)
$$

We now demonstrate that the combined operator $\vec S = \vec s+ \vec \tau$
depends
only on three Majorana components of the f-electron.
Suppose we decompose the complex Fermi operators into
their real and imaginary Majorana components as follows
$$
f_{j} = {1 \over \sqrt 2}\left( f_o + i \vec \si \cdot\vec f\right)z_o
\qquad\qquad z_o = \pmatrix{0 \cr i \cr}
\eqno(appx)
$$
In terms of these components, the ``spin'' and ``isospin'' operators are
$$
\eqalign{
s_{j}^{i} &= {i \over 2} \left[ \eta_{j}^{0} \eta_{j}^{i} -
{1 \over 2} \epsilon_{ilm} \eta_{j}^{l} \eta_{j}^{m} \right],\cr
\tau_{j}^{i} &= - { i \over 2} \left[ \eta_{j}^{0} \eta_{j}^{i} +
{1 \over 2} \epsilon_{ilm} \eta_{j}^{l} \eta_{j}^{m} \right].\cr
}\eqno(appy)
$$
The sum of these two is then
$$
s_{j}^{i} + \tau_{j}^{i} = - { i \over 2} \epsilon^{ilm}
\eta_{j}^{l} \eta_{j}^{m},\eqno(appz)
$$
which is precisely our Majorana representation.
With this choice of $z$, the zeroth component of the Majorana fermions at each
site
does not enter into the Hamiltonian. This component can
therefore be explicitly traced out of the partition function.
Formally, this may be done by pairing the zeroth Majorana fermions
throughout the lattice
$$
a_{\lambda} = {1 \over \sqrt 2} \left( \eta_{i_{\lambda}}^{0} - i \eta_{j
_{\lambda}}^{0}
\right), \qquad\qquad(\lambda=1,2\dots N/2)\eqno(pairsite)
$$
where each site $l$ belongs to one pair:
$l\in \{(i_{\lambda},j_{\lambda}),\lambda = 1,N\}$.
The set of
$N/2$ {\sl complex} fermions are independent and form
a completely
decoupled zero energy Fock space of dimension $2^{N/2}$. Hence
$$
Z= {1 \over 2^{N/2}} { \rm Tr}\biggl[\exp (-\beta H[\vec S_j
\rarrow - { i \over 2}
\vec \eta_{j} \times \vec \eta_{j}]
)\biggr],\eqno(partito)
$$
where the remaining  unconstrained
trace is over the $l=1,2,3$ components of each
Majorana fermion, and the other real electron states of the system.

The overcompleteness of our representation is closely related to
a residual discrete local
$Z_2$ symmetry of the Majorana spin representation under
the transformation
$$
\vec \eta_{j} \rarrow - \vec \eta_{j}.\eqno(z22)
$$
In this respect, the Majorana representation is similar to the
pseudofermion representation.
However, in this case, the canonical
and grand-canonical ensembles have precisely the
same partition function, up to a simple normalization factor.
In the pseudofermion
representation, the Gibbs partition function for
each conserved subspace is not constant (for if n$_j=0$ or n$_j=2$ there is no
spin
at site $j$), and the projection of the unwanted spaces is
an unavoidable necessity.

Majorana fermions are easily treated in a momentum space representation.
The Fourier transformed operators
$$
\vec \eta_{\vk} = {1 \over \sqrt N} \sum_{i} e^{i\vk \cdot \vec R_{i}}
\; \vec \eta_{i}, \qquad \qquad
\eqno(app3)
$$
Since the original Majorana fermions are real $\eta^a_j=\eta^{a\dagger}_j$,
their complex Fourier transforms satisfy $\eta^a_{\vk} =
\eta^{a\dagger}_{-\vk}$, forming
a set of independent complex fermions that span half of the
Brillouin Zone
$$
\left\{ \eta^{a}_{\vk}, \eta^{b \dagger}_{\vk^{'}} \right\} =
\delta_{a,b} \; \delta_{\vk,\vk^{'}} \qquad \qquad
\vk , \vk^{'} \in {\rm half \; the \; Brillouin \; Zone}.\eqno(app3b)
$$
The inverse transformation can be written
$$
\vec \eta_j = {1 \over \sqrt{ N}}
 \sum_{\vk \in {\rm {1/2} \
 B. Z.}}
\left\{\vec \eta_{\vk} e^{i \vec k \cdot \vec R_j}+
\vec \eta\dg_{\vk} e^{-i \vec k \cdot \vec R_j}\right\}
\eqno(app3c)
$$
The corresponding Lagrangian for the Majoranas is then
$$
{\cal L} = \sum_{\vk \in {1 \over 2} {\rm B. \; Z.}} \vec \eta_{\vk}\dg \cdot
\partial_{\tau} \vec \eta_{\vk} + {\cal H}\eqno(app4)
$$
or in terms of the original site representation
$$
{\cal L} = {1 \over 2} \sum_{i} \vec \eta_{i} \cdot
\partial_{\tau} \vec \eta_{i} + {\cal H}.\eqno(app5)
$$
Note that for each momentum $\vec k$, we can choose either $\eta_{\vk}$ or
$\eta_{-\vk} = \eta\dg_{\vk}$ as the independent destruction operator.
This has an important consequence for broken symmetry solutions, for
there are $2^{N/2}$ equivalent ways of making the choice of
the vacuum state: by making one particular choice,
the normalization constant in front of the partition function is absorbed.

\noindent{\bf APPENDIX B. Some simple examples}

\taghead{B.}
In this Appendix, we illustrate the use of the Majorana fermion
representation of spin 1/2, by means of some specific examples.

Consider first the Heisenberg Model for 2 spins 1/2 (where we take the
exchange coupling to be $1$), written in terms of the Majorana fermions
$$
H = \vec S_1 \cdot \vec S_2 = -{1 \over 2}\bigl[ (\vec \eta_1
\cdot \vec \eta_2)^2+{3 \over 4}\bigr].\eqno(one)
$$
One can now define 3 complex fermions, by taking appropriate linear
combinations of the 6 Majorana fermions
$$
\eqalign{
\vec f \equiv {1 \over \sqrt 2} (\vec \eta_1 - i \vec \eta_2),\cr
{\vec f}\dg  \equiv {1 \over \sqrt 2} (\vec \eta_1 + i \vec \eta_2).\cr
}\eqno(two)
$$
These operators satisfy the usual fermionic anticommutation algebra
$$
\{ f^i, {f^j}\dg \} = \delta_{ij}\qquad\qquad(i,j=1,3)\eqno(three)
$$
and act on a Hilbert space of dimension $2^3 = 8$. As pointed out in
the first section, the dimensionality of the original space has been
increased by a factor of $2^{({\cal N}/2)} = 2$, where ${\cal N}=2$
is the number of spins.

By using
$$ -i \vec \eta_1 \cdot \vec \eta_2 = \vec f\dg \cdot \vec f -
{3 \over 2},\eqno(four)
$$
one can now write the Hamiltonian in terms of the f-operators
$$ H = {3 \over 8} - {1 \over 2}
( \vec f\dg \cdot \vec f - {3 \over 2} ) ^2.
$$

The spectrum can now be easily worked out
$$
\eqalign{
& E_o = - 3/4, \cr
& E_1 = 1/4. \cr
}\eqno(5)
$$
The first level is doubly degenerate and the second one is 6-fold
degenerate. The exact eigenenergies are correctly obtained, as
expected. Besides, the singlet ground state and the triplet excited
state are replicated by the same factor of 2. This additional
degeneracy can be traced back to the invariance of the Majorana
representation with respect to the $Z_2$ transformations $\vec \eta_i
\rightarrow - \vec \eta_i$, which is reflected in a particle-hole
symmetry of the Hamiltonian $\vec f \rightarrow \vec f\dg$.

A mean-field treatment of this model can be performed by the following
decoupling procedure
$$
H_{MF} = {3 \over 8} + iV (\vec \eta_1 \cdot \vec \eta_2) + {V^2 \over
2}
= {3 \over 8} - V (\vec f\dg \cdot \vec f - {3 \over 2}) + {V^2 \over 2}.
\eqno(six)$$

A static order parameter $V$ breaks the aforementioned $Z_2$ symmetry
and there are two stable solutions related to each other by the
transformation
$$
V \rightarrow - V, \qquad \vec f \rightarrow \vec f\dg.\eqno(seven)
$$
If $V$ is positive, the ground state corresponds to $\vec f\dg \cdot
\vec f - {3 \over 2} = V = {3 \over 2}$ and its energy is $ - 3/4 $,
which is the exact value.

Consider now the one-dimensional XY model, which can be solved exactly
through a mapping to a free fermion model. This mapping is
conventionally performed by means of a Jordan-Wigner transformation.
It will now be shown that an analogous mapping can be achieved through
the Majorana representation of spins.

The 1D XY Model Hamiltonian is given by
$$ H_{XY} = \sum_{i=1}^{\cal N} \left( S^{i}_{x} S^{i+1}_{x} +
S^{i}_{y} S^{i+1}_{y} \right) = {1 \over 2} \sum_{i=1}^{\cal N} \left(
S^{i}_{+} S^{i+1}_{-} + S^{i}_{-} S^{i+1}_{+} \right).\eqno(eight)
$$

In terms of the Majorana fermions, one can write
$$
\eqalign{
S^{i}_{+} \equiv S^{i}_{x} + i S^{i}_{y} = \eta^{i}_{3} (\eta^{i}_{1}
+ i \eta^{i}_{2}), \cr
S^{i}_{-} \equiv S^{i}_{x} - i S^{i}_{y} = (\eta^{i}_{1}
- i \eta^{i}_{2}) \eta^{i}_{3}. \cr
}\eqno(nine)
$$
At each site one can define a complex
fermion
$$
\eqalign{
c^{i} \equiv {1 \over \sqrt 2} ( \eta_1 - i  \eta_2),\cr
c^{i}\dg \equiv {1 \over \sqrt 2} ( \eta_1 + i  \eta_2),\cr
}\eqno(ten)
$$
satisfying
$$
\{ c^{i}, c^{j}\dg \} = \delta_{ij} ; \qquad \{ c^{i}, \eta^{j}_{3} \}
= 0 ; \qquad
\{ c^{i}\dg, \eta^{j}_{3} \}
= 0.\eqno(eleven)
$$

The Hamiltonian can now be written as
$$ H_{XY} = \sum_{i=1}^{\cal N} \left( c^{i}\dg \eta^{i}_{3}
\eta^{i+1}_{3} c^{i+1} + c^{i} \eta^{i}_{3}
\eta^{i+1}_{3} c^{i+1}\dg \right).\eqno(twelve)
$$

Let $U^{i}$ be an operator acting on site $i$
$$
U^{i} \equiv P^i_{o} + {\sqrt 2} \eta^{i}_{3} P^{i}_{1},\eqno(thirteen)
$$
where
$$
P^i_{o} \equiv 1 - c^{i}\dg c^{i}, \qquad P^i_{1}
\equiv c^{i}\dg c^{i},\eqno(fourteen)
$$
are projectors onto the vacant and occupied states of site $i$. The
operator $U^{i}$ is both hermitian and unitary
$$
\eqalign{
& U^{i}\dg = P^i_{o} + {\sqrt 2} P^{i}_{1} \eta^{i}_{3} = U^{i}, \cr
& U^{i} U^{i}\dg = U^{i}\dg U^{i} = {P^{i}_{o}}^2 + 2 {\eta^{i}_{3}}^2
{P^{i}_{1}}^2 = P^{i}_{o} + P^{i}_{1} = 1. \cr
}\eqno(fifteen)
$$
One can now easily prove that
$$
\eqalign{
U^{i} (\eta^{i}_{3} c^{i}\dg ) U^{i}\dg = {c^{i}\dg \over \sqrt
{2}},\cr
U^{i} (c^{i} \eta^{i}_{3} ) U^{i}\dg = {c^{i} \over \sqrt {2}}.\cr
}\eqno(sixteen)
$$
The canonical transformation generated by $U^{i}$ transforms away the
third Majorana component.

Let $U$ be the ordered product
$$
U \equiv \prod_{i=1}^{\cal N} U^{i}\eqno(seventeen)
$$
Under the action of $U$, the transformed operator now acquires a
non-local phase factor
$$
\eqalign{
U (\eta^{i}_{3} c^{i}\dg ) U\dg & = \prod_{j=1}^{i} U^{j} (\eta^{i}_{3}
c^{i}\dg ) \prod_{k=i}^{1} U^{k} = \prod_{j=1}^{i-1} U^{j} \left({c^{i}\dg
\over \sqrt {2}}\right) \prod_{k=i-1}^{1} U^{k} = \cr
& = (-1)^{\sum_{j=1}^{i-1} n_{j}} \left( {c^{i}\dg
\over \sqrt {2}}\right), \cr
}
\eqno(canonical)
$$
where $n_{j} = c^{j}\dg c^{j}$, and, in the last step, use has been made
of
$$\eqalign{
U^{j} ( c^{i}\dg ) U^{j} &= c^{i}\dg (P^{j}_{o} - \sqrt {2}
\eta^{j}_{3} P^{j}_{1}) (P^{j}_{o} + \sqrt {2} \eta^{j}_{3} P^{j}_{1}) \cr
&=
c^{i}\dg (-1)^{n_{j}} \quad (i \neq j).
}
\eqno(eighteen)
$$

Using relation \(canonical) and its complex conjugate, one can
transform the Hamiltonian into
$$
\eqalign{
H'_{XY} &=  U H_{XY} U\dg \cr
&= \sum_{i=1}^{\cal N} \left( U c^{i}\dg
\eta^{i}_{3} U\dg U
\eta^{i+1}_{3} c^{i+1} U\dg  + U c^{i} \eta^{i}_{3} U\dg U
\eta^{i+1}_{3} c^{i+1}\dg U\dg \right) \cr
&=  {1 \over 2} \sum_{i=1}^{\cal N} \left( c^{i}\dg
c^{i+1} (-1)^{n_{i}} + c^{i} c^{i+1}\dg (-1)^{n_{i}} \right) \cr
&=  {1 \over 2} \sum_{i=1}^{\cal N} \left( c^{i}\dg
c^{i+1} - c^{i} c^{i+1}\dg \right), \cr
}\eqno(nineteen)
$$
which is the usual free fermion expression obtained by the
Jordan-Wigner transformation. The third Majorana components have been
transformed out of the problem. Tracing over these variables will
cancel out the overall factor of $2^{{\cal N}/2}$ and one is left
with a free fermion theory.

\references

\refis{steglich}F. Steglich , J. Aarts. C. D. Bredl, W. Leike,
D. E. Meshida, W. Franz \& H. Sch\"afer, \prl 43, 1892, 1976.

\refis{steglich2}C. Giebel. S. Thies, D. Kacrowski, A. Mehner,
A. Granel, B. Seidel, U. Ahnheim, R. Helfrich, K. Peters,
C. Bredl and F. Steglich, \zpb 83, 305, 1991;
C. Giebel, C. Shank, S. Thies, H. Kitazawa, C. D. Bredl, A. B\" ohm,
A. Granel, R. Caspary, R. Helfrich, U. Ahlheim, G. Weber and
F. Steglich, \zpb 84, 1, 1991.

\refis{ott}K. Andres   , J. Graebner \& H. R. Ott., \prl 35, 1779, 1975.

\refis{meyer}M. Goeppart Meyer, \pr 60, 184, 1941.

\refis{mott}N. F. Mott, \phl 30, 402, 1974.

\refis{lonz}L. Taillefer and G. G. Lonzarich, \prl 60, 1570, 1988.

\refis{spring}P. H. P. Reinders, M. Springford et al, \prl 57, 1631, 1986.

\refis{blount}E. I. Blount, \prb 60, 2935, 1985.

\refis{blandin}A. Blandin \& J. Friedel, {\sl J. Phys. Radium} ,
{\bf 19} , 573 , (1958).

\refis{anda}P. W. Anderson, \pr 124, 41, 1961.

\refis{langreth}D. Langreth, \pr 150, 516, 1966.

\refis{haldane}F. D, M. Haldane \prl 40, 416 , 1978.

\refis{wilson}K. G. Wilson, \rmp 47, 773, 1976.

\refis{Martin}R. M. Martin, \prl 48, 362, 1982.

\refis{krish}H. R. Krishnamurthy, J. Wilkins and K. G. Wilson, \prb 21 , 1003,
1980.

\refis{nunes}B. Jones and C: M. Varma, \prl 58, 842, 1987;
V. L. Lib\'ero and L. N. Oliviera, \prl 65,
2042, 1990;  \prb 42, 3167, 1990.

\refis{Phil}P. W.  Anderson, \jpc 3, 2436, 1970.

\refis{NCA}Y. Kuromoto, \zpb 52, 37, 1986 ; W. E. Bickers, D. Cox \&
J. Wilkins, \prl 54, 230, 1985.

\refis{nick}N. Read \&  D. M. Newns, \jpc 29, L1055, 1983 ;
N.Read, \jpc 18, 2051, 1985.

\refis{me}P. Coleman, \prb 28, 5255, 1983.

\refis{long}P. Coleman, \prb  35, 5072, 1987.

\refis{Noz}P. Nozi\`eres,\jdc 37, C1-271, 1976 ;
P. Nozi\`eres and A. Blandin, \jdp 41, 193, 1980.

\refis{noz}P. Nozi\`eres and A. Blandin, \jdp 41, 193, 1980.

\refis{morel}P. Morel \& P. W. Anderson, \pr 125, 1263, 1962.

\refis{usound}D. Bishop et al, \prl 65, 1263, 1984.

\refis{broholm}C. Broholm et al., \prl 65, 2062, 1990.

\refis{anderson}P. W. Anderson, \prb 30, 1549, 1984.

\refis{varma}C. M. Varma, {\sl Comments in Solid State Physics}
{\bf 11} 221, 1985.

\refis{volovic}G. E. Volovik and L.P.  Gorkov, \jetl 39, 674, 1984;
\jetp 61, 843, 1984.

\refis{lev} L. P. Gorkov, {\sl Europhysics Lett.} , Nov (1991).

\refis{miyake}K. Miyake, S. Schmitt Rink and C. M. Varma, \prb 34 , 7716, 1986.

\refis{bealmonod}M. T. B\'eal Monod, C. Borbonnais \& V.J.  Emery,
\prb 34, 7716, 1986.

\refis{momentUBE13}R. H. Heffner, D. W. Cooke \& D. E. Maclaughlin,
{\sl 5th Int. Conf. Valence Fluctuations} (1988).
\refis{blount}
E. I. Blount, \prb 32, 2935 , 1985.

\refis{Upt3theory}E. I. Blount et al, \prl 64, 3074, 1990 ;
R. Joynt, {\sl   S. Sci. Technol.} 1, 210, 1988 ;
W. Puttika \& R. Joynt, \prb 37, 2377, 1988 ;
T. A. Tokuyasi et al , \prb 41, 891, 1990 ;
K. Machida et al, {\sl J. Phys. Soc. Japan}{\bf 58} 4116, 1989.

\refis{us}A. P. Ramirez, P. Coleman,   P. Chandra et al, {\sl Phys. Rev.
Lett.}, {\bf 68}, 2680, 1992.

\refis{Cox}D. L. Cox, \prl 59, 1240, 1987 .

\refis{cox2}C. L. Seanan, M. B. Maple et al , \prl 67 , 2892, 1991.

\refis{tsvelik}B. Andraka \& A. M. Tsvelik, \prl 67, 2886, 1991.

\refis{gan}P. Coleman and J. Gan,
 {\sl Physica B} {\bf 171}, 3 (1991);
J. Gan and P. Coleman to be published.

\refis{sms}S. D. Bader, N. E. Phillips, D. B. McWhan, \prb 7, 4686, 1973.

\refis{takab}M. Kyogaku, Y. Kitaoka, K. Asayama, T. Takabatake and
H. Fujii, \jpjap 61, 43, 1992.

\refis{takab2}T. Takabatake, M. Nagasawa, H. Fujii, G. Kido, M. Nohara,
S. Nishigori, T. Suzuki, T. Fujita, R. Helfrich, U. Ahlheim, K. Fraas, C.
Geibel, F. Steglich, \prb 45, 5740, 1992.

\refis{ybb10}K. Sugiyama, H. Fuke, K. Kindo, K. Shimota, A. Menovsky,
J. Mydosh and M. Date, \journal J. Jap Phys. Soc, 59, 3331, 1990.

\refis{earlyins}T. Kasuya, M. Kasuya, K. Takegahara, \journal J. Less Common
Met, 127, 337, 1987.

\refis{rice} T. M. Rice and K. Ueda, \prb 34, 6420, 1986; C. M. Varma,
W. Weber and L. J. Randall, \prb 33, 1015, 1986.

\refis{franse} J. J. M. Franse, K. Kadowaki, A. Menovsky, M. Van Sprang
and A. de Visser, {\sl J. Appl. Phys.} {\bf 61}, 3380, (1987).

\refis{horn}S. Horn, \physica 171, 206, 1991.

\refis{visser}A. de Visser, J. Floquet, J.J.M. Franse, P. Haen,
K. Hasselbach, A. Lacerda and L. Taillefer, \physica   171, 190, 1991.

\refis{mignod}J. Rossat-Mignod, L. P. Regnault, J. L. Jacoud, C. Vettier,
P. Lejay and J. Floquet, \jmmm 76-77, 376, 1988.

\refis{aeppli} G. Aeppli, C. Broholm, E. Bucher and D. J. Bishop,
\physica 171, 278, 1991.

\refis{kuramoto}K. Miyake and Y. Kuramoto, \physica 171, 20 , 1991.

\refis{kuramoto2}Y. Kuramoto \& T. Watanabe, \physica 148B, 80, 1987.

\refis{aepplins}G. Aeppli, E. Bucher and  T. E. Mason,{\sl Proc. National High
Magnetic Field Conference}, eds E. Manousakis, P. Schlottmann,
P. Kumar, K. Bedell and F. M. Mueller
(Addison Wesley),  175, (1991).

\refis{aps1}Y. Dlichaoch, M. A. Lopez de la Torre, P. Visani, B. W. Lee amd
M. B. Maple, \journal Bull Am. Phys Soc, 37, 60, 1992.

\refis{aps2} J. G. Luissier,\journal Bull Am. Phys Soc, 37, 739, 1992.

\refis{allen}L. Z. Liui, J. W. Allen, C. L. Seaman, M. B. Maple,
Y. Dalichaouch, J. S. Kang, M. S. Torikachvili, M. A. Lopez de la
Torre, \prl 68, 1034, 1992.

\refis{ramirez2} A. Ramirez, to be published (1992).

\refis{andrei}P. Coleman and N. Andrei, \jpc 19, 3211, 1986.

\refis{andrei2}C. Destri and N. Andrei, \prl 52, 364, 1984

\refis{allen}J. W. Allen and R. M. Martin, \jdc 41, C5, 1980.

\refis{batlogg}J. W. Allen, R. M. Martin, B. Batlogg and P. Wachter,
{\sl Appl. Phys.} {\bf 49}, 2078, (1978).

\refis{smb6}A. Menth, E. Buehler and T. H. Geballe, \prl 22, 295, 1969.

\refis{fazekas}S. Doniach and P. Fazekas, {\sl Phil. Mag.} to be published
in Phil. Mag. (1992).

\refis{insul1}M. F, Hundley, P. C. Canfield, J. D. Thompson, Z. Fisk
and J. M. Lawrence, \prb 42, 6842, 1990.

\refis{insul2} S. K. Malik and D. T. Adroja, \prb 43, 6295, 1991.

\refis{aliev}F. G. Aliev, V. V. Moschalkov,
V. V. Kozyrkov, M. K. Zalyalyutdinov,
V. V. Pryadum and R. V. Scolozdra, \jmmm 76-76, 295, 1988.

\refis{lonzarich}G. G. Lonzarich, \jmmm 76-77, 1, 1988.

\refis{martinins}R. Martin and J. W. Allen, \journal J. Appl. Phys., 50, 11,
1979.

\refis{lacroix}C. Lacroix and M. Cyrot, \prb, 43, 12906, 1991.

\refis{kotliar} M. Rozenberg, X. Y. Zhang,  and G. Kotliar, to be  published
(1991)

\refis{auerbach}
A. Auerbach and K.Levin,\prl 57, 877, 1986.

\refis{millis}
A.J. Millis and P.A. Lee, \prb 35, 3394, 1986.

\refis{Crag79}
D.M. Cragg and P.Lloyd, \jpc 12, L215, 1979.

\refis{aepplins}
T. E. Mason, G. Aeppli, A. R. Ramirez, K. M. Glausen, , C. Broholm, N.
St\"ucheli, E. Bucher \& T. M. M. Pasltra, Bell Labs preprint (1992).

\refis{maple}\refis{6}

\refis{palstra}
 T.T.M. Palstra, A.A. Menovsky. J. van den Berg, A.J.
Dirkmaat, P.H. Kes, G.J. Nieuwenhuys and J.A. Mydosh, \prl 55, 2727, 1985.

\refis{maple} M.B. Maple, J.W. Chen, Y. Dalichaouch, T. Kohara, C. Rossel,
M.S. Torikachvili, M. McElfresh and J.D. Thompson, \prl 56, 185, 1986.

\refis{broholmhfafm}H. J. Kjems and C. Broholm, \jmmm 76\&77, 371, 1988.

\refis{hfmags}$URu_2Si_2$, $U(Pt_{1-x}Pd)_3$,  $U_2Zn_{17}$ and $CeB_6$
are examples of commensurate afms;
the large moment systems $Ce(Cu_{1-x}Ni_x)_2Ge_2$ and $CeGa_2$ are examples
of incommensurate order.

\refis{hfmags2}J. Rossat-Mignod, L. P. Regnault, J. L. Jacoud, C. Vettier,
P. Lejay, J. Floquet, E. Walker, D. Jaccard and A. Amato \jmmm 76\&77, 376,
1988.

\refis{summary}For a general review
of heavy fermion physics, see N. Grewe and F. Steglich, {\sl Handbook on the
Physics and Chemistry of Rare Earths}, eds. K. A. Gschneider and L. Eyring),
{\bf 14},  343, (1991) (Elsevier, Amsterdam).

\refis{theory}For a
review of the theory of the normal phase, see e.g. P. A. Lee, T. M. Rice, J. W.
Serene, L. J. Sham and J. W. Wilkins,
\journal Comm. Cond. Mat. Phys., 12, 99, (1986); also
P. Fulde, J. Keller and G. Zwicknagl, {\sl  Solid. State Physics} {\bf 41},
 1 (1988).

\refis{rauchs}For the most recent comparison of the magnetic properties of all
known
heavy fermion superconductors, see M. Kyogaku, Y. Kitaoka, K. Asayama, C.
Geibel, C. Schank and F. Steglich, \jpjap 61, 2660, 1992.

\refis{steglich}F. Steglich , J. Aarts. C. D. Bredl, W. Leike,
D. E. Meshida, W. Franz \& H. Sch\"afer, \prl 43, 1892, 1976.

\refis{steglich2}C. Giebel. S. Thies, D. Kacrowski, A. Mehner,
A. Granel, B. Seidel, U. Ahnheim, R. Helfrich, K. Peters,
C. Bredl and F. Steglich, \zpb 83, 305, 1991;
C. Giebel, C. Shank, S. Thies, H. Kitazawa, C. D. Bredl, A. B\" ohm,
A. Granel, R. Caspary, R. Helfrich, U. Ahlheim , G. Weber and
F. Steglich, \zpb 84, 1, 1991.

\refis{ott}K. Andres   , J. Graebner \& H. R. Ott.\prl 35, 1779, 1975.

\refis{meyer}M. Goeppart Meyer, \pr 60, 184, 1941.

\refis{mott}N. F. Mott, \phl 30, 402, 1974.

\refis{lonz}L. Taillefer and G. G. Lonzarich, \prl 60, 1570, 1988.

\refis{spring}P. H. P. Reinders, M. Springford et al, \prl 57, 1631, 1986.

\refis{blount}E. I. Blount, \prb 60, 2935, 1985.

\refis{blandin}A. Blandin \& J. Friedel, {\sl J. Phys. Radium} ,
{\bf 19} , 573 , (1958).

\refis{anda}P. W. Anderson, \pr 124, 41, 1961.

\refis{langreth}D. Langreth, \pr 150, 516, 1966.

\refis{haldane}F. D, M. Haldane \prl 40, 416 , 1978.

\refis{wilson}K. G. Wilson, \rmp 47, 773, 1976.

\refis{Martin}R. M. Martin, \prl 48, 362, 1982.

\refis{krish}H. R. Krishnamurthy, J. Wilkins and K. G. Wilson, \prb 21 , 1003,
1980.

\refis{nunes}B. Jones and C: M. Varma, \prl 58, 842, 1987;
V. L. Lib\'ero and L. N. Oliviera, \prl 65,
2042, 1990;  \prb 42, 3167, 1990.

\refis{Phil}P. W.  Anderson, \jpc 3, 2436, 1970.

\refis{NCA}Y. Kuromoto, \zpb 52, 37, 1986 ; W. E. Bickers, D. Cox \&
J. Wilkins, \prl 54, 230, 1985.

\refis{kuromoto}Y. Kuromoto, \zpb 52, 37, 1986.

\refis{wilkins} W. E. Bickers, D. Cox \&
J. Wilkins, \prl 54, 230, 1985.

\refis{nick}N. Read \&  D. M. Newns, \jpc 29, L1055, 1983 ;
N.Read, \jpc 18, 2051, 1985.

\refis{me}P. Coleman, \prb 28, 5255, 1983.

\refis{long}P. Coleman, \prb  35, 5072, 1987.

\refis{Noz}P. Nozi\`eres,\jdc 37, C1-271, 1976 ;
P. Nozi\`eres and A. Blandin, \jdp 41, 193, 1980.

\refis{noz}P. Nozi\`eres and A. Blandin, \jdp 41, 193, 1980.

\refis{morel}P. Morel \& P. W. Anderson, \pr 125, 1263, 1962.

\refis{usound}D. Bishop et al, \prl 65, 1263, 1984.

\refis{broholm}C. Broholm et al., \prl 65, 2062, 1990.

\refis{anderson}P. W. Anderson, \prb 30, 1549, 1984.

\refis{varma}C. M. Varma, {\sl Comments in Solid State Physics}
{\bf 11} 221, (1985).

\refis{volovic}G. E. Volovik  \& L.P.  Gorkov, \jetl 39, 674, 1984 ;
\jetp 61, 843, 1984.

\refis{lev} L. P. Gorkov, {\sl Europhysics Lett.} , Nov (1991).

\refis{miyake}K. Miyake, S. Schmitt Rink and C. M. Varma, \prb 34 , 7716, 1986.

\refis{bealmonod}M. T. B\'eal Monod, C. Borbonnais \& V.J.  Emery,
\prb 34, 7716, 1986.

\refis{momentUBE13}R. H. Heffner, D. W. Cooke \& D. E. Maclaughlin,
{\sl 5th Int. Conf. Valence Fluctuations} (1988).
\refis{blount}
E. I. Blount, \prb 32, 2935 , 1985.

\refis{Upt3theory}E. I. Blount et al, \prl 64, 3074, 1990 ;
R. Joynt, {\sl   S. Sci. Technol.} 1, 210, 1988;
W. Puttika \& R. Joynt, \prb 37, 2377, 1988;
T. A. Tokuyasi et al , \prb 41, 891, 1990;
K. Machida et al, {\sl J. Phys. Soc. Japan}{\bf 58} 4116, 1989.

\refis{norman}M. R. Norman, \journal Physica, C194, 203, 1992.

\refis{machida}K. Machida and M. Ozaki, \prl 66, 3293, 1991.

\refis{us}A. P. Ramirez, P. Coleman,   P. Chandra et al, \prl 68, 2680, 1992.

\refis{colemantrans}P. Coleman, \prl 59, 1026, 1987.

\refis{Cox}D. L. Cox, \prl 59, 1240, 1987 .

\refis{cox2}C. L. Seanan, M. B. Maple et al , \prl 67 , 2892, 1991.

\refis{tsvelik}B. Andraka \& A. M. Tsvelik, \prl 67, 2886, 1991.

\refis{gan}P. Coleman and J. Gan,
 {\sl Physica B} {\bf 171}, 3 (1991);
J. Gan and P. Coleman to be published.

\refis{sms}S. D. Bader, N. E. Phillips, D. B. McWhan, \prb 7, 4686, 1973.

\refis{takab}M. Kyogaku, Y. Kitaoka, K. Asayama, T. Takabatake and
H. Fujii, \jpjap 61, 43, 1992.

\refis{takab2}T. Takabatake, M. Nagasawa, H. Fujii, G. Kido, M. Nohara,
S. Nishigori, T. Suzuki, T. Fujita, R. Helfrich, U. Ahlheim, K. Fraas, C.
Geibel, F. Steglich, \prb 45, 5740, 1992.

\refis{ybb10}K. Sugiyama, H. Fuke, K. Kindo, K. Shimota, A. Menovsky,
J. Mydosh and M. Date, \journal J. Jap Phys. Soc, 59, 3331, 1990.

\refis{earlyins}T. Kasuya, M. Kasuya, K. Takegahara, \journal J. Less Common
Met, 127, 337, 1987.

\refis{rice} T. M. Rice and K. Ueda, \prb 34, 6420, 1986; C. M. Varma,
W. Weber and L. J. Randall, \prb 33, 1015, 1986.

\refis{franse} J. J. M. Franse, K. Kadowaki, A. Menovsky, M. Van Sprang
and A. de Visser, {\sl J. Appl. Phys.} {\bf 61}, 3380, (1987).

\refis{horn}S. Horn, \physica 171, 206, 1991.

\refis{visser}A. de Visser, J. Floquet, J.J.M. Franse, P. Haen,
K. Hasselbach, A. Lacerda and L. Taillefer, \physica   171, 190, 1991.

\refis{mignod}J. Rossat-Mignod, L. P. Regnault, J. L. Jacoud, C. Vettier,
P. Lejay and J. Floquet, \jmmm 76-77, 376, 1988.

\refis{aeppli} G. Aeppli, C. Broholm, E. Bucher and D. J. Bishop,
\physica 171, 278, 1991.

\refis{kuramoto}K. Miyake and Y. Kuramoto, \physica 171, 20 , 1991.

\refis{kuramoto2}Y. Kuramoto \& T. Watanabe, \physica 148B, 80, 1987.

\refis{aepplins}G. Aeppli, E. Bucher and  T. E. Mason,{\sl Proc. National High
Magnetic Field Conference}, eds E. Manousakis, P. Schlottmann,
P. Kumar, K. Bedell and F. M. Mueller
(Addison Wesley),  175, (1991).

\refis{aeppliins}T. Mason, G. Aeppli, A. P. Ramirez, K. N. Clausen ,
C. Broholm, N. Stucheli, E. Bucher and T. T. M. Palstra , \prl 69, 490, 1992.

\refis{aps1}Y. Dlichaoch, M. A. Lopez de la Torre, P. Visani, B. W. Lee amd
M. B. Maple, \journal Bull Am. Phys Soc, 37, 60, 1992.

\refis{aps2} J. G. Luissier,\journal Bull Am. Phys Soc, 37, 739, 1992.

\refis{allen}L. Z. Liui, J. W. Allen, C. L. Seaman, M. B. Maple,
Y. Dalichaouch, J. S. Kang, M. S. Torikachvili, M. A. Lopez de la
Torre, \prl 68, 1034, 1992.

\refis{ramirez2} A. Ramirez, to be published (1992).

\refis{andrei}P. Coleman and N. Andrei, \jpc 19, 3211, 1986.

\refis{andrei2}C. Destri and N. Andrei, \prl 52, 364, 1984.

\refis{allen}J. W. Allen and R. M. Martin, \jdc 41, C5, 1980.

\refis{batlogg}J. W. Allen, R. M. Martin, B. Batlogg and P. Wachter,
{\sl Appl. Phys.} {\bf 49}, 2078, (1978).

\refis{smb6}A. Menth, E. Buehler and T. H. Geballe, \prl 22, 295, 1969.

\refis{fazekas}S. Doniach and P. Fazekas, {\sl Phil. Mag.} to be published
in Phil. Mag. (1992).

\refis{insul1}M. F. Hundley et al. \prb 42, 6842, 1990.

\refis{insul2} S. K. Malik and D. T. Adroja, \prb 43, 6295, 1991.

\refis{aliev}F. G. Aliev et al.\jmmm 76-76, 295, 1988.

\refis{lonzarich}G. G. Lonzarich, \jmmm 76-77, 1, 1988.

\refis{martinins}R. Martin and J. W. Allen, \journal J. Appl. Phys., 50, 11,
1979.

\refis{lacroix}C. Lacroix and M. Cyrot, \prb, 43, 12906, 1991.

\refis{kotliar} M. Rozenberg, X. Y. Zhang,  and G. Kotliar, to be  published
(1991)

\refis{auerbach}
A. Auerbach and K.Levin,\prl 57, 877, 1986.

\refis{millis}
A.J. Millis and P.A. Lee, \prb 35, 3394, 1986.

\refis{Crag79}
D.M. Cragg and P.Lloyd, \jpc 12, L215, 1979.

\refis{aepplins}
T. E. Mason, G. Aeppli, A. R. Ramirez, K. M. Glausen, , C. Broholm, N.
St\"ucheli, E. Bucher \& T. M. M. Pasltra, Bell Labs preprint (1992).

\refis{maple}\refis{6}


\refis{palstra}
 T.T.M. Palstra, A.A. Menovsky. J. van den Berg, A.J.
Dirkmaat, P.H. Kes, G.J. Nieuwenhuys and J.A. Mydosh, \prl 55, 2727, 1985.

\refis{maple} M.B. Maple, J.W. Chen, Y. Dalichaouch, T. Kohara, C. Rossel,
M.S. Torikachvili, M. McElfresh and J.D. Thompson, \prl 56, 185, 1986.

\refis{broholmhfafm}H. J. Kjems and C. Broholm, \jmmm 76\&77, 371, 1988.

\refis{hfmags}$URu_2Si_2$, $U(Pt_{1-x}Pd)_3$,  $U_2Zn_{17}$ and $CeB_6$
are examples of commensurate afms;
the large moment systems $Ce(Cu_{1-x}Ni_x)_2Ge_2$ and $CeGa_2$ are examples
of incommensurate order.

\refis{hfmags2}J. Rossat-Mignod, L. P. Regnault, J. L. Jacoud, C. Vettier,
P. Lejay, J. Floquet, E. Walker, D. Jaccard and A. Amato \jmmm 76\&77, 376,
1988.

\refis{pethick}C. J. Pethick  and D. Pines, \prl 57, 118, 1986.

\refis{uthbe13cv}J. S. Kim, B.  Andraka and G. Stewart, \prb 44, 6921, 1991.

\refis{upt3cv}R. A. Fisher, S. Kim, B. F. Woodford, N. E. Phillips,
L. Taillefer, K. Hasselbach, J. Flouquet, A. L. Georgi and J. L. Smith,
\prl 62, 1411, 1989.

\refis{upt3phase}S. Adenwalla, S. W. Lin, Z. Zhao, Q. Z. Ran, J. B. Ketterson,
J. A. Sauls, L. Taillefer, D. G. Hinks, M. Levy and B. K. Sarma,
\prl 65, 2298, 1990.

\refis{maechida}K . Maechida and M. Ozaki, \prl 66, 3293, 1991.

\refis{trappmann}T. Trappmann, H.V. L\"ohneysen and L. Taillefer,
\prb 43, 13714, 1991.

\refis{steglich}F. Steglich , J. Aarts. C. D. Bredl, W. Leike,
D. E. Meshida, W. Franz \& H. Sch\"afer, \prl 43, 1892, 1976.

\refis{steglich2}C. Giebel. S. Thies, D. Kacrowski, A. Mehner,
A. Granel, B. Seidel, U. Ahnheim, R. Helfrich, K. Peters,
C. Bredl and F. Steglich, \zpb 83, 305, 1991;
C. Giebel, C. Shank, S. Thies, H. Kitazawa, C. D. Bredl, A. B\" ohm,
A. Granel, R. Caspary, R. Helfrich, U. Ahlheim, G. Weber and
F. Steglich, \zpb 84, 1, 1991.

\refis{upd3al2mom}A. Krimmel, P. Fisher, B. Roessli, H. Maletta,
C. Giebel, C. Schank, A. Grauel, A. Loidl and F. Steglich,
\zpb 86, 161, 1992.

\refis{berezinskii}V. L. Berezinskii, \journal JETP Lett. , 20, 287, 1974.

\refis{balatsky}For recent interest in odd-frequency
pairing, see
E. Abrahams, A. V. Balatsky,
\prb 45, 13125, 1992;
F. Mila and E. Abrahams, \prl 67, 2379, 1991.

\refis{miranda}P. Coleman, E. Miranda and A. Tsvelik, proceedings of
SCES92 conference, Sendai, to appear in {\it Physica B}(1993); Rutgers
University
preprint, to be published.

\refis{abrik}A. A. Abrikosov, \journal Physics, 2, 5, 1965.

\refis{suhl} H. Suhl, \pr 138A, 515, 1965.

\refis{morin}P. Morin and D. Schmitt, \pl 73A, 67, 1979.

\refis{cox3}D. Cox and A. E. Ruckenstein, private communication (1993).

\refis{fisk2}Z. Fisk, G. Aeppli et al, {\sl Solid State Comm.}, to be
published (1993).

\refis{schlessinger}Z. Schlessinger, Z. Fisk, G. Aeppli et al, preprint (1993).

\refis{grunertrans}W. P. Beyerman, A. M. Awasthi, J. P. Carini and G. Gruner,
\jmmm 76\&77, 207, 1988.

\refis{luthi}B. Luthi, B. Wolf, P. Thalmeier, W. Sixl and G. Bruls, preprint,
to be published (1993).

\refis{buyers}C. Broholm, H. Lin, P. T. Mathews, T. E. Mason, W. J. L. Buyers,
M. F. Collins, A. A. Menovsky , J. A. Mydosh and J. K. Kjems, \prb 43, 12809,
1991.


\refis{felten}R. Felten, F. Steglich, et al, \epl 2, 323, 1986.

\refis{uthbe13nmr}D. MacLaughlin, Cheng Tien, W. G. Clark, M. D. Lan,
Z. Fisk, J. L. Smith and H. R. Ott, \prl 51, 1833, 1984.

\refis{schloussky}K. Maki in ``Superconductivity'', editor R. Parks (1964).

\refis{maki2}K. Maki and P. Fulde, \pr 140, A1586, 1965.

\refis{hirschfeld}P. J. Hirschfeld, P. Wolfle and D. Einzel, \prb 37, 83, 1988.

\refis{upt3nmr}K. Asayama, Y. Kitaoka and Y. Kohori, \jmmm 76-77, 449, 1988.

\refis{hayden2}P. A. Midgley, S. M. Hayden, L. Taillefer and
H. v. L\" ohneysen, \prl 70, 678, 1993.

\refis{yosida}K. Yosida and Y. Yamada, \journal Prog. Theo. Phys., 46, 244,
1970;  {\bf 53}, 1286, (1975).

\refis{multi}  J. Gan, N. Andrei and P. Coleman,
{\sl Phys. Rev. Lett.} {\bf 70}, 686, (1993).


\refis{zuber}J. M. Drouffe and J. B. Zuber, {\it Phys. Reps} {\bf 102},
1, (1983).

\refis{cege2si2}D. Jaccard, K. Behnia, R. Cibin,L. Spendeler, J. Sierro,
R. Calemzuk, C. Marcenat, L. Schmidt, J. Flouquet, J. P. Brison and
P. Lejay, {\sl
Proc. Int. Conf. on Strongly Correlated Electron Systems}, Sendai 1992,
to be published in {\sl Physica B}, 1993.

\refis{doniach}S. Doniach, {\it Valence Instabilities and Narrow Band
Phenomena}, edited by R. Parks, 34, (Plenum 1977); S. Doniach, {\it Physica
}{\bf B91}, 231 (1977).

\refis{pwa1}P.W. Anderson, \journal Mater. Res. Bull., 8, 153, 1973.

\refis{steglich}F. Steglich , J. Aarts. C. D. Bredl, W. Leike,
D. E. Meshida, W. Franz \& H. Sch\"afer, \prl 43, 1892, 1976.

\refis{steglich2}C. Giebel. S. Thies, D. Kacrowski, A. Mehner,
A. Granel, B. Seidel, U. Ahnheim, R. Helfrich, K. Peters,
C. Bredl and F. Steglich, \zpb 83, 305, 1991;
C. Giebel, C. Shank, S. Thies, H. Kitazawa, C. D. Bredl, A. B\" ohm,
A. Granel, R. Caspary, R. Helfrich, U. Ahlheim , G. Weber and
F. Steglich, \zpb 84, 1, 1991.

\refis{ott}K. Andres   , J. Graebner \& H. R. Ott.\prl 35, 1779, 1975.

\refis{meyer}M. Goeppart Meyer, \pr 60, 184, 1941.

\refis{mott}N. F. Mott, \phl 30, 402, 1974.

\refis{lonz}L. Taillefer and G. G. Lonzarich, \prl 60, 1570, 1988.

\refis{spring}P. H. P. Reinders, M. Springford et al, \prl 57, 1631, 1986.

\refis{kurom}K. Miyake and Y. Kuramoto, \physica 171, 20 , 1991

\refis{blount}E. I. Blount, \prb 60, 2935, 1985.

\refis{blandin}A. Blandin \& J. Friedel, {\sl J. Phys. Radium} ,
{\bf 19} , 573 , (1958).

\refis{zwiknagel}G. Zwicknagl, \journal Adv. Phy., 41, 203, 1992;
N. d' Abrumenil and P. Fulde,  \jmmm  47-48,  1,  1985.

\refis{anda}P. W. Anderson, \pr 124, 41, 1961.

\refis{langreth}D. Langreth, \pr 150, 516, 1966.

\refis{haldane}F. D, M. Haldane \prl 40, 416 , 1978.

\refis{wilson}K. G. Wilson, \rmp 47, 773, 1976.

\refis{Martin}R. M. Martin, \prl 48, 362, 1982.

\refis{krish}H. R. Krishnamurthy, J. Wilkins and K. G. Wilson, \prb 21 , 1003,
1980.

\refis{nunes}B. Jones and C: M. Varma, \prl 58, 842, 1987;
V. L. Lib\'ero and L. N. Oliviera, \prl 65,
2042, 1990;  \prb 42, 3167, 1990.

\refis{Phil}P. W.  Anderson, \jpc 3, 2436, 1970.

\refis{NCA}Y. Kuromoto, \zpb 52, 37, 1986 ; W. E. Bickers, D. Cox \&
J. Wilkins, \prl 54, 230, 1985.

\refis{nick}N. Read \&  D. M. Newns, \jpc 29, L1055, 1983 ;
N.Read, \jpc 18, 2051, 1985.

\refis{me}P. Coleman, \prb 28, 5255, 1983.

\refis{long}P. Coleman, \prb  35, 5072, 1987.

\refis{Noz}P. Nozi\`eres,\jdc 37, C1-271, 1976 ;
P. Nozi\`eres and A. Blandin, \jdp 41, 193, 1980.

\refis{noz}P. Nozi\`eres and A. Blandin, \jdp 41, 193, 1980.

\refis{paradox}
P.~Nozi\`eres, \journal Ann. Phys. Fr., 10, 19, 1985.

\refis{morel}P. Morel \& P. W. Anderson, \pr 125, 1263, 1962.

\refis{usound}D. Bishop et al, \prl 65, 1263, 1984.

\refis{broholm}C. Broholm et al., \prl 65, 2062, 1990.

\refis{anderson}P. W. Anderson, \prb 30, 1549, 1984.

\refis{varma}C. M. Varma, {\sl Comments in Solid State Physics}
{\bf 11} 221, 1985.

\refis{volovic}G. E. Volovik  \& L.P.  Gorkov, \jetl 39, 674, 1984 ;
\jetp 61, 843, 1984.

\refis{lev} L. P. Gorkov, {\sl Europhysics Lett.} , Nov (1991).

\refis{miyake}K. Miyake, S. Schmitt Rink and C. M. Varma, \prb 34 , 7716, 1986.

\refis{bealmonod}M. T. B\'eal Monod, C. Borbonnais \& V.J.  Emery,
\prb 34, 7716, 1986.

\refis{ube13}H. R. Ott, H. Rudgier,  Z. Fisk and J. L. Smith, \prl 50, 1595,
1983.

\refis{upt3}G. R. Stewart, Z. Fisk, J. O. Willis and J. L. Smith, \prl 52, 697,
1984.

\refis{uru2si2}W. Schlabitz, J. Baumann, B. Pollit, U. Rauchschwalbe,
H. M. Mayer, U. Ahlheim and C. D. Bredl, \journal Z. Phys, B62, 171, 1986;

\refis{momentUBE13}R. H. Heffner, D. W. Cooke \& D. E. Maclaughlin,
{\sl 5th Int. Conf. Valence Fluctuations} (1988).
\refis{blount}
E. I. Blount, \prb 32, 2935 , 1985.

\refis{Upt3theory}E. I. Blount et al, \prl 64, 3074, 1990 ;
R. Joynt, {\sl   S. Sci. Technol.} 1, 210, 1988 ;
W. Puttika \& R. Joynt, \prb 37, 2377, 1988 ;
T. A. Tokuyasi et al , \prb 41, 891, 1990 ;
K. Machida et al, {\sl J. Phys. Soc. Japan}{\bf 58} 4116, 1989.

\refis{us}A. P. Ramirez, P. Coleman,   P. Chandra et al, {\sl Phys. Rev.
Lett.}, {\bf 68}, 2680, 1992.

\refis{jon}P. A. Lee, T. M. Rice, J. W. Serene, L. J. Sham and J. W. Wilkins,
\journal Comm. Cond. Mat. Phys., 12, 99, 1986; see also
P. Fulde, J. Keller and G. Zwicknagl, \journal Solid. State Physics, 41, 1,
1988;
for a more general review
of heavy fermion physics, see N. Grewe and F. Steglich, {\sl Handbook on the
Physics and Chemistry of Rare Earths}, eds. K. A. Gschneider and L. Eyring),
{\bf 14},  343, 1991 (Elsevier, Amsterdam).

\refis{science} P.W. Anderson, \journal Science, 235, 1196, 1987

\refis{Cox}D. L. Cox, \prl 59, 1240, 1987 .

\refis{cox2}C. L. Seanan, M. B. Maple et al , \prl 67 , 2892, 1991.

\refis{tsvelik}B. Andraka \& A. M. Tsvelik, \prl 67, 2886, 1991.

\refis{gan}P. Coleman and J. Gan,
 {\sl Physica B} {\bf 171}, 3 (1991);
J. Gan and P. Coleman to be published.

\refis{sms}S. D. Bader, N. E. Phillips, D. B. McWhan, \prb 7, 4686, 1973.

\refis{takab}M. Kyogaku, Y. Kitaoka, K. Asayama, T. Takabatake and
H. Fujii, \jpjap 61, 43, 1992.

\refis{takab2}T. Takabatake, M. Nagasawa, H. Fujii, G. Kido, M. Nohara,
S. Nishigori, T. Suzuki, T. Fujita, R. Helfrich, U. Ahlheim, K. Fraas, C.
Geibel, F. Steglich, \prb 45, 5740, 1992.

\refis{ybb10}K. Sugiyama, H. Fuke, K. Kindo, K. Shimota, A. Menovsky,
J. Mydosh and M. Date, \journal J. Jap Phys. Soc, 59, 3331, 1990.

\refis{earlyins}T. Kasuya, M. Kasuya, K. Takegahara, \journal J. Less Common
Met, 127, 337, 1987.

\refis{rice} T. M. Rice and K. Ueda, \prb 34, 6420, 1986; C. M. Varma,
W. Weber and L. J. Randall, \prb 33, 1015, 1986.

\refis{franse} J. J. M. Franse, K. Kadowaki, A. Menovsky, M. Van Sprang
and A. de Visser, {\sl J. Appl. Phys.} {\bf 61}, 3380, (1987).

\refis{horn}S. Horn, \physica 171, 206, 1991.

\refis{visser}A. de Visser, J. Flouquet, J.J.M. Franse, P. Haen,
K. Hasselbach, A. Lacerda and L. Taillefer, \physica   171, 190, 1991.

\refis{mignod}J. Rossat-Mignod, L. P. Regnault, J. L. Jacoud, C. Vettier,
P. Lejay and J. Flouquet, \jmmm 76-77, 376, 1988.

\refis{aeppli} G. Aeppli, C. Broholm, E. Bucher and D. J. Bishop,
\physica 171, 278, 1991.

\refis{kuramoto}K. Miyake and Y. Kuramoto, \physica 171, 20 , 1991.

\refis{kuramoto2}Y. Kuramoto \& T. Watanabe, \physica 148B, 80, 1987.

\refis{aepplins}G. Aeppli, E. Bucher and  T. E. Mason,{\sl Proc. National High
Magnetic Field Conference}, eds E. Manousakis, P. Schlottmann,
P. Kumar, K. Bedell and F. M. Mueller
(Addison Wesley),  175, (1991).

\refis{aps1}Y. Dlichaoch, M. A. Lopez de la Torre, P. Visani, B. W. Lee amd
M. B. Maple, \journal Bull Am. Phys Soc, 37, 60, 1992.

\refis{aps2} J. G. Luissier,\journal Bull Am. Phys Soc, 37, 739, 1992.

\refis{allen}L. Z. Liui, J. W. Allen, C. L. Seaman, M. B. Maple,
Y. Dalichaouch, J. S. Kang, M. S. Torikachvili, M. A. Lopez de la
Torre, \prl 68, 1034, 1992.

\refis{ramirez2} A. Ramirez, to be published (1992).

\refis{colemantrans}P. Coleman, \prl 59, 1026, 1987.

\refis{andrei}P. Coleman and N. Andrei, \jpc 19, 3211, 1986.

\refis{andrei2}C. Destri and N. Andrei, \prl 52, 364, 1984.

\refis{allen}J. W. Allen and R. M. Martin, \jdc 41, C5, 1980.

\refis{batlogg}J. W. Allen, R. M. Martin, B. Batlogg and P. Wachter,
{\sl Appl. Phys.} {\bf 49}, 2078, (1978).

\refis{smb6}A. Menth, E. Buehler and T. H. Geballe, \prl 22, 295, 1969.

\refis{fazekas}S. Doniach and P. Fazekas, {\sl Phil. Mag.} to be published
in Phil. Mag. (1992).

\refis{insul1}M. F, Hundley, P. C. Canfield, J. D. Thompson, Z. Fisk
and J. M. Lawrence, \prb 42, 6842, 1990.

\refis{insul2} S. K. Malik and D. T. Adroja, \prb 43, 6295, 1991.

\refis{aliev}F. G. Aliev, V. V. Moschalkov,
V. V. Kozyrkov, M. K. Zalyalyutdinov,
V. V. Pryadum and R. V. Scolozdra, \jmmm 76-76, 295, 1988.

\refis{hundley}P. C. Canfield, M. F. Hundley, A. Lacerda, J. D. Thompson and
Z. Fisk, to be published (1992).

\refis{lonzarich}G. G. Lonzarich, \jmmm 76-77, 1, 1988.

\refis{martinins}R. Martin and J. W. Allen, \journal J. Appl. Phys., 50, 11,
1979.

\refis{lacroix}C. Lacroix and M. Cyrot, \prb, 43, 12906, 1991.

\refis{kotliar} M. Rozenberg, X. Y. Zhang,  and G. Kotliar, to be  published
(1991)

\refis{auerbach}
A. Auerbach and K.Levin,\prl 57, 877, 1986.

\refis{millis}
A.J. Millis and P.A. Lee, \prb 35, 3394, 1986.

\refis{Crag79}
D.M. Cragg and P.Lloyd, \jpc 12, L215, 1979.

\refis{poor}P. W. Anderson, \journal Comm. S. St. Phys., 5, 72, 1973;
\jpc 3, 2346, 1970.

\refis{yuval}Anderson P. W. \& G. Yuval, \prl 45, 370, 1969;
Anderson P. W. \& G. Yuval, \prb 1, 1522, 1970;
Anderson P. W. \& G. Yuval, \jpc 4, 607, 1971.

\refis{swolf}J. R. Schrieffer and P. Wolff, \pr 149, 491, 1966.

\refis{cschr}B. Coqblin and J. R. Schrieffer, \pr 185, 847, 1969.

\refis{rkky}M. A. Ruderman and C. Kittel, \pr 78, 275, 1950;
T. Kasuya, \journal Prog. Theo. Phys., 16, 45, 1956;
K. Yosida, \pr 106, 896, 1957.

\refis{aepplins}
T. Mason, G. Aeppli, A. P. Ramirez, K. N. Clausen, C. Broholm,
N. Stucheli, E. Bucher, T. T. M. Palstra, \prl 69, 490, 1992.

\refis{maple}\refis{6}

\refis{palstra}
 T.T.M. Palstra, A.A. Menovsky. J. van den Berg, A.J.
Dirkmaat, P.H. Kes, G.J. Nieuwenhuys and J.A. Mydosh, \prl 55, 2727, 1985.

\refis{maple} M.B. Maple, J.W. Chen, Y. Dalichaouch, T. Kohara, C. Rossel,
M.S. Torikachvili, M. McElfresh and J.D. Thompson, \prl 56, 185, 1986.

\refis{broholmhfafm}H. J. Kjems and C. Broholm, \jmmm 76\&77, 371, 1988.

\refis{hfmags}$URu_2Si_2$, $U(Pt_{1-x}Pd)_3$,  $U_2Zn_17$ and $CeB_6$
are examples of commensurate afms;
the large moment systems $Ce(Cu_{1-x}Ni_x)_2Ge_2$ and $CeGa_2$ are examples
of incommensurate order.

\refis{hfmags2}J. Rossat-Mignod, L. P. Regnault, J. L. Jacoud, C. Vettier,
P. Lejay, J. Flouquet, E. Walker, D. Jaccard and A. Amato \jmmm 76\&77, 376,
1988.

\refis{sokol}L. Gorkov and A. V. Sokol, {\sl JETP Lett} { \bf 52} 1103 (1990).

\refis{azaria} P. Azaria, B. Delamotte and T. Jolicoeur, {\sl Phys. Rev. Lett},
{\bf 64}, 3175 (1990); K. B. Efetov, D. E. Khelminitskii and A. I. Larkin,
{\sl J. E. T. P. Lett.}, {\bf 52} , 568 (1980).

\refis{tsvelik}A. M. Tsvelik, {\sl Phys. Rev. B.} to be published (1990);
A. M. Polyakov and P. Weigmann {\sl Phys. Lett.} {\bf 131B}, 121 (1983);
H. M. Babujian and A. M. Tsvelik, {\sl Nucl. Phys. } {\bf B265}, 24 (1986).

\refis{toulouse}G. Toulouse and M. Kl\'eman, {\sl J. Phys. Lett. (Paris)}
{\bf 37}, L149 (1976).

\refis{volovik}G. E. Volovik and V. P. Mineev, {\sl J. E. T. P. } {\bf 45 (6)},
1186 (1977).

\refis{saslow}B. I. Halperin and W. M.  Saslow, {\sl Phys. Rev.} {\bf B 16},
2154
(1977).

\refis{japs}H. Kawamura and S. Miyashita, {\sl J. Phys. Soc. Jap.} {\bf 53},
4138 (1984).

\refis{liang2}S. Liang, B. Doucot, I. Ritchey, P. Chandra
and P. Coleman, to be published (1990).

\refis{ritchey2} I. Ritchey and P. Coleman, {\sl J. Cond. Matt.}, in
press(1990).

\refis{trieste}P. Chandra and P. Coleman,  {Int. J. of Mod. Phys.},
{\bf 3} 1729 (1989).

\refis{devries}H. de Vries, {\sl Acta Crystallograph} {\bf 4},
219 (1951); P. G. de Gennes, {\sl Physics of Liquid Crystals},
Oxfd. Univ. , pp 222-235 (1974).

\refis{blume2} M. Blume and D. Gibbs, {\sl Phys. Rev. } {\bf B 37},
1779, 1988;
M. Blume, {\sl J. Appl. Phys.}{\bf  57}, 3615, 1985.

\refis{optact} D. P. Siddons, M. Hart, Y. Amemiya and J. B.
Hastings, {\sl Phys. Rev. Lett.} {\bf 64}, 1967 (1990).

\refis{luther}A. Luther and I. Peschel, {\sl Phys. Rev.} {\bf B   } (1975).

\refis{inui}Inui and Doniach TO BE SUPPLIED.

\refis{levy}H.H. Chen and P.M. Levy, \prl 27, 1383, 1971
and \prl 27, 1385, 1971.

\refis{bhatt}R. Bhatt and S. Sachdev, AT\&T Bell Laboratory preprint,
to be published (1990).

\refis{gulayev}Y. V. Gulayev, {\sl J.E.T.P. Lett}, {\bf 2}, 1 (1965).

\refis{halp}B. I. Halperin and P. C. Hohenberg. {\sl Phys. Rev. } {\bf 188},
898 (1969).

\refis{enz}C. P. Enz, {\sl Rev. Mod. Phys} {\bf 46}, 704 (1974).

\refis{obradors}X. Obradors, A. Labarta, A. Isalgu\'e, J. Tejada,
J. Rodriguez and M. Pernet {\sl Sol. St. Comm.} {\bf 65},189 (1988).

\refis{ramirez}A. P. Ramirez, G. P. Espinosa and
A. S. Cooper, {\sl Phys. Rev. Lett.} {\bf 64} , 2070 (1990).

\refis{aeppli} C. Broholm, G. Aeppli, A. P. Ramirez,
G.P. Epinosa and A.S. Cooper, to be published (1990).

\refis{ch}S. Chakravarty et al, private communication.

\refis{and1} P.W. Anderson, \journal Science, 235, 1196, 1987;
P.W. Anderson in ``Frontiers in Many Particle Physics'', International
School of Physics, edited by J.R. Schrieffer and R. A. Broglia
(North-Holland, Amsterdam, 1988).

\refis{ry} J.D. Reger and A.P. Young, \prb 37, 5493, 1988;
\prb 37, 5978, 1988; M. Gross, E. Sanchez-Velasco, and E. Siggia,
\prb 39, 2484, 1989; J.D. Reger, J.A. Riera and A.P. Young,
\journal J. Phys. C. , I, 1955, 1989.

\refis{huse} D.A. Huse and V. Elser, \prl 60, 2531, 1988.

\refis{chak} S. Chakravarty, B.I. Halperin and D.R. Nelson, \prl 60,
1057, 1988.

\refis{cd} P. Chandra and B. Doucot, \prb 38, 9335, 1988.

\refis{il} L.B. Ioffe and A.I. Larkin, \journal Mod. Phys. B, 2, 203,
1988.

\refis{dm} E. Dagotto and A. Moreo, \prb 39, 4744, 1989.

\refis{read} N. Read and S. Sachdev, {\sl Nuclear Physics} {\bf B316},
609 (1989); \prl 62, 1694, 1989.

\refis{villain1} J. Villain, \journal J. Physique,  38, 26, 1977;
J. Villain, R. Bidaux, J. P. Carton and R. Conte,
\journal J. Physique,  41, 1263, 1980.

\refis{villain} J. Villain, \journal J. Physique,  38, 26, 1977;
J. Villain, R. Bidaux, J. P. Carton and R. Conte,
\journal J. Physique,  41, 1263, 1980.

\refis{henley} C.L. Henley, \prl 62, 2056, 1989; M.W. Long, \journal J. Phys.
Cond. Matt., 1, 2857, 1989.

\refis{raman} P.A. Fleury and R. Loudon, \pr 166, 514, 1968.

\refis{shendar1} E. Shendar, \journal Sov. Phys. JETP , 56, 178, 1982;
A. G. Gukasov, et al. \journal Europhys. Letter., 7, 83, 1988.

\refis{raste} E. Rastelli, L. Reatto and A. Tassi, \journal J. Phys.
C., 16, L331, 1983;

\refis{raste2}A. Pimpanelli, E. Rastelli
and A. Tassi,\journal J. Phys. C., 1 , L2131, 1989.

\refis{raste3} E. Rastelli, L. Reatto and A. Tassi, \journal J. Apply.
Phys. , 55, 1871, 1984.

\refis{poly}A. M. Polyakov, \journal Phys. Lett. , 59B , 97, 1975.

\refis{hald1}F. D. M. Haldane,  {\sl Phys. Lett.} {\bf A 93}, 464 (1983).

\refis{hald}F. D. M. Haldane     , \prl 61, 1029, 1988.

\refis{Doucot}B. Doucot, private communication.

\refis{bask}G. Baskaran, ITP preprint (1989) points out that similar
results might be also be obtained if ``chiral order'' rather than collinear
order exists at this point, provided vortices can be ignored.

\refis{bask2}G. Baskaran, {\sl Phys.
Rev. Lett.} {\bf 63   }, 2524 (1989).

\refis{elser}V. Elser, {\sl Phys. Rev. Lett.}, $\bf 20$, 2405,
(1989).

\refis{blume}M. Blume and Y.Y. Hsieh, \journal J. Appl. Phys., 40,
1249, 1969.

\refis{ritchey}I.  Ritchey and P. Coleman, to be published (1990).

\refis{andreev1}A. F. Andreev and V. I. Marchenko,
{\sl Sov. Phys. Uspekhi} { \bf 23} , 21, (1980).

\refis{andreev2}A. F. Andreev and I. A. Grishchuk,
{\sl Sov.  Phys. J.E.T.P. } { \bf 60} 267 ,  (1984).

\refis{weigman}P. B. Weigman, A. I. Larkin and V. M. Filev, {\sl J.E.T.P.}
{\bf 41} 944 (1975).

\refis{biax}P. B. Weigman, A. I. Larkin and V. M. Filev, {\sl J.E.T.P.}
{\bf 41} 944 (1975).

\refis{krishnamurthy}S. Sarker,
C. Jayaprakash, H. R. Krishnamurthy and
M. Ma, \prb 40, 5028, 1989.

\refis{ons} L.  Onsager,  {\sl J. Am. Chem. Soc.} {\bf 58}, 1486  (1936).

\refis{brout}R. Brout and H. Thomas, {\sl Physics} {\bf 3}, 317 (1967).

\refis{hohn}D. Scalapino and D. Hone, private communication, to be
published.

\refis{affl}I. Affleck, T. Kennedy, E. Leib and H. Tanaka {\sl Phys.
Rev. Lett.} {\bf 59}, 799, (1987).

\refis{twistev1}J. Orenstein, G. A. Thomas, A. J. Millis, S. L. Cooper
et al, {\sl AT\& T Bell Labs preprint}, (1989); T. Imai, T. Shimuzi,
H. Yasuoka, Y. Ueda, K. Yoshimori and K. Kosuge, {\sl Univ Tokyo Solid State
Preprint}, (1989).

\refis{twistev2}G. Shirane,
R. J. Birgenau, Y. Endoh, P. Gehring, M. A. Kastner,
K. Kitazawa, H. Kojime, I. Tanaka, T. R.  Thurston and K. Yamada,
{\sl Phys. Rev. Lett.} { \bf 63 } , 330 (1989).

\refis{gel} M. Gelfand, R. Singh and D. A. Huse, Bell Laboratories preprint
(1989).

\refis{gel2} M. Gelfand, R. Singh and D. A. Huse,
{\sl Phys. Rev. }{\bf B40}, 10801 (1989); R.  Singh and R. Narayan,
{\sl Phys. Rev. Lett.}  { \bf 65 } , 1072 (1990);
E. Dagotto and A. Moreo, {\sl Phys. Rev. Lett.} {\bf 63}, 2148 (1989).

\refis{twist1} The existence of twisted magnetic structures was
predicted simultaneously by Villain, Kaplan and Yoshimori in
the references below:  J. Villain, \journal J. Phys. Chem. Soids, 11,
303, 1959; T. A. Kaplan, \journal Phys. Rev. , 116, 88, 1959; and
A. Yoshimori, \journal J. Phys. Soc. Japan, 14, 807, 1959.

\refis{yoshi}D. Yoshioka, {\sl J. Phys. Soc. Japan} {\bf 58}, 1516 (1989);
D. Yoshioka, in {\sl ``Strong Correlation and Superconductivity''}, editors
H. Fukeyama, S. Makeawa and A. P. Malezemoff, 124 (Springer Verlag 1988);

\refis{yoshi2}D. Yoshioka, {\sl J. Phys. Soc. Japan} {\bf 58}, 3733 (1989).

\refis{twist2}  For a detailed study of twisted helicoidal magnetic
structures see I.E. Dzyaloshinski, \journal JETP, 19, 1960, 1964;
I. E. Dzyaloshinski, \journal JETP, 20, 223, 1965; and
I.E. Dzyaloshinski, \journal JETP, 20, 665, 1965.

\refis{hydro}Note that $S=1/2$ Bose fluids are feasible in
non-relativistic systems, a spin polarized hydrogen is such an example.
The
long wavelength magnetic modes of this triplet paired
fluid are be described by  an SO(3)
sigma model. See E.D. Siggia and A.E. Ruckenstein, \prl 44, 1474, 1980.

\refis{lda} S. Liang, B. Doucot and P.W. Anderson, \prl 61, 365, 1988.

\refis{cdw} R.E. Peierls, {\sl Quantum Theory of Solids}, Oxford Press,
(1955), p. 108; H. Fr\"ohlich, \journal Proc. Royal Soc. , A223, 296, 1954;
A.W. Overhauser, \journal Phys. Rev. , 167 , 691 , 1963.

\refis{cooper}B. R. Cooper, R. J. Elliott, S. J. Nettel and H. Suhl,
\journal Phys. Rev. , 127 , 57 , 1962.

\refis{forster}D. Forster, {\sl ``Hydrodynamics, Fluctuations,
Broken Symmetry,  and Correlation Functions''}, (Benjamin),
p. 163, (1975).

\refis{brinkman}P. W. Anderson and W. F. Brinkman, in {\sl
``Proc. 15th Scottish Summer School in Physics''},
Academic Press(eds J. Armytage and I. Farquhar), 317 (1975).

\refis{kane}C. Kane, P. Lee,  Ng  , N. Read and B. Chakraborty,
M.I.T. preprint (1989).

\refis{dombre}
Note, that if one takes a long-wavelength average of the susceptibilities,
in the magnetization plane,
setting $\gamma^{\mu}_{-}=0$, the large non-
uniform response is effectively removed from the
linear response theory. See
T. Dombre and N. Read, {\sl Phys. Rev.}{ \bf B39},
6797 (1989).

\refis{mila}F. Mila, P. Coleman and P. Chandra, to be published.

\refis{ioffe} L.B. Ioffe and A.I. Larkin, \journal Mod. Phys. B, 2, 203,
1988.

\refis{tak}M. Takahashi, {\sl Prog. Theo. Phys. Suppl.} {\bf 87} 233 (1986).

\refis{tak2}M. Takahashi, {\sl Phys. Rev. } { \bf B40} , 2494 ,(1989).

\refis{cdil} P. Chandra and B. Doucot, \prb 38, 9335, 1988;
L.B. Ioffe and A.I. Larkin, \journal Mod. Phys. B, 2, 203,
1988.

\refis{ccl} P. Chandra, P. Coleman and A. I. Larkin, {\sl J. Phys
Cond. Matt.} {\bf 2} 7933, (1990).

\refis{ising}P. Chandra, P. Coleman and A. I. Larkin,
{\sl Phys. Rev. Lett.} {\bf 64}, 88 (1990).

\refis{siggia}B. I. Shraiman and E. D. Siggia,
{\sl Phys Rev Lett.} {\bf 62}, 1564 (1989).

\refis{shendar} E. Shendar, \journal Sov. Phys. JETP , 56, 178, 1982;
A. G. Gukasov, et al. \journal Europhys. Letter., 7, 83, 1988.

\refis{chubukov} A.I. Chubukov, \journal J. Phys. C., 17, L991, 1984.

\refis{dirty} P.W. Anderson, \journal J. Phys. Chem. Solids, 11, 26, 1959.

\refis{schwinger} J. Schwinger ``On Angular Momentum'' {\sl U.S.
Atomic Energy Comm.} NYO-3071 (1952), unpublished report.

\refis{chandra2} P. Chandra, P. Coleman and A. I. Larkin, {\sl J. Phys
Cond. Matt.} {\bf 2} 7933, (1990).

\refis{dag} E. Dagotto and A. Moreo, {\sl Phys. Rev. Lett.} {\bf 63}, 2148
(1989).

\refis{arovas}D.P.  Arovas and A. Auerbach, Phys. Rev. B. 38, 316 (1988).

\refis{mal}S. V. Mal\'eev, {\sl J. E. T. P} {\bf 59}, 366, (1984);
A. V. Lazuta, S. V. Mal\'eev, and B. P. Toperverg, {\sl Phys. Lett.} {\bf
A65} , 438 (1978);{\sl J.
E. T. P.} {\bf 48} 386, (1978) ; {\sl J.
E. T. P.} {\bf 54} 1113, (1981).

\refis{kl}V. Kalmeyer and R. Laughlin,
{\sl Phys. Rev. Lett.} {\bf 59   }, 2095     (1988).

\refis{wwz} X. G. Wen, F. Wilczek and A. Zee , \prb 39, 11413, 1989.

\refis{dynes}See e.g. M. Gurvitch et al. \prl 63, 1008,  1989. For a
review see M. Lee, A. Kapitulnik \& M. Beasley, in {\sl Mechanisms
of High $T_c$ Superconductivity}, eds. H. Kamimura and A. Oshiyama,
(Springer Verlag), (1989).

\refis{taraphder}A. Taraphder and
P. Coleman,  \prl 66, 2814, 1991.

\refis{fivefriends}C. M. Varma, P. B. Littlewood, S. Schmitt Rink, E.
Abrahams and A.E. Ruckenstein, \prl 63, 1996 , 1989.

\refis{tunnel}M. Lee, A. Kapitulnik, M. R. Beasley, in {\sl ``
Mechanisms of High Temperature Superconductivity''} , (editors
Kawimuar and Oshiyama), Springer (1989).

\refis{us2}P. Chandra and P. Coleman, to be published (1992).

\refis{tsvelik}A. Tsvelik, \prb 42, 10499, 1991.
A. Tsvelik, to be published (1992).

\refis{smalley}R. E. Smalley et al, \journal J. Phys. Chem., 95, 7568, 1991.

\refis{pradeep}T. Pradeep , G. Kulkami, K. Kannan, T. Rau \& C. N. R. Rao
, \journal J. Am. Chem. Soc. , 114,  2272, 1992.

\refis{c60}A. Rosen \& B. W\" astberg, \journal Z. Phys. D., 12, 387, 1989.

\refis{hebard}A. Hebard et al., \journal Nature, 350, 660, 1991.

\refis{pressides}K. Pressides et al, \journal Nature, 354, 462, 1991.

\refis{ramirez}A. Ramirez et al, \prl, 68, 1058, 1992.

\refis{weissman}M. B. Weissman, N. E. Israeloff and G. B. Alers,
U. Illinois preprint, to be published (1991).

\refis{berezin}F. A. Berezin and F. S. Marinov,
\journal Annals of Physics, 104, 336, 1900.

\refis{fradkin} D. Withoff \& E. Fradkin, \prl 64, 1835, 1990.

\refis{science}P. W. Anderson, \journal Science, 235, 1196, 1987.

\refis{micnas}R. Micnas, J. Ranninger and S. Robaszkiewicz, \rmp 62, 113, 1990,
and references therein.

\refis{ranninger}J. Ranninger and S. Robaszkiewicz,
\journal Physica, 135B, 468, 1985; J. Ranninger, S. Robaszkiewicz, A. Sulpice
and
R. Tournier, \journal Europhys. Lett., 3, 3353, 1991.

\refis{andersonmargin}P. W. Anderson, Kathamndu Lectures 1989, Editor Yu Lu
(World Scientific) 1990.

\refis{rev1}Reviews of high temperature superconductivity:
{\sl ``Physical
Properties of High Temperature Superconductors''}, Vols I \& II,
D. M. Ginsberg ed. , World Scientific, Singapore (1989,1990);
{\sl ``High Temperature Superconductivity''} , Proc. Los Alamos Symposium,
1989, eds K . S. Bedell, D. Coffey, D. E. Meltzner, D. Pines and J. R.
Schrieffer, (Addison Wesley), 1990.

\refis{mechanism1}A. Fetter, C. Hanna and R. B. Laughlin, \prb, 39, 9679, 1989.

\refis{mechanism2}C. M. Varma, S. Schmitt-Rink and E. Abrahams,
\journal Solid State Comm., 62, 681, 1987.

\refis{mechanism3}J. M. Wheatley, T. C. Hsu \& P. W. Anderson,
\journal Nature, 333, 121, 1988; P. W. Anderson, G. Baskaran, Z. Zou
and T. Hsu, \prl 58, 2790, 1987.

\refis{mechanism5}J. R. Schrieffer, X. G. Wen \& S. C. Zhang, \prl 60, 944,
1988.

\refis{mechanism4}V. J. Emery, \prl 58, 2794, 1987.

\refis{normalstate1}P. W. Anderson, \prl 64, 1839, 1990 ; P. W. Anderson,
\prl 65, 2306, 1990.

\refis{normalstate2}N. Nagaosa and P. A. Lee, \prl 64, 2450, 1990.

\refis{normalstate3}V. Kalmeyer and R. B. Laughlin, \prl 59, 2095, 1987.

\refis{normalstate4}P. Lederer, D. Poilblanc and T. M. Rice, \prl 63, 1519,
1989.

\refis{normalstate5}A. Virosztek \& J. Ruvalds \prb  42, 4064, 1990;
A. Virosztek \& J. Ruvalds \prl  67, 1657, 1991.

\refis{normalstate6}C. C. Tsuei, D. M. Newns, C. C. Chui and P. C. Pattnaik,
\prl 65 2724, 1990.

\refis{hfreviews}For general reviews,
cf. P. A. Lee, T. M. Rice, J. W. Serene, L. J. Sham and J. W. Wilkins,
\journal Comm. Cond. Matt. Phys, B12, 99, 1986;
P. Fulde, J. Keller, G. Zwicknagel,
\journal Solid State Physics, 41, 1, 1988; H. R. Ott,
\journal Prog. Low. Temp. Physics, Vol X1, 215, 1987.

\refis{millisins}A. J. Millis, {\sl Proc. National High Magnetic Field
Conference}, eds E. Manousakis, P. Schlottmann,
P. Kumar, K. Bedell and F. M. Mueller
(Addison Wesley),  146, (1991).

\refis{doniachins}S. Doniach and P. Fazekas, {\sl Phil. Mag.} , to be
published (1992).

\refis{lacroixins}C. Lacroix and M. Cyrot, \prb 20, 1969, 1970.

\refis{uedains}K. Ueda, H. Tsunetsunga and M. Siegrist,
\prl 68, 1030, 1992.

\refis{fyeins}R. M. Fye and D. J. Scalapino, \prb 44, 7486, 1991.

\refis{coherence1}e.g. T. K. Worthington, W. J.. Gallagher, T. R. Dinger ,
\prl 59, 1160, 1987.

\refis{gorkov2}L. Gorkhov, {\sl Proc. National High Magnetic Field Conference},
eds E. Manousakis, P. Schlottmann , P. Kumar, K. Bedell and F. M. Mueller
(Addison Wesley),  133, (1991).

\refis{gan1} J. Gan, P. Coleman \&
N. Andrei, {\sl J. Cond. Matt} {\bf 20}, 3396, 1991.

\refis{halsey}T. C. Halsey, \prl 55, 1018, 1985.

\refis{kawakami}N. Kawakami, \prb 45, 7525, 1992.

\refis{marchenko} V. I. Marchencko, \journal J.E.T.P. Lett.,  48, 428, 1988.

\refis{coleman}P. Coleman, {\sl Phys. Rev.} {\bf B 29}, 3035, (1984).

\refis{colemanlong}P. Coleman, {\sl Phys. Rev. }
{\bf B 35}, 5073, (1987).

\refis{bickers}G. Bickers, \rmp 59, 845, 1987.

\refis{grewe}N. Grewe, \ssc 50, 19, 1989;
N. Grewe, T. Pruschke \& H. Keiter, \journal Z. Phys. B, 71, 75 , 1988;   N.
Grewe, \journal Z. Phys. B, 67, 323, 1987.

\refis{germans1}P. van Dongen and D. Vollhardt, \prl 65, 1663, 1990;
D. Vollhardt, \journal Physica B, 169, 277, 1991.

\refis{germans2}E. Muller Hartmann, {\sl Int. J. Mod. Phys.}, {\bf 3}, 2169
(1989).

\refis{frogskrauts}A. Georges and W. Krauth, Ecole Normale Superieur Preprint
(1992).

\refis{kotliar1}A. Georges and G. Kotliar, \prb  45, 6479, 1992.

\refis{kotliar2}M. J. Rozenberg, X. Y. Zhang \& G. Kotliar, Rutgers Univ.
Preprint, 1992.

\refis{mills}D. R. Harshman and A. P. Mills, {\sl ``On the nature of
High $T_c$ Superconductivity: Survey of Experimental Properties and
Implications for Interlayer Coupling''}, Bell Labs preprint, 1992.

\refis{haldaneg1}F. D. M. Haldane, \prl 60, 635, 1988;
F. D. M. Haldane, \prl 66, 1529, 1991.

\refis{shastryg}B. S. Shastry, \prl 60, 639, 1988.

\refis{usnow}I. Ritchey, P. Coleman and P. Chandra, submitted to Physical
Review (1992).

\refis{lobb}C. J. Lobb, \journal Physica, B152, 1, 1988.

\refis{pseudo}P. W. Anderson, \journal Phys. Rev. , 112, 1900, 1958.

\refis{iche}G. Iche and A. Zawadowksi, \ssc 10, 1001, 1972.

\refis{nickcat}N. Read, \jpc 18, 2651, 1985.

\refis{shivaram}B. S. Shivaram, Y. H. Jeong, T. F. Rosenbaum and D. G. Hicks,
\prl 56, 1078, 1986.

\refis{louis} G. Goll, H. v. L\"ohneysen, I. K. Yanson and L.
Taillefer, \prl 70, 2008, 1993.

\refis{previous}Short summaries of
this work have been published in outline form in two prior
publications: P. Coleman, E. Miranda and  A. Tsvelik,
{\sl Physica B}, in press (1993); P. Coleman, E. Miranda and  A. Tsvelik,
{\sl Phys. Rev. Lett.}, {\bf 70}, 2960 (1993).

\refis{chevrel}``Superconductivity in ternary compounds'',
Ed. by M. B. Maple and O. Fischer, Berlin, Springer-Verlag, 1982;
O. Fischer, \journal Appl. Phys., 16, 1, 1978.

\refis{wachtersmb6}G. Travagllini and P. Wachter, \prb 29, 893, 1984.

\refis{proximity}S. Han, K. W. Ng, E. L. Wolf, A. Millis. J. L Smith and
Z. Fisk, \prl 57, 238, 1986.

\refis{aeppli3}For a review of Kondo insulators see e.g. G. Aeppli and Z. Fisk,
\journal Com. Mod. Phys. B,
 16, 155, 1992.

\refis{tsvelik}A. M. Tsvelik, \prl 69, 2142, 1992.

\refis{private}E. Abrahams (private communication).

\refis{xtal}Many heavy fermion compounds clearly show Schottky
anomalies in their specific heat where the entropy integral
beneath corresponds to the suppression of magnetic fluctuations
into the higher crystal field states. See for example,
F. Rietschel et al,
\jmmm 76\&77, 105, 1988, R. Felten et al, \journal Eur. Phys. Let.,
2 , 323, 1986.

\refis{berezinskii}V. L. Berezinskii, \journal JETP Lett. , 20, 287, 1974.

\refis{abrahams}A. V. Balatsky and E. Abrahams, \prb 45, 13125, 1992;
E. Abrahams, A. V. Balatsky, J. R. Schrieffer and P. B. Allen,
\prb 47, 513, 1993.

\refis{emery} V.J.  Emery and S. Kivelson, \prb 46, 10812, 1992.

\refis{kirkpatrick}In the context of disordered systems, the possibility
of odd-frequency triplet pairing has been discussed by D. Belitz and
T. R. Kirkpatrick, \prb 46, 8393, 1992.

\refis{mermin}N.D. Mermin, \journal Rev. Mod. Phys., 51, 591, 1979.

\refis{ins1}M. F, Hundley, P. C. Canfield, J. D. Thompson, Z. Fisk
and J. M. Lawrence, \prb 42, 6842, 1990.

\refis{ins12}M. F. Hundley et al, \prb 42, 6842, 1990.

\refis{ins2}F. G. Aliev, V. V. Moschalkov,
V. V. Kozyrkov, M. K. Zalyalyutdinov,
V. V. Pryadum and R. V. Scolozdra, \jmmm 76-77, 295, 1988.

\refis{ins22}F. G. Aliev et al, \jmmm 76-77, 295, 1988.

\refis{ins3}S. Doniach and P. Fazekas, {\sl Phil. Mag. }, {\bf 65B}
1171 (1992).

\refis{weyl}See e.g.
R. Brauer and H. Weyl, \journal Amer. J. Math, 57, 425, 1935.

\refis{kuramoto}Y. Kuramoto and K. Miyake,
\journal Prog. Theo. Phys. Suppl., 108, 199, 1992.

\refis{norman}M. R. Norman, \journal Physica, C194, 203, 1992.

\refis{aeppli}T. Takabatake, M. Nagasawa, H. Fujii, G. Kido,
M. Nohara, S. Nishigori, T. Suzuki, T. Fujita, R, Helfrich, U. Ahlheim,
K. Fraas, C. Geibel, and F. Steglich, \prb, 45, 5740, 1992;
T. Mason, G. Aeppli, A. P. Ramirez, K. N. Clausen, C. Broholm,
N. Stucheli, E. Bucher, T. T. M. Palstra, \prl 69, 490, 1992.

\refis{aeppli2}T. Takabatake et al, \prb, 45, 5740, 1992;
T. Mason et al, \prl 69, 490, 1992.

\refis{history}J. L. Martin, \journal Proc. Roy. Soc., A 251, 536, 1959;
R. Casalbuoni, \journal Nuovo Cimento, 33A, 389, 1976;
F. A. Berezin \& M. S. Marinov, \journal Ann. Phys., 104, 336, 1977.

\refis{linear}U. Rauschwalbe, U. Ahlheim, C. D. Bredl, H. M. Meyer and
F. Steglich \jmmm 63\&64, 447, 1987; R. A. Fisher, S. Kim, B. F. Woodfield,
N. E. Phillips, L. Taillefer, K. Hasselbach, J. Floquet, A. L. Giorgi and
J. L. Smith, \prl 62, 1411, 1989.

\refis{linear2}U. Rauschwalbe et al,\jmmm 63\&64, 447, 1987; R. A. Fisher et
al,  \prl 62, 1411, 1989.

\refis{joynt2}R. Joynt, V. P. Mineev, G. E. Volovik and M. E.
Zhitomirsky, \prb 42, 2014, 1990.

\refis{mermin}N.D. Mermin, \journal Rev. Mod. Phys., 51, 591, 1979.

\refis{ins1}M. F, Hundley, P. C. Canfield, J. D. Thompson, Z. Fisk
and J. M. Lawrence, \prb 42, 6842, 1990.

\refis{ins2}F. G. Aliev, V. V. Moschalkov,
V. V. Kozyrkov, M. K. Zalyalyutdinov,
V. V. Pryadum and R. V. Scolozdra, \jmmm 76-77, 295, 1988.

\refis{ins3}S. Doniach and P. Fazekas, {\sl Phil. Mag. }, {\bf 65B}
1171 (1992).

\refis{kuramoto}Y. Kuramoto and K. Miyake,
\journal Prog. Theo. Phys. Suppl., 108, 199, 1992.

\refis{norman}M. R. Norman, \journal Physica, C194, 203, 1992.

\refis{aeppli}T. Takabatake, M. Nagasawa, H. Fujii, G. Kido,
M. Nohara, S. Nishigori, T. Suzuki, T. Fujita, R, Helfrich, U. Ahlheim,
K. Fraas, C. Geibel, and F. Steglich, \prb, 45, 5740, 1992;
T. Mason, G. Aeppli, A. P. Ramirez, K. N. Clausen, C. Broholm,
N. Stucheli, E. Bucher, T. T. M. Palstra, \prl 69, 490, 1992.

\refis{linear}U. Rauschwalbe, U. Ahlheim, C. D. Bredl, H. M. Meyer and
F. Steglich \jmmm 63\&64, 447, 1987; R. A. Fisher, S. Kim, B. F. Woodfield,
N. E. Phillips, L. Taillefer, K. Hasselbach, J. Floquet, A. L. Giorgi and
J. L. Smith, \prl 62, 1411, 1989.

\refis{stegrev}N. Grewe and F. Steglich in {\sl Handbook on the Physics and
Chemistry of Rare Earths}, (Editor K. A. Gschneider Jr.
and L. Eyring), {\bf 14}, 343, (Elsevier, 1991).

\refis{jon}P. A. Lee, T. M. Rice, J. W. Serene, L. J. Sham and J. W. Wilkins,
\journal Comm. Cond. Mat. Phys., 12, 99, 1986; see also
P. Fulde, J. Keller and G. Zwicknagl, \journal Solid. State Physics, 41, 1,
1988;
for a more general review
of heavy fermion physics, see N. Grewe and F. Steglich, {\sl Handbook on the
Physics and Chemistry of Rare Earths}, eds. K. A. Gschneider and L. Eyring),
{\bf 14},  343, 1991 (Elsevier, Amsterdam).

\refis{affleck1} I. Affleck, Lecture at the Nato Advanced
Study Institute on {\sl Physics, Geometry and Topology}, Banff, August
1989.

\endreferences

\figurecaptions

\noindent {\bf Fig. 1. }Illustrating the elementary ``$Z_2$'' vortex
for a charge $e$ spinor order parameter. Around the vortex, the
phase change of the order parameter is $\pi$. The supercurrent around
this vortex can be removed by introducing a
magnetic flux for which ${e \Phi \over \hbar} = {e \over \hbar} \int \vec A.
d\vec l = \pi$,  so the flux quantum for a charge $e$ spinor is
$\Phi = {\hbar \pi \over e}= {h \over 2 e}$, which corresponds to
the flux quantum of a charge $2e$ complex scalar order parameter.

\noindent {\bf Fig. 2. }(i) Bare Majorana and conduction  propagators, (ii)
Interaction between local moment and conduction electrons.

\noindent {\bf Fig. 3. }Diagrammatic illustration of the pairing equations
showing (i) the spinor vertex between conduction and Majorana spins
(ii) the self consistent equation for the conventional and anomalous
conduction electron propagators.

\noindent{\bf Fig. 4. }Schematic illustration of the quasiparticle
excitation spectrum. The gapful spin excitations are separated
from the gapless band of neutral singlet Majorana excitations by
half the Brillouin zone.

\noindent {\bf Fig. 5. }Quasiparticle spectrum
within mean field theory
for $\mu/D=1/6$. Bold line: gapped ``up'' excitations; dashed
line gapped ``down'' excitations; dotted line: neutral
singlet Majorana band. Inset,  density of
states for up electron bands.

\noindent {\bf Fig. 6. }Illustrating the distinction between
(i) the Hartree/RPA decoupling scheme used and (ii) a Hartree-Fock
decoupling procedure. The ``Fock'' part of the vertex equation is
absent from the Hartree approximation, but is reincorporated as
a leading term in the RPA.

\noindent {\bf Fig. 7. }Temperature dependence of the specific
heat, calculated at low temperatures for a variety of $\mu/D$.
Note that as $\mu/D$ increases, the gapless band becomes ``heavy''.

\noindent {\bf Fig. 8. }Mean field free energy of the odd-frequency
paired state plotted as a function of the ``twist'' wave vector
of the triplet pairing field $\hat d_c(\vec R) = e^{i \vec Q \cdot \vec R}
\hat d_c(0)$. The uniform state ($Q=0$) is unstable with respect to the state
with a staggered phase ($Q=\pi$).

\noindent {\bf Fig. 9. }Conduction electron
NMR relaxation
rate, computed using our toy model for the same sequence of
of $\mu/D$ values shown in Fig. 7.
Even though the model
predicts a linear specific heat(Fig. 7.),  the coherence factors give rise to
an NMR relaxation rate normally be associated with
lines, rather than surfaces of gapless excitations.
There is no
Hebel-Slichter peak below the mean-field transition temperature.

\noindent {\bf Fig. 10. }Local dynamic spin susceptibility of the odd frequency
state for various values of $\mu/D$. Inset details
mid-gap response that grows quadratically
with the frequency, due to linear spin coherence factors.

\noindent {\bf Fig. 11. }Dependence of mean field free energy
on the magnetic wave vector $\vec P$ in three dimensions.
In three dimensions, the state at $\vec P=(\pi,0,0) $ is locally
stable.

\noindent {\bf Fig. 12. }Illustrating the energy dependence on the flux
through a ring of odd frequency triplet superconductor (no spin
anisotropy). The phase $\phi$ displayed here refers to the phase
of the charge $2e$ composite order parameter
$
g{\cal M}_{\alpha}^{\ \beta}(x)=\la \tau_{\a}(x)S^{\be}(x)\ra
$.
When  the flux through
the loop exceeds one flux quantum, the system can relax
the energy and supercurrent by rotating the axes of the order parameter
into the third dimension, creating two  $Z_2$ antivortices.
This reduces the effective flux through the ring by 2 flux quanta, changing
the sign of the supercurrent $J= - \partial E/\partial \phi$ and
producing a saw-tooth dependence of current on flux.
The ground-state energy is
a periodic, rather than a quadratic function of the applied flux.
A macroscopic Meissner current can not develop in response to the
flux threading the loop unless spin anisotropy is added to prevent the
rotation of the order parameter into the third dimension.

\endfigurecaptions

\endit
\end